\documentclass[a4paper,fleqn,usenatbib,dvipdfmx]{mnras}
\usepackage{newtxtext,newtxmath}

\usepackage[T1]{fontenc}
\usepackage{ae,aecompl}
\usepackage{graphicx}	
\usepackage{amsmath}	
\usepackage{color}	
\usepackage{natbib}

\usepackage{pifont}
\usepackage{mathrsfs}
\usepackage{bm}

\usepackage {afterpage}
 \newcommand{ \Mearth }{M_{\oplus}}
 \newcommand{ \Rearth }{R_{\oplus}}

\usepackage{ulem}
\title[Hydrodynamic Escape of Mineral Atmosphere]{Hydrodynamic Escape of Mineral Atmosphere from Hot Rocky Exoplanet. I. Model Description}

\author[Y. Ito \& M. Ikoma]{
Yuichi Ito$^{1,2}$\thanks{E-mail: yuichi.ito.kkyr@gmail.com}
and Masahiro Ikoma$^{1,3}$
\\
$^{1}$Department of Earth and Planetary Science, The University of Tokyo, 7-3-1 Hongo, Bunkyo-ku, Tokyo 113-0033, Japan\\
$^{2}$Department of Physics and Astronomy, University College London, Gower Street, London, WC1E 6BT, United Kingdom \\
$^{3}$Research Center for the Early Universe (RESCEU), The University of Tokyo, 7-3-1 Hongo, Bunkyo-ku, Tokyo 113-0033, Japan\\
}
\date{Accepted XXX. Received YYY; in original form ZZZ}
\pubyear{2020}

\begin{document}
\label{firstpage}
\pagerange{\pageref{firstpage}--\pageref{lastpage}}
\maketitle

\begin{abstract}
Recent exoplanet statistics indicate that photo-evaporation has a great impact on the mass and bulk composition of close-in low-mass planets. While there are many studies addressing photo-evaporation of hydrogen-rich or water-rich atmospheres, no detailed investigation regarding rocky vapor atmospheres (or mineral atmospheres) has been conducted. 
Here, we develop a new 1-D hydrodynamic model of the 
UV-irradiated mineral atmosphere composed of Na, Mg, O, Si, their ions and electrons, includin
molecular diffusion, thermal conduction, photo-/thermo-chemistry, X--ray and UV heating, and radiative line cooling (i.e., the effects of the optical thickness and non-LTE). The focus of this paper is on describing our methodology but presents some new findings. 
Our hydrodynamic simulations demonstrate that almost all of the incident X-ray and UV
energy from the host-star is converted into and lost by the radiative emission of the coolant gas species such as Na, Mg, Mg$^+$, Si$^{2+}$, Na$^{3+}$ and Si$^{3+}$. 
For an Earth-size planet orbiting 0.02~AU around a young solar-type star, we find that the X-ray and UV heating efficiency is as small as $1 \times 10^{-3}$, which corresponds to 0.3~$\Mearth$/Gyr of the mass loss rate simply integrated over all the directions.
Because of such efficient cooling, the photo-evaporation of the mineral atmosphere on hot rocky exoplanets with masses 
of $1\Mearth$ is not massive enough to exert a great influence on the planetary mass and bulk composition. 
This suggests that close-in high-density exoplanets with sizes larger than the Earth radius survive in the high-UV environments. 

\end{abstract} 



\section{Introduction}\label{sec: int}
Atmospheric escape is an important process that changes the mass and bulk composition of close-in exoplanets, which are highly irradiated by stellar X-rays and ultra-violet (UV)  (collectively called X+UV, hereafter). 
For example, the California-{\it Kepler} exoplanet survey has revealed that there are two dominant clusters of exoplanets with periods of $<100$ days: The radii of planets in one cluster are $R_p<1.5 $~$\Rearth$ and those in another cluster are $R_p=2-3$~$\Rearth$ \citep{Fluton+2017}.
Some theories demonstrate that such a bimodal size distribution is an outcome of atmospheric escape  \citep{Owen+2017,Jin+2018}; that is, the former  cluster consists of bare rocky planets that have lost their primordial hydrogen-rich atmospheres.
Similar bimodality is predicted theoretically for water-rich planets \citep{Kurosaki+13,Lopez17}, although its boundary seems inconsistent with the observed distribution above.

An interesting question is whether rocky vapor atmospheres are stable or not in strong stellar X+UV environments.
Some super-Earths detected so far are dense and hot enough that they are likely to be rocky planets with rocky vapor atmospheres on top of magma oceans, which include CoRoT-7~b  \citep[][]{Queloz+09} and Kepler-78~b \citep[][]{Sanchis-Ojeda+13}.
If such hot rocky exoplanets (HREs, hereafter) have no highly volatile elements such as H, C, N, S, and Cl, their atmospheres are composed mainly of Na, O$_2$ and SiO \citep[e.g.,][]{Schaefer+09,Miguel+11,Ito+2015}  \citep[see][for the volatile--rich cases]{Schaefer+12}.
Same as in \citet{Ito+2015}, we call such an atmosphere the mineral atmosphere in this study.

Hydrodynamic escape of highly-irradiated atmospheres is known to occur in an energy-limited fashion \citep[][]{Sekiya+1980,Sekiya+1981,Watson+1981}; 
the mass loss rate, $\dot{M}$, is given by
\begin{eqnarray}
\dot{M} = \epsilon \frac{R_p}{GM_p} \pi F_\mathrm{X+UV}R_\mathrm{X+UV}^2,
\label{eq: els}
\end{eqnarray}
where $G$ is the gravitational constant, $M_p$ and $R_p$ are the planetary mass and radius, respectively, $F_\mathrm{X+UV}$ is the incident X+UV flux, $R_\mathrm{X+UV}$ is the effective radius for X+UV absorption and  $\epsilon$ is the net X+UV heating efficiency that is defined as the ratio of the net heating rate (i.e., stellar-energy absorption minus cooling) to the stellar-energy absorption rate.
Given $R_\mathrm{X+UV} \simeq R_p$, most of the physics and atmospheric properties are hidden in $\epsilon$, which is determined by the heating and cooling processes in the atmosphere. 
Combined simulations of atmospheric hydrodynamics and photo-chemistry demonstrate that $\epsilon$ $\sim$ 0.03--0.4 for hydrogen-rich atmospheres, including hot-Jupiter envelopes, and oxygen-rich terrestrial atmospheres \citep[e.g.,][references therein]{Tian2015}.

\citet{Valencia+10} examined the stability of a possible mineral atmosphere of CoRoT-7~b as an example of HREs. 
They first estimated that the equilibrium vapor pressure is high enough and, thus, the atmosphere is thick enough to absorb the incident stellar UV completely. 
Then, assuming the energy-limited hydrodynamic escape with $\epsilon = 0.4$, they demonstrated that CoRoT-7~b could have lost several Earth masses via the escape, provided the planet had been born as a bare rocky planet. 
However, such a substantial mass loss may not be consistent with the \textit{Kepler}'s discovery of numerous close-in Earth- and Super-Earth-sized exoplanets, which was made later.

As yet, however, we have a poor understanding of the heating and cooling processes in the mineral atmosphere and, thus, no exact idea about the correct value of $\epsilon$. 
When it comes to hydrogen-rich atmospheres, there are many studies about such heating/cooling processes done in the context of photo-evaporation of hot Jupiters.
In such atmospheres, the infra-red (IR) emission by H$^{3+}$ is the dominant cooling process, in addition to advective cooling, while heating occurs via photo-dissociation and ionization of hydrogen due to stellar EUV irradiation \citep[e.g.,][]{Yelle2004,GM2007b}.
\citet{Murray-Clay+2009} showed that hydrogen Lyman-$\alpha$ line (H-Ly$\alpha$) emission becomes the dominant cooling mechanism in hydrogen-rich atmospheres in extremely high EUV conditions like for close-in planets around T Tauri stars.

Similarly to the H-Ly$\alpha$ cooling for hydrogen-rich atmospheres,
radiative cooling of the sodium D line (Na-D) could be a dominant cooling mechanism in sodium-rich mineral atmospheres. 
Electronic transition from the 3p to the 3s level of sodium for 
Na-D is characterized by the large value of the Einstein coefficient for spontaneous emission, $A \simeq 6 \times 10^7$~s$^{-1}$, and the low excitation energy, $E_{\rm{ex}} \simeq 2.1$~eV, 
whereas $A$ $ \simeq 6 \times 10^8$~s$^{-1}$ and $E_{\rm{ex}}\simeq 10.2$~eV for H-Ly$\alpha$ (see NIST Database\footnote{\url{https://www.nist.gov/pml/atomic-spectra-database}}). 
Provided radiation is emitted by gas in collisional equilibrium,
namely, the atmospheric gas is optically thin
in a local thermal equilibrium (LTE),
the radiative cooling rate is simply given by 
$A E_{\rm{ex}}\exp(-E_{\rm{ex}}/T)$, where $T$ is the temperature in eV.
In such an ideal condition, the Na-D cooling rate is approximately $10^{12}$, $10^6$ and $10^2$ times as high as the H-Ly$\alpha$ cooling rate at $T=$ 3000, 5000, and 10000~K, respectively. 
This simple analysis suggests that the hydrodynamic escape of mineral atmospheres is less massive compared to that of hydrogen-rich atmospheres.
However, the condition of LTE and extremely small optical thickness is not always achieved in the atmospheric region where hydrodynamic escape is driven.

This study is aimed to clarify the cooling/heating processes and stability of the sodium-rich mineral atmosphere on top of a magma ocean of a volatile-free HRE.  
To do so, we develop a theoretical model of hydrodynamic escape,
incorporating the detailed photochemical processes. 
Then, we derive the values of the net X+UV heating efficiency, $\epsilon$, and the mass loss rate. 
The focus of this paper is on describing our methodology, although we show some new findings regarding the escape of the mineral atmosphere. 
In this paper we consider only 1-$\Mearth$ HREs; Dependence on planetary mass and other parameters will be investigated in detail in our forthcoming paper.

The rest of this paper is organized as follow:
In Sec.~\ref{sec: mod}, we describe the hydrodynamic model of the mineral atmosphere of an HRE.
In Sec.~\ref{sec: mv}, we perform some benchmark tests of our model for isothermal transonic escape and hydrodynamic escape of a hydrogen-rich atmosphere.
In Sec.~\ref{sec: r}, we present results of our new hydrodynamic calculations, which include the profiles of density, temperature and velocity as well as energy budget in the mineral atmosphere. 
Finally, we discuss the X+UV heating efficiency of the mineral atmosphere based on our findings and the caveats of our model in Sec.~\ref{sec: dis}, and summarize our results in Sec.~\ref{sec: c}

\section{Model}\label{sec: mod}
 \begin{figure}
\begin{center}
\includegraphics[width=\columnwidth]{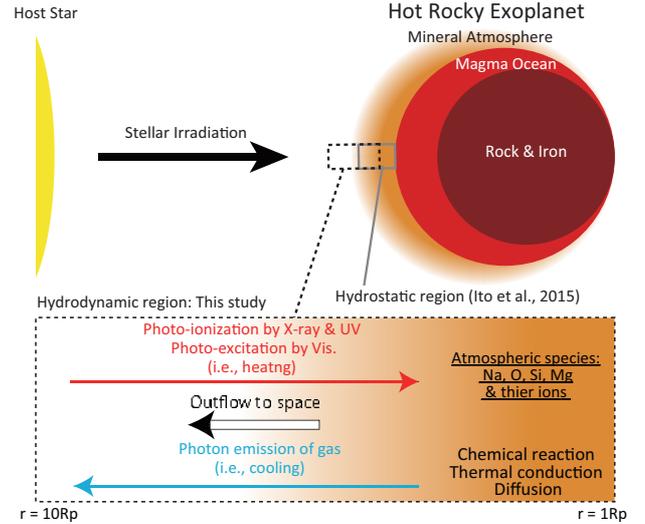}
\\
\caption{
Schematic illustration of the hot rocky exoplanet (upper panel) and processes considered in this study (lower panel). Note that we explore only the hydrodynamic part of the mineral atmosphere and give the lower boundary conditions according to \citet{Ito+2015}.
See the text for the details.
}
\label{fig:model}
\end{center}
\end{figure}

We construct a one-dimensional hydrodynamic model of an X+UV-irradiated, mineral atmosphere of an HRE orbiting a G-type star (see Fig.~\ref{fig:model}). 
We find steady-state solutions of transonic, hydrodynamic gas flow of the atmosphere in the stream-tube along the line connecting the planetary and stellar centers. 
We integrate the hydrostatic structure of the lower atmosphere separately, following \citet{Ito+2015}, to determine the lower boundary conditions at $r$ = $R_p$ for the hydrodynamic upper atmosphere.

The lower atmosphere is assumed to be always in the thermal and chemical equilibrium with the underlying magma ocean.
That is, the supply of gas from the magma ocean occurs immediately to make up for the loss by the atmospheric escape \citep{Valencia+10}. 
Following previous chemical-equilibrium calculations for mineral atmospheres \citep{Schaefer+09,Miguel+11,Ito+2015}, we assume that the magma is the same in composition as the bulk silicate Earth (BSE) without highly volatile elements such as H, C, N, S, and Cl.
Although such simulated mineral atmospheres contain various gas species, 
we consider only the major species in this study, for simplicity.
From the chemical point of view, 
monoatomic gases would be major species in the upper atmosphere where  
the absorption efficiency of X+UV and temperature are high enough that polyatomic molecules are completely dissociated and density is too low for two-body recombination to occur efficiently. 
Thus, we consider Na, O and Si, 
which are the most abundant chemical elements in the upper atmosphere with a substellar-point equilibrium temperature of $\gtrsim$ 2200~K \citep[see Fig. 5 in][]{Ito+2015}.
Also, in addition to the alkali metal Na,
we consider the alkaline-earth metal Mg, which possibly becomes a strong coolant like Na because its ions behave like alkali metal atoms in the sense that a single electron exists in the outermost shell.
In summary, 
we consider Na, O, Si, Mg, their ions (up to  $+$4 charge) and electrons as the constituents of the upper atmosphere. 
The calculations of energy budget shown below, which is of particular interest in this study, confirm that 
these components except for oxygen and its ions make important contribution to the heating and cooling rate in the mineral atmosphere.
Also, although K and Fe are more abundant than Mg in the atmosphere \citep{Ito+2015}, they are not considered in this model, for simplicity. 
Note that since their cooling effects in the atmosphere are  smaller than Na based on a simple analysis for cooling rates in LTE and optically thin conditions (see  Sec.~\ref{sssec: kfe}), these would affect the energy budget in the atmosphere only slightly.

While planetary and stellar magnetic fields would affect the motion of ionic gases and electrons, 
we focus on investigating the thermal-escape process and neglect the effects in this paper; 
that is, we assume that the atmosphere is neutral as a whole (ambipolar constraint). 
Also, all the gas species including neutrals, ions and electrons are assumed to be the same in temperature, which is also assumed in hydrodynamic simulations for hot Jupiters \citep[e.g.,][]{Yelle2004,GM2007b}.

In the rest of this section, we describe the basic equations (Section \ref{ssec:he}), chemical reactions (Section \ref{ssec:cr}), heating/cooling processes (Section \ref{ssec:hc}), diffusion/conduction (Section \ref{ssec:dc}), boundary conditions (Section \ref{ssec: bc}) and numerical scheme (Section \ref{ssec:ns}) in this model.

\subsection{Hydrodynamic Equations}\label{ssec:he} 
Following previous models of hydrodynamically escaping atmospheres \citep[e.g.,][]{Watson+1981,GM2007b,Tian+2008}, 
we integrate the equations of continuity, motion and energy for a spherically symmetric flow of an inviscid, multi-component gas:
\begin{equation}
     \frac{\partial}{\partial t} \left( r^2 \rho_s \right)
     =  -\frac{\partial}{\partial r}\left[ r^2 \rho_s (u+u_{s}) \right] +r^2 \dot{\rho_s},\  
               \label{eq: hyd1}
\end{equation}
\begin{equation}
     \frac{\partial}{\partial t} \left( r^2 \rho u \right)
     =  -\frac{\partial}{\partial r}\left[ r^2 (\rho u^2 + P) \right] +r^2 \rho f_{\rm{ext}} +2Pr,
               \label{eq: hyd2}
\end{equation}
\begin{equation}
     \frac{\partial}{\partial t} \left( r^2 \rho E \right)
     =  -\frac{\partial}{\partial r}\left[r^2 (\rho E+ P)u + r^2q \right] + r^2 (\rho u f_{\rm{ext}}+Q_{\rm{net}}),
          \label{eq: hyd3}
\end{equation}
where $t$ is time and $r$ is the radial distance from the planetary center.
All the other quantities are functions of $r$ and $t$: 
$\rho_s$ and $\dot{\rho}_s$ are respectively the mass density and net mass production rate per unit volume of gas species $s$, 
$\rho$ is the total mass density (i.e.,
$\rho= \sum_{s\in \mathscr{S}}\rho_s$ 
$=\sum_{s\in \mathscr{S}} m_s n_s$,
where
$m_s$  and $n_s$ are the mass and number density of species $s$, respectively, and 
$\mathscr{S}$ represents the set of gas species, 
namely, $\mathscr{S}$ = Na, Na$^+$, Na$^{2+}$, etc.),
$u$ is the bulk velocity, $u_s$ is the diffusion velocity of species $s$ (or the component of a velocity difference caused by diffusion),
$P$ and $E$ are the pressure and
the total energy density (or the sum of the kinetic energy density and the specific internal energy of the bulk gas flow), respectively,
$f_{\rm{ext}}$ is the external force, $q$ is the heat flux, and $Q_{\rm{net}}$ is the net energy deposition rate.

We assume that the atmosphere consists of perfect monoatomic gases.
The total energy density is given by
\begin{equation}
	E = \frac{1}{2}u^2+\frac{1}{\gamma-1}\frac{P}{\rho}, 
	\label{eq: total energy}
\end{equation}
and the sound speed is $c_S = \sqrt{\gamma {P}/{\rho}}$ with  
the heat capacity ratio $\gamma=$ 5/3. 

The external force $f_{\rm{ext}}$ is given by
\begin{equation}
f_{\rm{ext}}= - \frac{GM_p}{{r}^2}+\frac{GM_*}{(a-r)^2}
-\left(\frac{M_*}{M_*+M_p} a-r  \right) \frac{G(M_*+M_p)}{a^3},
\label{eq: fext}
\end{equation}
where $a$ is 
the separation between the planetary and stellar centers, 
$M_p$ and $M_*$ are the planetary and stellar masses, respectively. 
The first, second, and third terms on the right-hand side represent the planetary gravity, the stellar gravity, and the centrifugal force,  respectively. 
The last two terms are collectively termed the tidal force.

Finally,
the quantities $\dot{\rho_s}$, $Q_{\rm{net}}$, $u_s$ and $q$ are determined from the models of chemical reaction, heating/cooling processes and diffusion/thermal conduction, which are described below.

\subsection{Chemical Reaction}\label{ssec:cr} 
Since the atmosphere considered in this study is composed of atoms and their ions, 
ionization and recombination dominate the atmospheric chemistry. 
Here, we consider reactions relevant both to radiative and thermal processes.
All those processes except inverse ones are listed in Table~\ref{tbl:CHEMT}:
Such inverse processes are the thermal recombination of each ion (i.e., three-body recombination opposite to the thermal ionization reactions named TI 1--16 in Table~\ref{tbl:CHEMT}).
For the thermal recombination, we assume an electron as the third body that receives the reaction heat and 
calculate the reaction rate constant of thermal recombination from that of thermal ionization and the equilibrium constant given by the Saha's ionization equation \citep[Eq. 9.45 in][]{Rybicki+86}. 
We ignore direct exchange of electric charges between atoms and ions for simplicity, since those reaction rates are unavailable for almost all of the species considered in this model.
Note that, instead of the charge exchange, the thermal ionization and recombination reactions of each atom or ion with electrons lead to removing and adding their electrons, and then these two processes consequently exchange the charge between different atoms and ions in this model.

Consider a photo-ionization process such that an $i$-times charged ion of chemical element $Y$, which is denoted by $Y^i$, leads to a further loss of $x$ electrons and produces an ion with an $i+x$ charge (i.e., $Y^{i+x}$). 
Using the symbol $n(Y^i)$, instead of $n_s$, the rate of photo-ionization from $Y^i$ to $Y^{i+x}$ is given by 
\begin{eqnarray}
\frac{dn(Y^{i+x})}{dt} = {n(Y^i)} \int F^*_\lambda \exp(-\tau_\lambda) \,
\sigma_{Y^i, \lambda} \, \eta_{i \rightarrow (i+x), \, Y, \lambda} \, 
 d\lambda ,
  \label{eq:photoi}
\end{eqnarray}
where  
$F^*_\lambda$ is the emergent photon flux from the host star, 
$\tau_\lambda$ is the optical depth at wavelength $\lambda$ measured from infinity,
$\sigma_{Y^i, \lambda}$ is the photo-ionization cross section of species $Y^i$,
$\eta_{i \rightarrow (i+x), Y, \lambda}$ is the probability that a collision with a photon results in removing $x$ electrons from $Y^{i}$ and yielding $Y^{i+x}$.

\begin{figure}
\begin{center}
\includegraphics[width=\columnwidth]{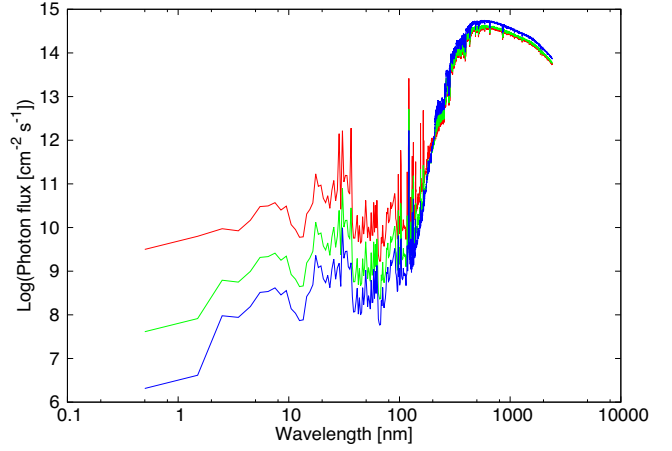}
\\
\caption{
Spectrum models of the host star from \citet{Claire+2012}.
The photon fluxes at 1~AU from the host star with age of 0.1~Gyr (red), 1~Gyr (green) and 4.56~Gyr (blue) are shown as functions of wavelength. The host star is assumed to be a solar-type star. 
}
\label{fig:SUV}
\end{center}
\end{figure}

As for the stellar spectrum $F^*_\lambda$, we adopt the solar-type star models that \citet{Claire+2012} developed by combining the observed spectra of the Sun and solar analogs of different ages. 
Three model spectra of different stellar ages 0.1, 1, and 4.56~Gyr are presented in Fig.~\ref{fig:SUV}. 
The photo-ionization cross section $\sigma_{Y^i, \lambda}$ is given by \citet{Verner+1995}. 
As for $\eta_{i \rightarrow (i+x), Y, \lambda}$, we consider single and multiple electron emission that occurs
when X+UV removes an electron from an atom and then such electron emission from an inner-shell of the atom brings about further emission of electrons (which is termed the Auger effect). 
Since we consider only up to 4$+$ ions in this study,
we use the form of $\eta_{i \rightarrow (i+x), Y ,\lambda}$ given by \citet{Kaastra+1993} for $i+ x \leq$ 4 
and assume $\eta_{i \rightarrow (i+x), Y ,\lambda}$ = 0 for $i+x \geq 5$. 
Note that the function is normalized so that the sum of $\eta_{i \rightarrow (i+x), Y, \lambda}$ is unity.

To integrate Eq.~(\ref{eq:photoi}),
we consider 23~spectral intervals in the wavelength region between 0.1 and 250~nm in such a way 
\begin{equation}
	\lambda_k = \left\{
	\begin{array}{ll}
		0.1~\mathrm{nm} & (k = 1) \\
		1.0~\mathrm{nm} & (k = 2) \\
		1.0 + 2.0 (k-2)~\mathrm{nm} & (3 \leq k \leq 14) \\
		25 \times 10^{0.1 (k-14)}~\mathrm{nm} & (15 \leq k \leq 24)
	\end{array} \right.
\end{equation}
In each bin of width $\Delta \lambda_k$ ($\equiv \lambda_{k+1} - \lambda_k$), we calculate the averaged value of the photo-ionization cross section,  
$\bar{\sigma}_{s, k}$, as
\begin{eqnarray}
\bar{\sigma}_{s, k} = \int_{\lambda \in \Delta \lambda_k} \sigma_{s,\lambda} 
\frac{F^*_{\lambda}}{\langle F_\lambda^\ast \Delta \lambda \rangle_k} d\lambda,
\label{eq:asig}
\end{eqnarray}
where $\langle F_\lambda^\ast \Delta \lambda \rangle_k$ is the photon flux in the $k$th bin given by 
\begin{eqnarray}
\langle F_\lambda^\ast \Delta \lambda \rangle_k
= \int_{\lambda \in \Delta \lambda_k} F^*_{\lambda} d\lambda.
\label{eq:fband}
\end{eqnarray}
Note that we simply use the value of $F^*_{\lambda}$ calculated at $\lambda$ = 0.5~nm over 
wavelength between 0.1 and 0.5 nm, where the data of $F^*_\lambda$ are unavailable in \citet{Claire+2012}.
Likewise, the averaged photon energy at the $k$th bin is given by 
\begin{eqnarray}
\frac{hc}{\bar{\lambda}_k} = \int_{\lambda \in \Delta \lambda_k} \frac{hc}{\lambda} 
{\frac{F^*_\lambda}
{\langle F_\lambda^\ast \Delta \lambda \rangle_k}
}
d\lambda,
\end{eqnarray}
where $c$ is the light velocity and $h$ is the Planck constant. 
Note that while we use the above averaged values, $\bar{\sigma}_{s, k}$ and $\bar{\lambda}_k$, at each bin below, we do not indicate so for the sake of shorthand.

Finally, the net mass production rate,
$\dot{\rho_s}$ in Eq.~(\ref{eq: hyd1}), is calculated from reactions tabulated in Table~\ref{tbl:CHEMT}. 
For instance, in the case of a reaction of two-body reactants $s_1$ and $s_2$ and a single product $s_3$ with a rate coefficient  $k_{r}$, 
the mass production rate of species $s_3$ is given by $\dot{\rho}_{s_3}=m_{s_3}k_r n_{s_1}n_{s_2}$.
 
\subsection{Heating and Cooling Processes}\label{ssec:hc} 
In this model, we consider the absorption of stellar X+UV by atoms and ions via photo-ionization.
The absorbed energy is used for removing the outermost electrons (i.e., normal ionization) and also the inner-shell electrons, which induces characteristic X-ray emission, and is also partitioned into the thermal energy.
In addition, we consider chemical reactions and radiative transitions as other heating and cooling processes.
Thus, the net energy deposition rate, $Q_{\rm{net}}$, is given by
\begin{eqnarray}
Q_{\rm{net}} = Q_{\rm{X+UV}} - Q_{\rm{Ei}} - Q_{\rm{X}}- Q_{\rm{chem}} - Q_{\rm{rad}},
\label{eq: qnet}
\end{eqnarray}
where 
$Q_{\rm{X+UV}}$ is the total energy of X+UV absorbed per unit time given by
\begin{eqnarray}
Q_{\rm{X+UV}}  =  \sum_Y \sum_{i=0}^3 n(Y^i) \int \frac{hc}{\lambda} F^*_\lambda \,
e^{-\tau_\lambda} \,
\sigma_{Y^i, \lambda} \, 
d\lambda ;
\end{eqnarray}
$Q_{\rm{Ei}}$  is the energy loss rate via photo-ionization, which is given by the sum of the photo-ionization rates given in Eq.~(\ref{eq:photoi}) times corresponding ionization energies;
$Q_\mathrm{X}$ is the energy loss rate via characteristic X-ray emission, the probability of which is calculated from \citet{Kaastra+1993} with the assumption that the characteristic X-ray is lost completely from the atmosphere to space, and, thus 
\begin{eqnarray}
Q_{\rm{X}} = 
	 \sum_Y \sum_{i=0}^3 \sum_{x=1}^{4-i} n(Y^i)
	 \int
	\left(\sum_k \eta_{\rm{X, \lambda}}^k E_{\rm{X}, \lambda}^{k}\right)	
 F^*_\lambda e^{-\tau_\lambda} \,
\sigma_{Y^i, \lambda} \, 
d\lambda,
\end{eqnarray}
where $\eta_{\rm{X, \lambda}}^k$ and $E_{\rm{X, \lambda}}^{k}$ are the  emission probability and energy  of  the $k$th characteristic X-ray \citep[see][]{Kaastra+1993};
$Q_{\rm{chem}}$ is  the endothermic/exothermic rate of thermo-chemical reactions (i.e., reactions except photo-ionization and radiative recombination), whose ionization energy data are taken from the NIST database\footnote{\url{https://physics.nist.gov/PhysRefData/ASD/ionEnergy.html}};
$Q_{\rm{rad}}$ is the net radiative energy absorbed/emitted per time by energy level transition. 

We define the primary X+UV heating rate, $Q'_{\rm{X+UV}}$, as
\begin{eqnarray}
Q'_{\rm{X+UV}} = Q_{\rm{X+UV}} - Q_{\rm{Ei}} - Q_{\rm{X}},
\label{eq:qdash}
\end{eqnarray} 
and, hereafter, use this quantity, instead of $Q_\mathrm{X+UV}$, namely,
\begin{eqnarray}
Q_{\rm{net}} = Q'_{\rm{X+UV}} - Q_{\rm{chem}} - Q_{\rm{rad}}.
\end{eqnarray}

\subsubsection{Energy level transition}
As for $Q_{\rm{rad}}$, we consider atomic-line radiation, taking the effects of non-local thermodynamic equilibrium (non-LTE) into account in the following way.
First, since energy-level transitions occur frequently enough on the hydrodynamic timescale of interest, we assume that the populations of energy levels are always
in statistical equilibrium such that both collisional-radiative excitation and de-excitation balance each other. 
We consider all the energy levels below 10~eV or below the one from which
a permitted radiative transition occurs (the Einstein coefficients being taken
from NIST MCHF/MCDHF database\footnote{\url{http://nlte.nist.gov/MCHF/}}).
Table~\ref{tbl:ELG} shows the energy levels of each species used in this model. 
Note that 
the radiative lines of Na$+$, Na$^{2+}$, Mg$^{2+}$ and Mg$^{3+}$ are ignored in this model (see Table~\ref{tbl:ELG}) since they would not be effective compared to the others. 
This is because their first excited levels have very high excitation energy over 20~eV as their electron configurations are like those of noble gases and halogens; that is, they are hardly excited via collisions and then hardly play a dominant role in cooling.

Then, for taking the effects of radiative transfer approximately into account, 
we adopt the escape probability (EP) method \citep[e.g.,][]{Iron1978,Rybicki1984}, 
which assumes that the local emission and absorption line profiles are equal to each other and 
that the source function is the same everywhere in a medium. 
The former assumption is generally valid in the cores of strong lines. 
The latter is not exactly valid in planetary atmospheres because of inhomogeneous
temperature and line profiles, although
being often adopted in modeling of interstellar clouds \citep[e.g.,][]{Osterbrock+1989} and  hydrodynamic simulations for the photo-evaporation of protoplanetary disks and hydrogen-rich atmospheres of close-in exoplanets \citep[e.g.,][]{Ercolano+2008,Owen+2010,Salz+2016}. 
The latter assumption, however, brings about no significant error.
\citet{Dumont+2003} compared
the EP method with the so-called accelerated lambda-iteration method \citep[e.g.,][]{Hubeny2001}, which is one of the most efficient and secure line transfer methods, for the typical AGN or X-ray binary emission medium. 
They showed that results obtained from the two methods differed by tens of percent in computed energy balance, 
indicating that the EP method would cause only about 10~\% error in temperature in LTE regions, because temperature is proportional to the fourth root of emission intensity in LTE. 
Also, the EP method is valid in wide, optically thin, non-LTE regions, because it yields the exact solution in extremely optically thin media. 
To estimate the error of the EP method in optically thick non-LTE regions,  
one has to compare it to other efficient and secure line transfer methods; evaluating such an error is beyond the scope of this study. 

In the EP method, 
the local mean intensity integrated over frequency around the atomic line, $J$ 
($=\int d\Omega \int I_\nu \, d\nu \, / 4\pi$), 
which is needed to obtain radiative transition rates, is then given by \citep{Hubeny2001}
\begin{equation}
J = (1 - {P_e^\prime}) S,
\label{eq: ep1}
\end{equation}
where 
$P_e^\prime$ is the probability for photons to escape from the atmospheric gas (termed the escape probability or EP), 
which depends on the optical depth, $\tau$, and 
$S$ is the source function given by
\begin{equation}
 S = \frac{2h\nu_0^3}{c^2} \left( \frac{g_u n_l}{g_l n_u}-1\right)^{-1}, 
\end{equation}
$n_u$ and $n_l$ are the number densities of the upper and lower energy levels, respectively, 
$g_u$ and $g_l$ are the degeneracies of the upper and lower levels, respectively, 
and $\nu_0$ is the frequency of the atomic line between the upper level $u$ and lower level $l$. 
In this study, we adopt the formula of $P_e^\prime$ that \citet{Chatzikos+2013} derived 
considering the absorption effect of an external radiation field as 
 \begin{eqnarray}
P_e'&=&P_e(1- J_{\rm{ext}}/S),
\label{eq: ep2}
\end{eqnarray}
where $P_e$ is the EP without external field 
and $J_{\rm{ext}}$ is the mean intensity in the external radiation field.
In this study, we assume that the line radiation from ions propagates both upwards and downwards, 
whereas that from neutral atoms propagates only upwards. 
This is because the lower atmosphere is composed mainly of neutral atoms, so that it is optically thick for line radiation from neutral atoms.
Thus, letting $p_e(\tau)$ and $p_e(\tau_\mathrm{tot}-\tau)$
be the EPs from $r$ to space and from $r$ to the lower boundary in the atmosphere, respectively, one can write
\begin{equation}
P_e = \left\{
	\begin{array}{cl}
		\displaystyle{
		\frac{p_e(\tau)+ p_e(\tau_\mathrm{tot}-\tau)}{2}
		}
		& \mbox{for ions}
		\vspace{2ex} \\ 
		p_e(\tau)
		& \mbox{for atoms}
	\end{array} \right.
\label{eq: ep3}
\end{equation}
where $\tau$ is the frequency-integrated optical depth. 

Since the atmospheric gas of interest in this study is relatively tenuous, 
we consider only the Doppler broadening for calculating $\tau$. 
Then, $P_e$ is given by \citep{Kwan+1981}
\begin{equation}
	P_e = \left\{
	\begin{array}{cl}
		\left[{\tau\sqrt{\pi}
		\left(1.2+\frac{\sqrt{\ln \tau}}{{1+10^{-5}\tau}}\right)
		}\right]^{-1}
		&
		(\tau \geq 1),
		\vspace{1ex} \\ 
		\left({2\tau}\right)^{-1} 
		\left[{1-\exp(-2\tau)}\right]
		&
		(\tau < 1).
	\end{array} \right.
\label{eq: ep4}
\end{equation}
If the optical thickness at the line center is denoted by $\tau_0$, $\tau = \sqrt{\pi}\tau_0$.
To be consistent with the EP model,
we assume that the line profile including the Doppler width is unchanged through a medium, provided temperature is constant. 
$\tau_0$ is much less sensitive to temperature than to  number density, 
because $\tau_0$ is inversely proportional just to the square root of temperature \citep[see][]{Rybicki+86}.
For calculating the Doppler width of ions, we adopt the typical temperature for highly ionized interstellar media, $1 \times 10^4$~K \citep[e.g.,][]{Osterbrock+1989}, for simplicity, since we do not know the temperature of the upper atmosphere a priori and have confirmed that choice of the temperature has little effect on the results. 
Also, as for neutral atoms, we use the temperature at the inner boundary of the calculated region in this model, which is described in Section \ref{ssec: bc}, 
since almost all of the neutrals are in the lower region of the atmosphere.

Also, $J_{\rm{ext}}$ is assumed as 
\begin{eqnarray}\label{eq: jex}
J_{\rm{ext}} = J^*_{\nu_0} + \frac{1}{2} B_{\nu0}(T_0) ,
\end{eqnarray}
where {$J^*_{\nu_0}$ (= $F^*_{\nu_0}/4\pi$)} is the mean intensity of the incident radiation at frequency $\nu_0$ from the host star and $B_{\nu0} (T_0)$ is the Plank function for the temperature at the atmospheric lower boundary, $T_0$. 
The first and second terms correspond, respectively, to the irradiation by the host star and to the thermal radiation from below.
We take the value of $F^*_{\nu_0}$ from \citet{Claire+2012}.

Using Eqs.~(\ref{eq: ep1})--(\ref{eq: jex}), 
one can write the equation for transition between energy levels in statistical equilibrium in the matrix form
\begin{eqnarray}
\label{eq: mrt1}
\bm{\chi} \bm{n_x} &=& \bm{0},
\end{eqnarray}
where $\bm{n_x}$ is the vector of the number densities of energy levels. 
The sum of the components of $\bm{n_x}$, $n_x$, is equal to the total number density of species $s$:
\begin{eqnarray}
\label{eq: mrt2}
\sum{n_x} = n_s.
\end{eqnarray}
The elements of the lower-triangle and upper-triangle matrices of $\bm{\chi}$ are written, respectively, as
\begin{equation}
\chi_{lu} = C_{lu}, 
\hspace{3ex} 
\chi_{ul} = C_{ul} + A_{ul} P_e',
\end{equation}
where $C_{lu}$, $C_{ul}$, and $A_{ul}$ are the collisional de-/excitation transition rates and the Einstein coefficient for the spontaneous de-excitation rate from the upper level $u$ to the lower level $l$, respectively.
We assume that only collision with electrons causes de-/excitation transition of atoms and ions; that is,
\begin{equation}
C_{ul} = q_{ul}n_e, 
\hspace{3ex}
C_{lu} = q_{lu}n_e,
\label{eq: cr1}
\end{equation}
where $n_e$ is the number density of electrons, 
$q_{ul}$ and $q_{lu}$ are the collisional de-/excitation transition rate coefficients and  
the mutual relationship is given by \citep{Osterbrock+1989} 
\begin{eqnarray}
q_{ul}=\frac{g_l}{g_u} q_{lu}\exp(\Delta E_{lu}/k_bT),
\label{eq: cr2}
\end{eqnarray}
where $T$ is the temperature, $k_b$ is the Boltzmann constant, and $\Delta E_{lu}$ is the transition energy between $u$ and $l$,
\begin{eqnarray}
q_{lu} = \frac{8.629\times 10^{-6}}{g_lT^{1/2}} \gamma_{lu} \exp(- \Delta E_{lu}/k_bT),
\end{eqnarray}
and $\gamma_{lu}$ is a dimensionless quantity called the effective collision strength.
We ignore its small temperature dependence in this study, for simplicity.
Table~\ref{tbl:RCT}  summarizes the values of $A_{ul}$ and  $\gamma_{lu}$ for each transition used in this model.
For $\Delta E_{lu}$, we use each different values of each excitation energy summarized in Table~\ref{tbl:ELG}.

From Eqs.~(\ref{eq: mrt1}) and (\ref{eq: mrt2}),
we determine the level populations in each gas species.
Finally, $Q_{\rm{rad}}$ is given by
\begin{eqnarray}
Q_{\rm{rad}}= \sum_{s\in \mathscr{S}} \sum_{u} \sum_{l} \Delta E_{lu}A_{ul} P_e' n_{s,u},
\end{eqnarray}
where $n_{s,u}$ is the number density of gas species $s$ with energy level $u$.

\subsection{Diffusion and Conduction}\label{ssec:dc} 
We consider multi-component diffusion, following the formula described by \citet{GM2007a}.
This formula is based on the momentum equations for multi-component gas derived by \citet{Burgers1969} and arranged in the classical Stefan--Maxwell form.
In this study, we adopt the first order approximation of this method for the mineral atmosphere in ambipolar constraint,
 which is also used in hydrodynamic simulations for hot Jupiters \citep{GM2007b}. 
For calculating the diffusion velocity of species $s$, $u_s$, from the approximate diffusion matrix \citep[see Appendix A in][for the details]{GM2007b},
one needs the binary diffusion coefficient between species $i$ and $k$, $\mathscr{D}_{i,k}$.
 Following \citet{GM2007a}, for the atom-atom and atom-electron binary diffusions,
we calculate $\mathscr{D}_{i,k}$ from the hard sphere model as 
\begin{eqnarray}
n\mathscr{D}_{i,k}= 1.95 \times 10^6 \frac{T^{1/2}}{{\bar{m}_{i,k}^{1/2}}} 
\hspace{1ex} \mbox{cm$^{-1}$ s$^{-1}$},
\label{eq: dik1}
\end{eqnarray}
where $\bar{m}_{i,k}$ = $m_im_k/(m_i+m_k)$.
For the atom-ion binary diffusion, the interaction between ions and atoms is assumed to be due to the induced polarization potential and $\mathscr{D}_{i,k}$ is given by
\begin{eqnarray}
n\mathscr{D}_{i,k}= 4.13 \times 10^{-8} \frac{T}{{\bar{m}_{i,k}^{1/2}}}\frac{1}{(\alpha_n Z_i^2)^{1/2}}
\hspace{1ex} \mbox{{cm$^{-1}$ s$^{-1}$}},
\label{eq: dik2}
\end{eqnarray}
where $\alpha_n$ is the polarizability of the atom and $Z_i$ is the charge number for the ion. 
The values of $\alpha_n$ are taken from \citet{CRC92}.
For the ion-ion binary diffusion, we use the formula of the Coulomb interaction as
\begin{eqnarray}
n\mathscr{D}_{i,k}= 1.29 \times 10^{-3} \frac{T^{5/2}}{{\bar{m}_{i,k}^{1/2}}}\frac{1}{Z_i^2 Z_k^2 \ln{ \beta}}
\hspace{1ex} \mbox{{cm$^{-1}$ s$^{-1}$}},
\label{eq: dik3}
\end{eqnarray}
where $\beta$ is given as $ {\beta} = 1.26 \times 10^{4} (T^3/n_e)^{1/2}$ \citep{Burgers1969}.
For the ion-electron binary diffusion,
following \citet{Schunk+2000}, 
we calculate $\mathscr{D}_{i,k}$ as
\begin{eqnarray}
n\mathscr{D}_{i,k}= 2.8 \times 10^{9} \frac{T^{5/2}}{Z_i^2}
\hspace{1ex} \mbox{{cm$^{-1}$ s$^{-1}$}}.
\label{eq: dik4}
\end{eqnarray}
Finally, we calculate $u_s$ from Eqs.~(A3), (A4) and (A7)--(A12) of \citet{GM2007a} with the binary diffusion coefficients described above.

We assume that heat transport occurs via thermal conduction and heat exchange due to chemical diffusion.
The heat flux $q$ is calculated as \citep{GM2007b}
\begin{eqnarray}
q=- \kappa \frac{\partial T} {\partial r} + \sum_{s\in \mathscr{S}} \rho_s h_s u_s,
\end{eqnarray}
where $\kappa$ is the thermal conductivity and 
$h_s$ is the specific enthalpy of species $s$ given by
$h_s$ = $\gamma k_b T /[(\gamma-1) m_s]$.
$\kappa$ is given as a sum of the thermal conductivities of species $s$, $\kappa_s$, namely
\begin{eqnarray}
\kappa = \frac{1}{n} \sum_{s\in \mathscr{S}} n_s \kappa_s;
\end{eqnarray}
for neutrals, 
$\kappa_s$ is given from Eqs.~(14.30) and (14.42) of \citet{Banks+1973} as
\begin{eqnarray}
\kappa_s =\frac{75}{64} \left(\frac{k_b^{3} T}{m_s \pi} \right )^{1/2} \frac{1}{d_s^2} 
\hspace{2ex} {\rm erg \, cm^{-1}K^{-1}s^{-1}},
\end{eqnarray}
where $d_s$ is the atomic diameter for which we adopt the van der Waals size; 
for ions, $\kappa_s$ 
 is given from Eq. (22.105) of \citet{Banks+1973}  as
\begin{eqnarray}
\kappa_s = 7.37 \times 10^{-8}\frac{T^{5/2}}{(m_s/m_p)^{1/2}}
\hspace{2ex} {\rm erg \, cm^{-1}K^{-1}s^{-1}},
\end{eqnarray}
where $m_p$ is the proton mass;
for electrons, $\kappa_s$ is given from Eq (22.122) of \citet{Banks+1973} as 
\begin{eqnarray}
\kappa_s = 1.23 \times 10^{-6}{T^{5/2}}
\hspace{2ex} {\rm erg \, cm^{-1}K^{-1}s^{-1}}.
\end{eqnarray}

\subsection{Boundary conditions}\label{ssec: bc} 

 \begin{figure}
   \begin{center}
  \includegraphics[width=\columnwidth]{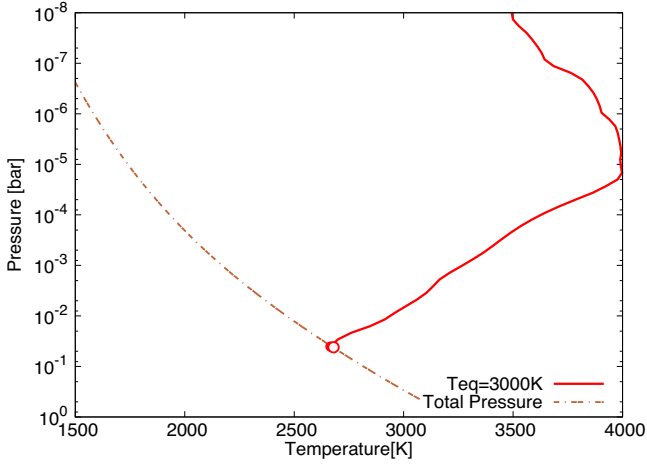}
   \end{center}
   \caption{
     Temperature-pressure profile in the hydrostatic part of the mineral atmosphere on top of the volatile-free bulk-silicate-Earth (BSE) magma ocean of a super-Earth with gravity of 25~m/s$^2$ for the substellar-point equilibrium temperature, $T_\mathrm{eq}$, of 3000~K (red solid), 
   which we have calculated in the same way as \citet{Ito+2015}. 
   The host star is assumed to be a Sun-like star with radius of 1~$R_\odot$ and effective temperature of 6000~K and emit the blackbody radiation of 6000~K.
    The open circle shows the temperature at the bottom of the atmosphere
     and the filled circle, which is almost overlapped with the open one, shows the temperature at the ground.
    The orange dash-dotted line represents the total vapor pressure for the BSE composition, which corresponds to the ground.}
\label{fig:HSE_sub}
\end{figure}

For simulating the hydrodynamic flow, we integrate Eq.~(\ref{eq: hyd1}) for $S$ gas species and Eqs.~(\ref{eq: hyd2})--(\ref{eq: hyd3}).  
Thus, we need $S+2$ boundary conditions.

The lower boundary of the escaping atmosphere is defined as
a spherical surface above which the incident stellar X+UV is completely absorbed. 
To be exact, the optical depths at all the wavelength bins in the X+UV are larger than ten at the lower boundary.
Since the hydrostatic equilibrium approximation is valid in such a deep atmosphere, we determine the lower-boundary temperature and mass density from our hydrostatic model of the mineral atmosphere developed in \citet{Ito+2015}. 
Figure~\ref{fig:HSE_sub} shows an example of the temperature-pressure profile in the hydrostatic atmosphere at the sub-stellar point of the planet with $T_\mathrm{eq}$ of $3000$~K, which corresponds to 0.02~AU around a Sun-like star \citep[see Eq.~(16) in][$A_p = 0$]{Ito+2015}.  

Also, the hydrostatic model provides the molar fractions of neutral atomic species at  the lower boundary.
For instance, in the above case, the molar fractions of Na, O, Si and Mg are 0.375, 0.484, 0.127 and 0.0139.
At the lower boundary, the molar fractions of all the ionic species are set to zero.
In addition, we set the velocity at the lower boundary, $u_1$, to the value extrapolated from the upper atmosphere at each time step in the following way:
\begin{eqnarray}
R_p^2 \rho_1 u_1 = \frac{ \int^{R_{o}}_{R_p} r^2 \rho u dr }{R_o-R_p},
\label{eq:uibc}
\end{eqnarray}
where $R_p$ and $R_o$ the radial distances of the lower and upper boundaries from the planet's center, respectively, and $\rho_1$ is the density  at $r$ = $R_p$.
This extrapolation prevents the mass flux from depending on the lower boundary condition.

The upper boundary conditions are less important in this study, since we 
focus on an escaping atmosphere with a super-sonic velocity, which means no information of the upper boundary is propagated to regions with subsonic velocities where the escaping flow is driven.
Thus, we set simply the quantities such as $n_s$, $T$ and $u$ at the upper boundary to values extrapolated from the interior of the computational domain, following \citet{GM2007b}. We set the upper boundary radius to 10~$R_p$.

\subsection{Numerical scheme}\label{ssec:ns}
To find the steady state of the hydrodynamic flow, 
we perform time integration of the conservative forms of Eqs.~(\ref{eq: hyd1})--(\ref{eq: hyd3}) 
based on spatial- and time-discretization:
\begin{eqnarray}
\label{eq: nhyd}
\frac{d}{dt} \bm{U} &:=&
\lim_{\Delta t \to 0} \frac{\bm{U}(t+\Delta t)-\bm{U}(t)}{\Delta t} \\
&=& \mathcal{L_H} +  \mathcal{L_R},
\end{eqnarray}
where
\begin{eqnarray}
 \mathcal{L_H} &=& -(\Delta \bm{F}+\Delta \bm{F_d})/ \Delta r +\bm{S_H},
 \\ 
  \mathcal{L_R} &=& \bm{S_R},
\end{eqnarray}
$\Delta r$ is the size of the non-overlapping spatial cell at $r$, and
\begin{equation}
\bm{U}=
    \begin{bmatrix}
      r^2 \rho_s \\
      r^2 \rho u \\
      r^2 \rho E \\
    \end{bmatrix}, 
\hspace{3ex}
\bm{F}=
    \begin{bmatrix}
      r^2 \rho_s u \\
      r^2 (\rho u^2 + P) \\
      r^2 (\rho E+ P)u  \\
    \end{bmatrix} , 
\hspace{3ex}
\bm{F_d}=
    \begin{bmatrix}
      r^2 \rho_s u_s \\
      0 \\
      r^2 q \\
    \end{bmatrix},
    \label{eq: nhyd1}
    \end{equation}
    \begin{equation}
\bm{S_H}=
    \begin{bmatrix}
      0 \\
      r^2 \rho f_{\rm{ext}} +2Pr  \\
      r^2 (\rho u f_{\rm{ext}}+Q_\mathrm{net}) \\
    \end{bmatrix},
\hspace{3ex}    
\bm{S_R}=
    \begin{bmatrix}
      r^2 \dot{\rho_s} \\
      0  \\
      0 \\
    \end{bmatrix}.
\end{equation}
Using these expressions,
we explain the spatial discretization and the numerical schemes for the advection term and marching time, below.

The computational region of the atmosphere is divided into $N_r$ layers. 
The thickness of each layer is assumed to increase with altitude in such a way that  the thickness ratio between two neighboring layers, $l_r$, is constant.
Following \citet{GM2007b}, we use $N_r=600$  and $l_r=1.014$. 
Also, all the variables in each cell are defined at the center of each cell.
For $\Delta F_d$, we calculate $F_d$ at the cell boundaries using the values of variables at the centers of the neighboring cells.
For $\Delta F$, we use the finite volume method with second-order accuracy. 

The above spatial discretization sometimes causes a numerical instability during
time integration of $\bm{\Delta F}$, 
because of unattenuated short-wavelength numerical disturbances 
\citep[see also][]{Hirsch1990}.
To ensure the numerical stability, 
we add an artificial dissipation flux $\bm{F_{\rm{AD}}}$ to $\bm{F}$ as \citep{Swanson+1992, Swanson+1997}
\begin{eqnarray}
\bm{F'}&=&\bm{F}+\bm{F_{\rm{AD}}},
\\
\bm{F_{\rm{AD}}}&=&\zeta \bm{A'} 
\triangle \nabla \triangle \bm{U_{AD}} \Delta r^3,
\label{eq:fad}
\end{eqnarray}
where $\triangle$ and $\nabla$ are the forward and backward spatial difference operators, respectively, $\zeta$ is a constant (= 0.01 in this study), $\bm{U_{AD}}$ is given by 
\begin{eqnarray}
\bm{U_{AD}}=
    \begin{bmatrix}
      r^2 \rho_s \\
      r^2 \rho u \\
      r^2 (\rho E+P) \\
    \end{bmatrix},
\end{eqnarray}
and  $\bm{A'}$ is a square matrix of $S+2$ rows and columns which is given by
\begin{eqnarray}
\bm{A'}=\bm{\Re}|\bm{\hat \lambda'}| \bm{\Re^{-1}}.
\end{eqnarray}
$\bm{\Re}$ is the eigenmatrix of $\bm{\partial F}/\bm{\partial U}$ defined as
\begin{eqnarray}
\frac{ \bm{\partial F}}{\bm{\partial U}}	&=&\bm{\Re}\bm{\hat \lambda} \bm{\Re^{-1}},
\label{eq: jaco}
\end{eqnarray}
where $\bm{\hat \lambda}$ is the eigenvalue matrix, which is a diagonal one with $S+2$ rows and columns given by 
\begin{eqnarray}
\bm{\hat \lambda'}=diag(\hat \lambda_1^\prime, \hat \lambda_2^\prime, \hat \lambda_2^\prime, \cdots, \hat \lambda_2^\prime, \hat \lambda_2^\prime, \hat \lambda_3^\prime).
\end{eqnarray}
The three dots mean that all the elements except the first one ($\hat{\lambda}_1^\prime$) and the final one ($\hat{\lambda}_3^\prime$) are $\hat{\lambda}_2^\prime$. 

In this matrix, 
\begin{eqnarray}
	\hat{\lambda}^\prime_j = \left\{
	\begin{array}{cl}
		\max \left( \left| \hat{\lambda}_j \right|, V_n \max(\hat{\lambda}_j) \right) 
		& (j =1, 3)
	\vspace{1ex} \\
		\max \left( \left| \hat{\lambda}_j \right|, V_l \max(\hat{\lambda}_j)  \right)
		& (j =2),
	\end{array} \right.
\end{eqnarray}
with reduction constants $V_n$ and $V_l$, and $\hat \lambda_{{j}}$ is the eigenvalue of $\bm{\partial F}/\bm{\partial U}$.
Here we use $V_n=0.25$ and $V_l=0.025$, which are their typical values \citep{Swanson+1992}.
From Eqs.~(\ref{eq: nhyd1}) and (\ref{eq: jaco}), 
$\hat \lambda_1 = u-c_S$, 
$\hat \lambda_2 = u$, 
and $\hat \lambda_3 = u+c_S$.
Note that when $\bm{F_{\rm{AD}}}$ makes a major contribution to $\bm{F'}$ throughout the atmosphere, this computation may be invalid.
Basically,
the steeper the gradients of physical quantities in the atmosphere are, the larger  $\bm{F_{\rm{AD}}}$ is, 
because $\bm{F_{\rm{AD}}}$ includes  the third order differential of $ \bm{U_{AD}}$.  
We have, however, made sure that the contribution of $\bm{F_\mathrm{AD}}$ to $\bm{F}^\prime$ is less than 1 \% in the simulation region above 1.1~$R_p$ where is the atmospheric region of interest in this study.

Equation~(\ref{eq: nhyd}) includes chemical reaction terms, $\mathcal{L_R}$, which differ greatly in magnitude 
(i.e., mathematically stiff for stable time-integration). 
 To time-integrate such a system of stiff differential equations, 
we use the implicit-explicit Runge-Kutta method with six-step and fourth-order accuracy named ARK4(3)6L[2]SA, which was developed by \citet{Kennedy+2003}.
Treating $\mathcal{L_R}$ implicitly and $\mathcal{L_H}$ explicitly, 
we calculate the time marching for $\mathbf{U}$ by the integration of Eq. (\ref{eq: nhyd}).
Then, for the user specific tolerance error, we set that the relative error for all elements of $\mathbf{U}$ is $10^{-4}$ and the absolute error for $\rho_s$ is $10^{-10} \rho$.
Also, we use the variable time step as \citep{Kennedy+2003}
\begin{eqnarray}
\Delta t = \min{( \Delta t_{\rm CFL}, \Delta t_{\rm PID} )},
\end{eqnarray}
where  $\Delta t_{\rm CFL}$ is the time step limited by the Courant-Friedrichs-Lewy (hereafter, CFL) condition, and $\Delta t_{\rm PID}$ is the time step controlled by Proportional-Integral-Differential controllers. 
$\Delta t_{\rm CFL}$ is given by 
\begin{eqnarray}
 \Delta t_{\rm CFL} &=& \zeta_{\rm CFL} \min \left( \frac{\Delta r}{|c_S+u|}\right),
\end{eqnarray}
where $\zeta_{\rm CFL}$ is the CFL number that equals 2.01.
And, $\Delta t_{\rm PID}$ is given from Eq. (30) of \citet{Kennedy+2003} as
\begin{eqnarray}
 \Delta t_{\rm PID}^{\left( n+1 \right)} = 0.9  \Delta t^{\left( n \right)} 
 \left\|  \frac{\jmath}{\delta^{\left( n+1 \right)} } \right\| ^{0.49/p}
  \left\| \frac{\delta^{\left( n \right)} }{\jmath} \right\|^{0.34/p}
   \left\| \frac{\jmath}{\delta^{\left( n-1 \right)} } \right\|^{0.1/p},
\end{eqnarray}
where $\jmath$ is the user specific tolerance error, $p$ is the value that equals the order of accuracy minus one, thus three in this model, and $\delta^{\left( n+1 \right)}$ is the numerical error of $\bf{U}(t+\Delta t^{\left( n \right)})$. The error is estimated by the difference between the fourth-order accuracy values and the third-order accuracy values \citep[see][for the details]{Kennedy+2003}.

This numerical method is similar to that used in \citet{GM2007b}. 
The main difference between their and our numerical schemes  
is whether the artificial dissipative terms of the discrete equations are scaled by the eigenvalues of the Jacobian matrix or by the spectral radius \citep[see][for the details]{Swanson+1992}. The time integration based on a Runge-Kutta method is also used in \citet{GM2007b}.
Following \citet{GM2007b}, we decide that the system has converged to a steady state, when the relative changes in $\rho$, $\rho u$ and $\rho E$ averaged over the $N_r$ layers are below $10^{-6}$ for 1000 seconds.
It is also required that the sum of the mass flux terms, including those of the artificial dissipation, from the continuity equations over all species fluctuate within 1~\% throughout the calculated region.

\section{Benchmark Tests}
\label{sec: mv}
We perform two benchmark tests of the simulation code for atmospheric escape that we have newly developed.
As described below, we confirm that our numerical simulations correctly reproduce the analytical hydrodynamic solution of an isothermal atmosphere (section~\ref{ssec: iso})
and numerical solutions for a highly UV-irradiated hydrogen-dominated atmosphere 
derived by \citet{GM2007b} (section~\ref{ssec: hj}).

\subsection{Isothermal transonic escape}
\label{ssec: iso}
\begin{figure}
   \begin{center}
  \includegraphics[width=\columnwidth]{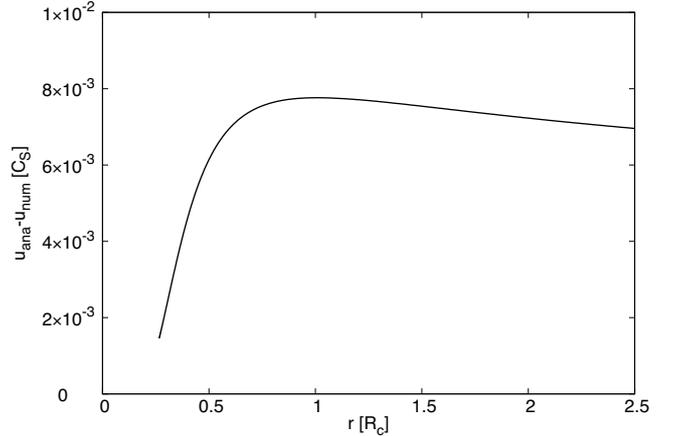}
   \end{center}
   \caption{
   Comparison of our numerical solution  
   with the analytical solution for an isothermal atmosphere that is escaping hydrodynamically. The velocity difference between the analytical one, $u_{\rm{ana}}$, and our numerical one, $u_{\rm{num}}$, in the unit of the sound velocity, $c_S$, is
  	shown as a function of radial distance, $r$, in the unit of the critical radius, $R_c$ (see Eq.~[\ref{eq:rc}] for its definition). 
Here we have assumed that the atmosphere consists only of hydrogen atoms and its temperature is 5000~K. 
	The planetary masses and radii are 10~$\Mearth$ and 2~$\Rearth$ , respectively.
	We have ignored the effect of the tidal force.
	}
\label{fig:iso}
\end{figure}

For an isothermal transonic atmosphere, the velocity is analytically given by \citep[][]{Parker1964} 
\begin{eqnarray}
-\frac{u^2}{c^2_s} + \ln\frac{u^2}{c^2_s}= -4\frac{R_c}{r}-4\ln{\frac{r}{R_c}}+3,
\label{eq: iso1}
\end{eqnarray}
where 
\begin{eqnarray}
R_c = \frac{GM_p\bar{m}}{2k_bT},
\label{eq:rc}
\end{eqnarray}
which is termed the critical radius, and $\bar{m}$ is the mean mass per gaseous particle. 
Our numerical solution is compared with Eq.~(\ref{eq: iso1}) in Fig.~\ref{fig:iso} for a super-Earth of mass 10~$\Mearth$ and radius 2~$\Rearth$ having an atmosphere composed only of atomic hydrogen whose temperature is 5000~K.
Here we have ignored the tidal force, namely, the last two terms of Eq.~(\ref{eq: fext}). 
Figure~\ref{fig:iso} shows that the numerically calculated velocity 
reproduces the analytical solution with enough accuracy
(at most one percent uncertainty of sound velocity).

\subsection{Hydrodynamic escape from hot Jupiter}
\label{ssec: hj}
\begin{figure}
 \begin{minipage}{\columnwidth}
  \begin{center}
   (a) Velocity \\
   \includegraphics[width=\columnwidth]{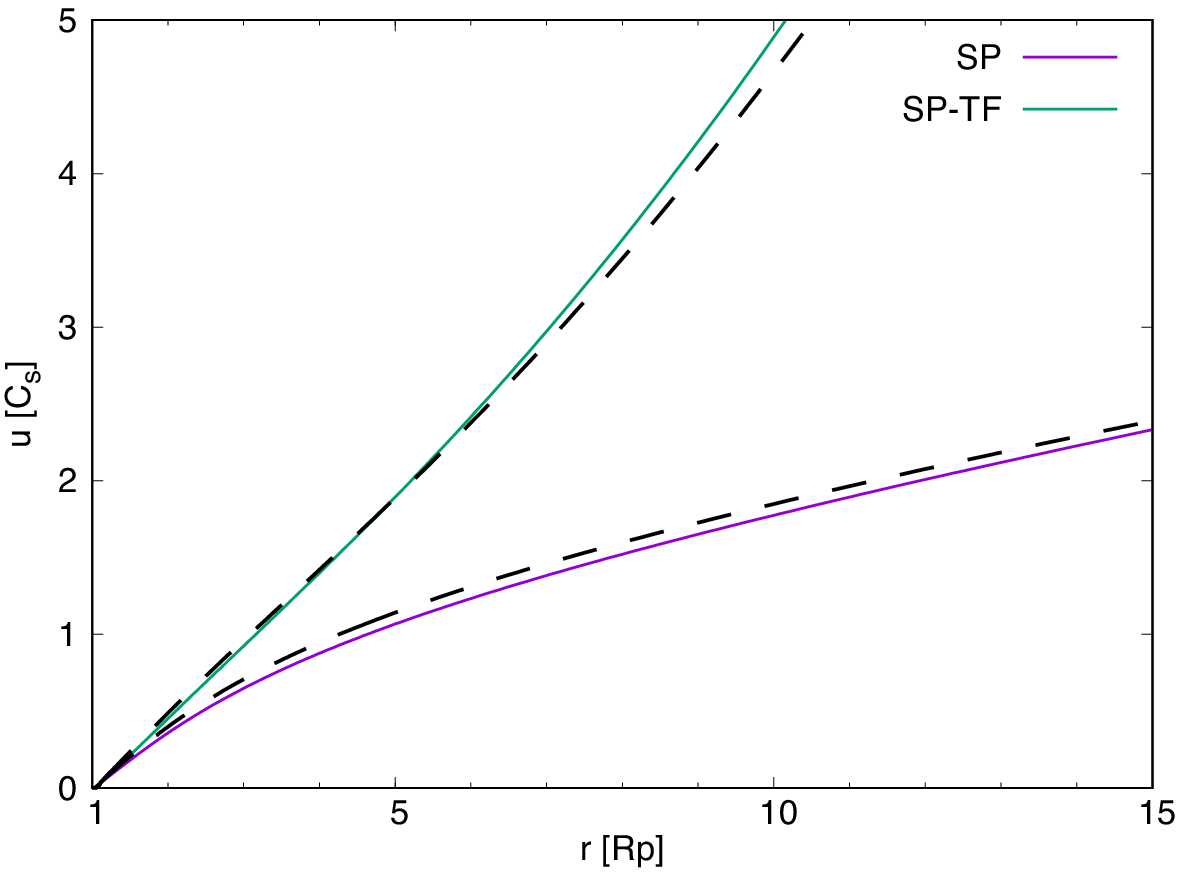}
  \end{center}
  \label{fig:GMU}
 \end{minipage}
 \\
 \begin{minipage}{\columnwidth}
  \begin{center}
   (b) Pressure \\
   \includegraphics[width=\columnwidth]{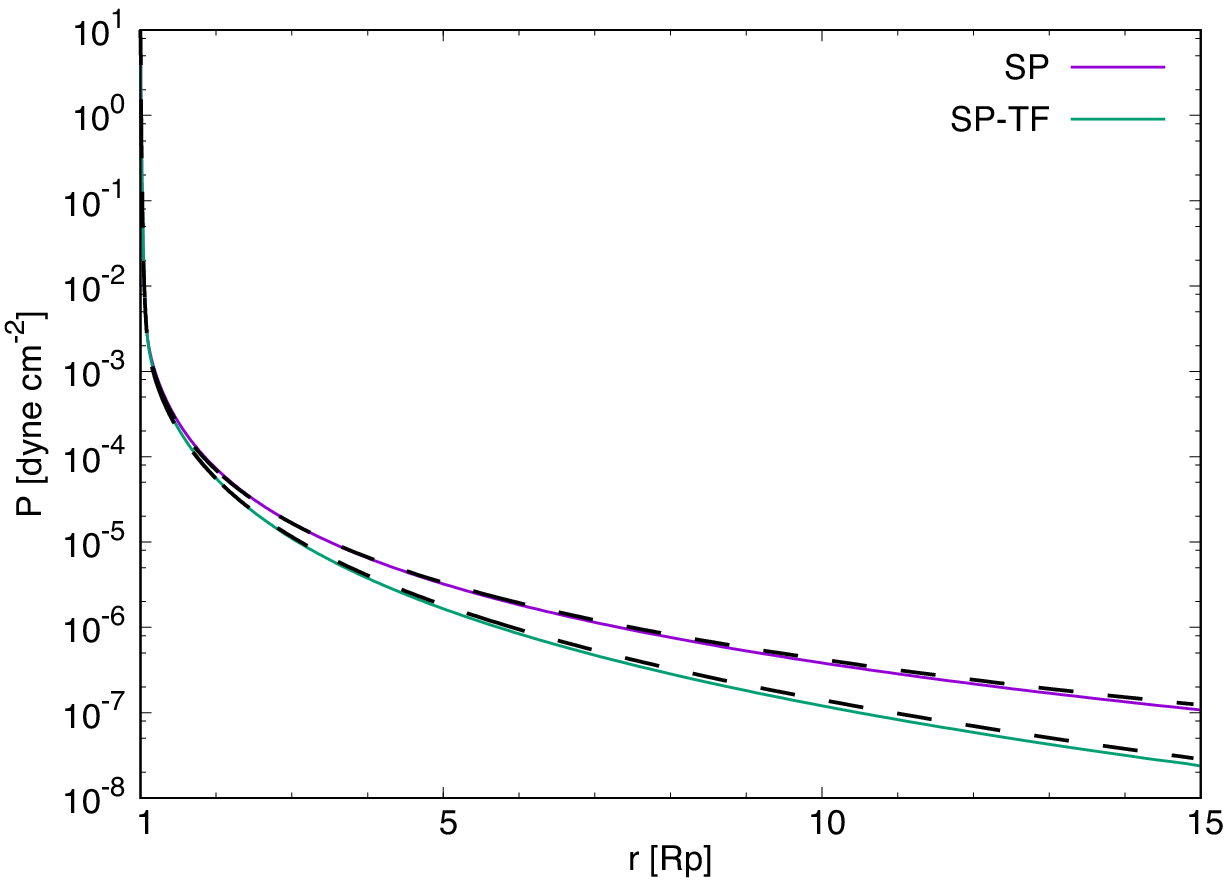}
  \end{center}
  \label{fig:GMP}
 \end{minipage} 
 \\
  \begin{minipage}{\columnwidth}
  \begin{center}
    (c) Temperature \\
   \includegraphics[width=\columnwidth]{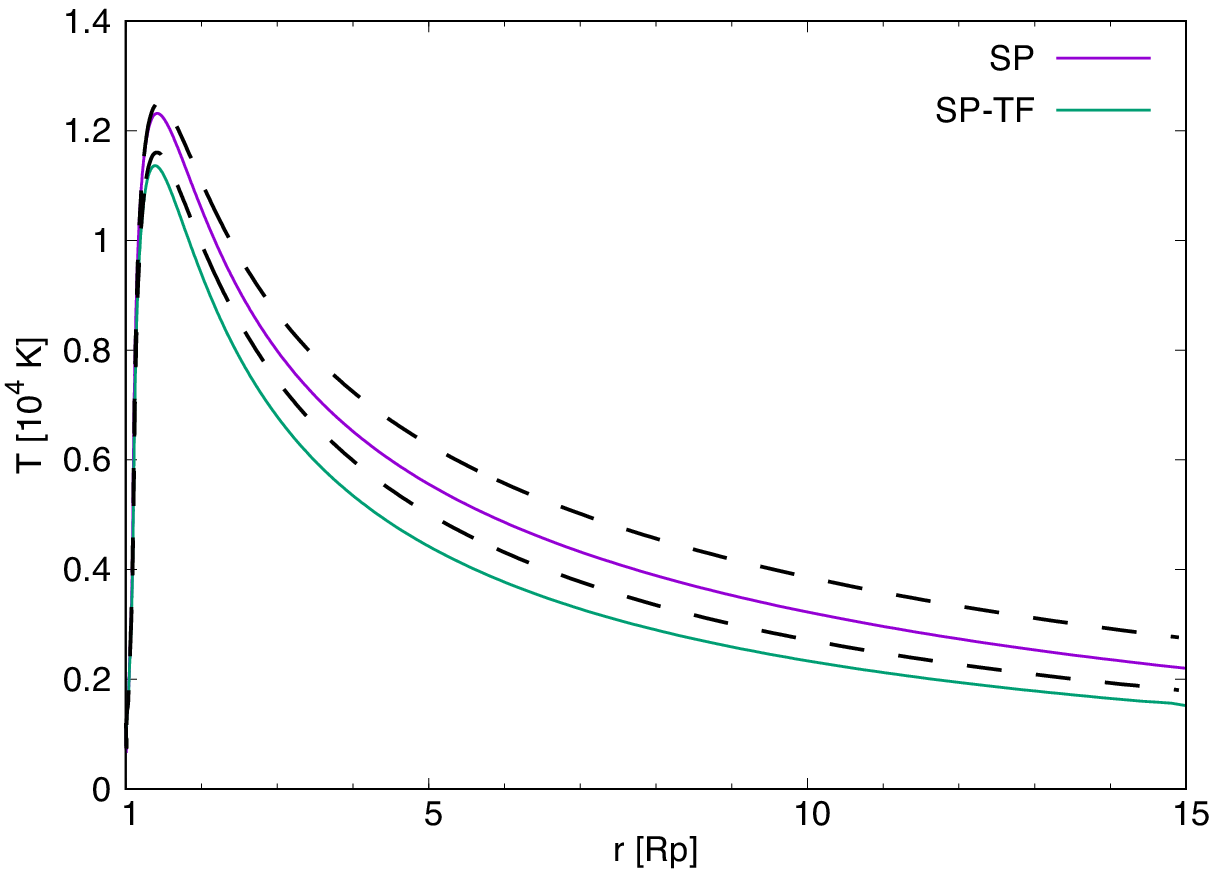}
  \end{center}
  \label{fig:G<T}
 \end{minipage}
  \caption{
  Reproduction of hydrodynamic simulations of \citet{GM2007b} (dashed lines) that assume a highly UV-irradiated hydrogen-dominated atmosphere of the hot Jupiter HD 209458~b (see the text for the details).
  Velocity (a), pressure (b) and temperature (c) are shown as functions of the radial distance (in the unit of planetary radius, $R_p$). 
  We consider two cases with (SP-TF; green) and without (SP; violet) the tidal effect given by the last two terms of Eq.~(\ref{eq: fext}). 
}
 \label{fig:GM2007UPT}
\end{figure}

\begin{figure}
 \begin{minipage}{\columnwidth}
  \begin{center}
   (a) Distribution of gas species  \\
   \includegraphics[width=\columnwidth]{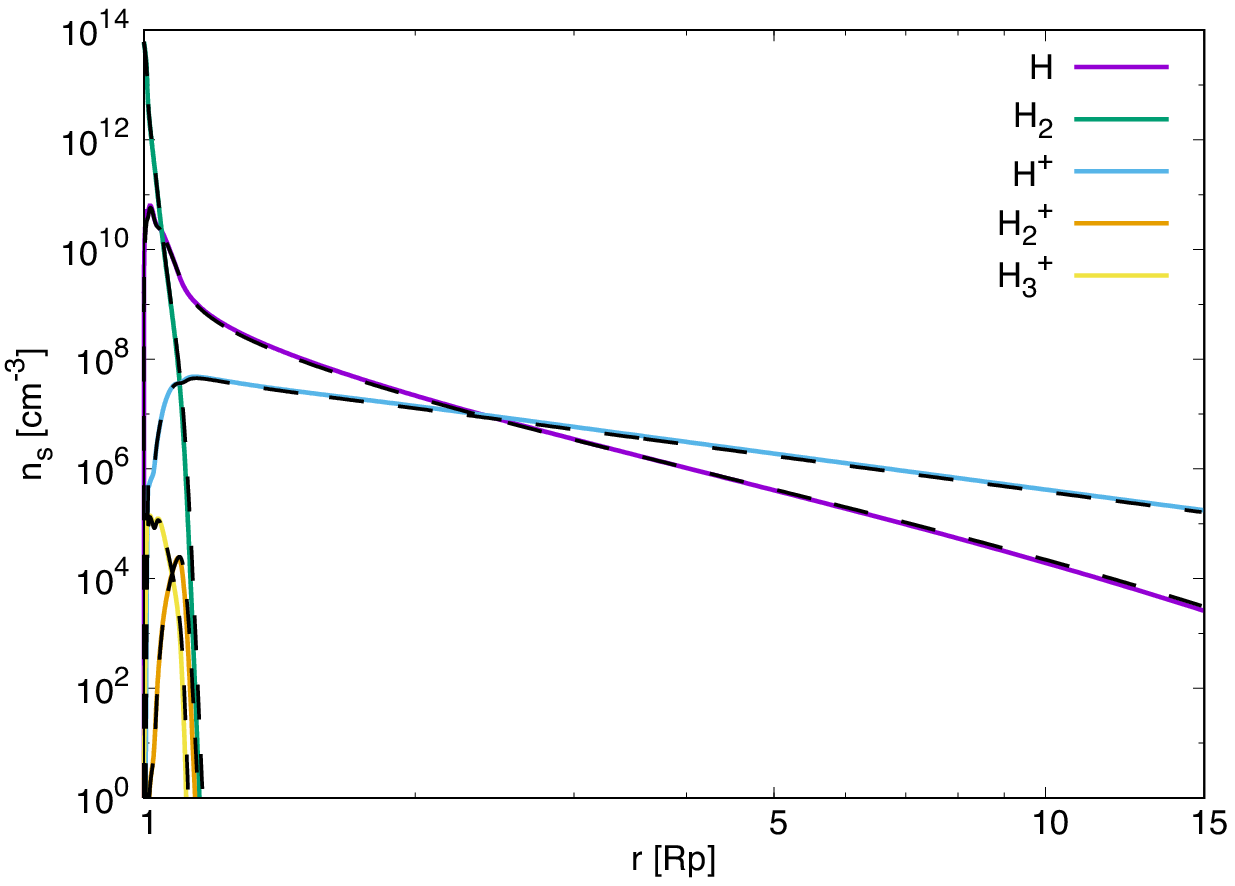}
  \end{center}
  \label{fig:GMN}
 \end{minipage}
 \begin{minipage}{\columnwidth}
  \begin{center}
   (b) Heating and cooling rate   \\
   \includegraphics[width=\columnwidth]{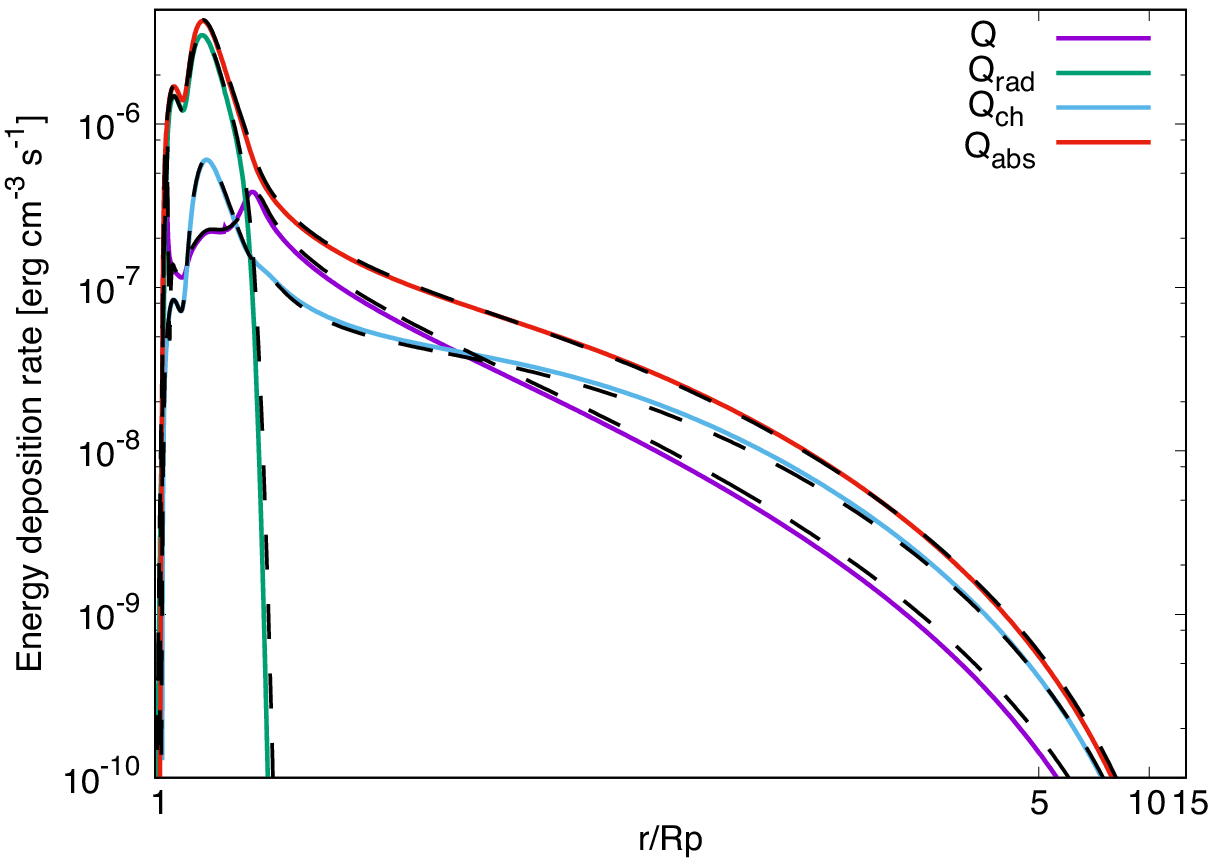}
  \end{center}
  \label{fig:GME}
 \end{minipage} 
  \caption{
Reproduction of hydrodynamic simulations of \citet{GM2007b} (dashed lines) that assume a highly UV-irradiated hydrogen-dominated atmosphere of the hot Jupiter HD 209458~b with no tidal effect (SP). 
The top panel (a) shows the distributions of the number densities of the five gas species, H (violet), H$_2$ (green), H$^+$ (cyan), H$_2^+$ (orange) and H$_3^+$ (yellow); 
  the bottom panel (b) shows the distributions of the heating and cooling rates, $Q$ (violet), $Q_{\rm{rad}}$ (light blue),  $Q_{\rm{chem}}$ (cyan) and  $Q_{\rm{abs}}$ (red), following the definitions of these terms described in \citet{GM2007b}.
The radial distance is measured in the unit of planetary radius, $R_p$.
}
 \label{fig:GM2007NE}
\end{figure}

\citet{GM2007b} performed numerical simulations of hydrodynamic escape of a highly UV-irradiated hydrogen-dominated atmosphere, supposing the hot Jupiter orbiting at 0.047~AU, HD 209458~b \citep[see also][]{Yelle2004}. 
With the same elementary processes and conditions as those considered in \citet{GM2007b},
we calculate the atmospheric structure.
Note that we use the fitting polynomial function of temperature \citep{Koskinen+2007} for the radiative cooling rate of H$^{3+}$ shown in \citet{Neale+1996}, although \citet{GM2007b} 
gives no clear description of how to use the cooling rate.

In Figs.~\ref{fig:GM2007UPT} and \ref{fig:GM2007NE}, we reproduce some of the results of \citet{GM2007b}: The former
 shows the velocity, pressure and temperature profiles, which can be compared directly with the previous study's results (dashed lines) that are shown in Figs.~3 and 4 of \citet{GM2007b}. 
Here we consider two cases without and with the tidal effect; the former and latter are referred to as the  "SP" case and the "SP-TF" case, respectively.
Figure \ref{fig:GM2007NE} shows the radial distributions of the number densities of five gaseous species (top panel) and the heating and cooling rate (bottom panel) calculated in the SP case; The latter panel is shown in the same form as Fig.~2 of \citet{GM2007b}. 

Our results are quite similar to those of \citet{GM2007b}, 
although small differences have arisen due to differences in the numerical scheme and the fitting function for the radiative cooling rate of H$^{3+}$.
For instance, in the SP case, the sonic point is located at $r$ = 4.6~${R_p}$ in our model, but 4.3~$R_p$ in \citet{GM2007b}.
Also, our model estimates the escaping mass flux to be 1.42 $\times 10^{11}$~g/s and 1.47 $\times 10^{11}$~g/s in the SP and SP-TF cases, respectively, 
while the estimates by \citet{GM2007b} are 1.48 $\times10^{11}$ g/s and 1.52 $\times10^{11}$ g/s, respectively; that is, the differences are less than 5~\%.
Thus, it would be fair to say that our model is in good agreement with \citet{GM2007b}.

\section{Results}\label{sec: r}
      \begin{table*}
          \caption{Simulation settings and calculated mass loss rates}
    \label{tbl:SBC} 
  \begin{tabular}{ c c c c c c c } 
    \hline \hline 
    Case & Planetary mass \& radius  & Stellar age & Pressure \& Temperature at the bottom & Composition & X+UV heating efficiency & Mass loss rate  \\ 
    & [$\Mearth$], [$\Rearth$] & [Gyr]  &  [dyne cm$^{-2}$], [K] &  & & [$\Mearth/$Gyr] \\
    \hline \hline
    A & 1, 1 & 0.1 & 300,  3500 & w/ Na & 1.6$\times10^{-3}$ & 3.0$\times10^{-1}$ \\
    B & 1, 1 & 1 & 300, 3500 & w/ Na & 2.2$\times10^{-4}$ & 3.7$\times10^{-2}$ \\
    C & 1, 1 & 0.1 & 20, 3800 &w/o Na & 3.8$\times10^{-3}$ & 4.4$\times10^{-1}$ \\
    \hline
  \end{tabular}  
  \begin{flushleft}
\end{flushleft}
    \end{table*}
    
           \begin{figure*}
            \begin{minipage}{\columnwidth}
   \begin{center}
      ($a$) HRE around 0.1-Gyr-old star \\
  \includegraphics[width=\columnwidth]{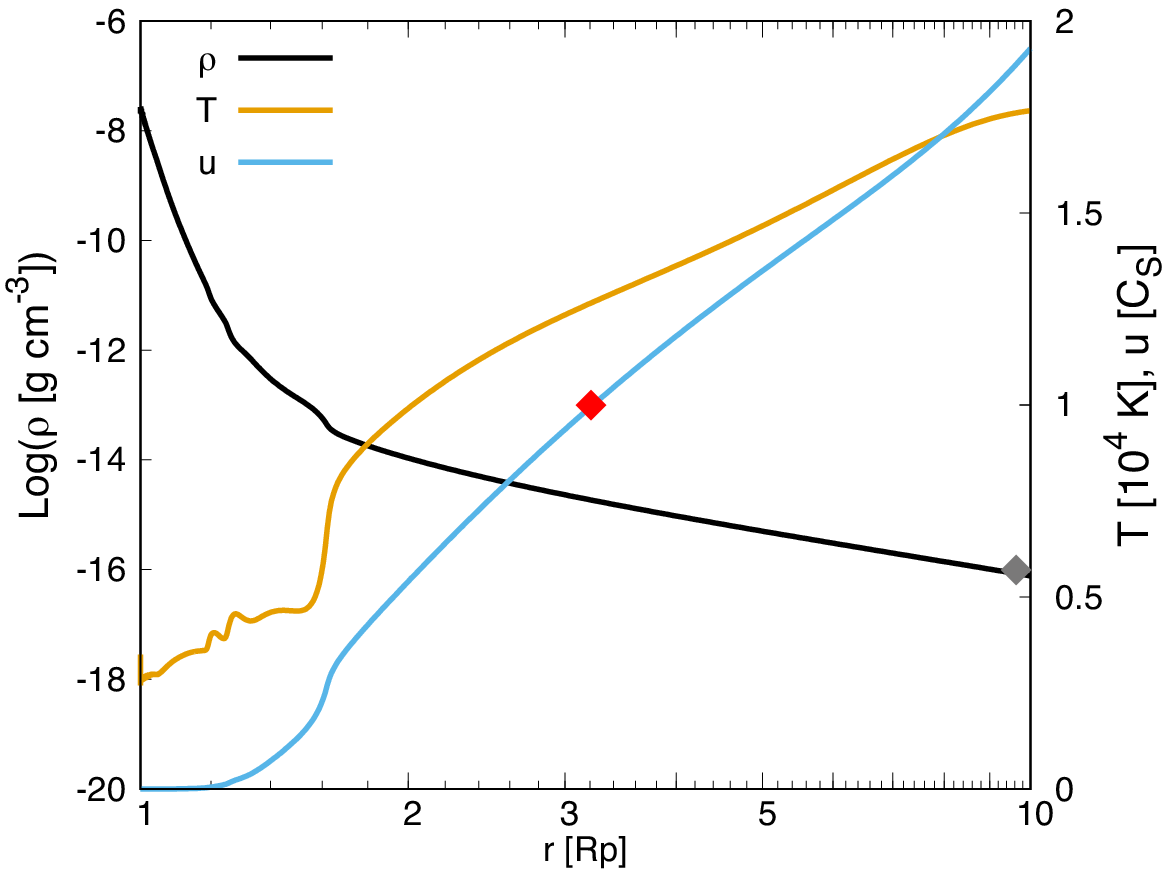}
   \end{center}
  \end{minipage}
            \begin{minipage}{\columnwidth}
   \begin{center}
        ($b$) HRE around 1-Gyr-old star  \\
  \includegraphics[width=\columnwidth]{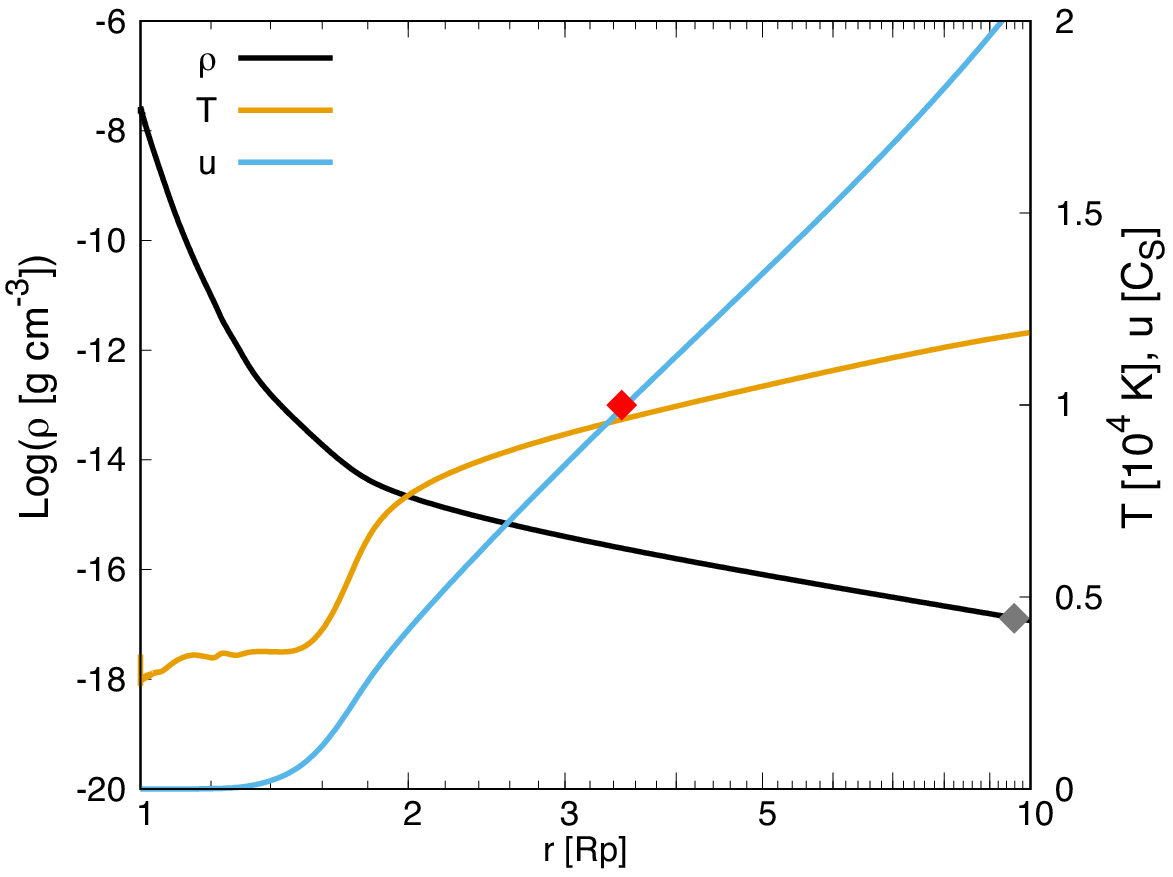}
   \end{center}
    \end{minipage}
 \\
      \caption{
   Profiles of mass density $\rho$ (black), temperature $T$ (orange), and velocity $u$ 
   (cyan) in the mineral atmosphere on the hot rocky exoplanet (HRE) with mass $M_p$ of 1~$\Mearth$ and inner radius $R_p$ of 1~$\Rearth$ that is orbiting at 0.02 AU around the solar-type host-star with age of ($a$) 0.1 Gyr and ($b$) 1 Gyr. 
   The unit of $u$ is the local sound velocity, meaning the cyan line represents the Mach number.
   The red and grey diamonds indicate the sonic point and exobase, respectively.
   }
 \label{fig:rut1-1}
 \end{figure*}
We show 
results of our hydrodynamic simulations for the mineral atmosphere, including the radial profiles of density, temperature, and velocity 
and the distribution of atoms and ions. 
Also, we investigate the energy budget at each altitude in the atmosphere.
From those results, 
we estimate the planetary mass loss rate as
\begin{eqnarray}
\dot{m} = 4 \pi r^2 \rho u, 
\label{eq:mlr}
\end{eqnarray}
assuming a steady, spherically symmetric escape. 
Note that Equation~(\ref{eq:mlr}) places an upper limit to the mass loss rate, because
this model considers the stream-tube of the atmosphere above the substellar point, where the tidal effect and X+UV irradiation are greatest.

 \begin{figure*}
 \begin{minipage}{1.0\hsize}
  \begin{center}
   ($a$) All species \\
   \includegraphics[width=0.50\linewidth]{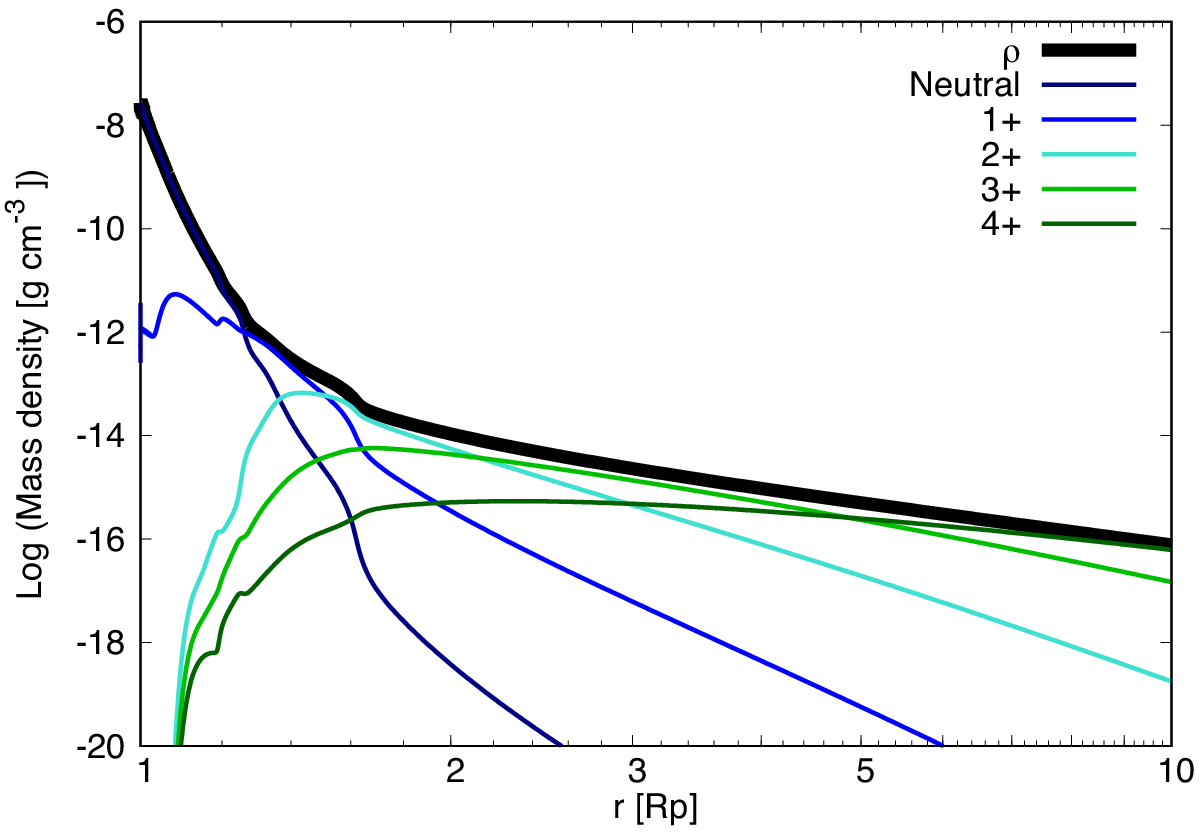}
  \end{center}
 \end{minipage}
 \\ \ 
 \begin{minipage}{\columnwidth}
  \begin{center}
   ($b$) Sodium \\
   \includegraphics[width=1.0\columnwidth]{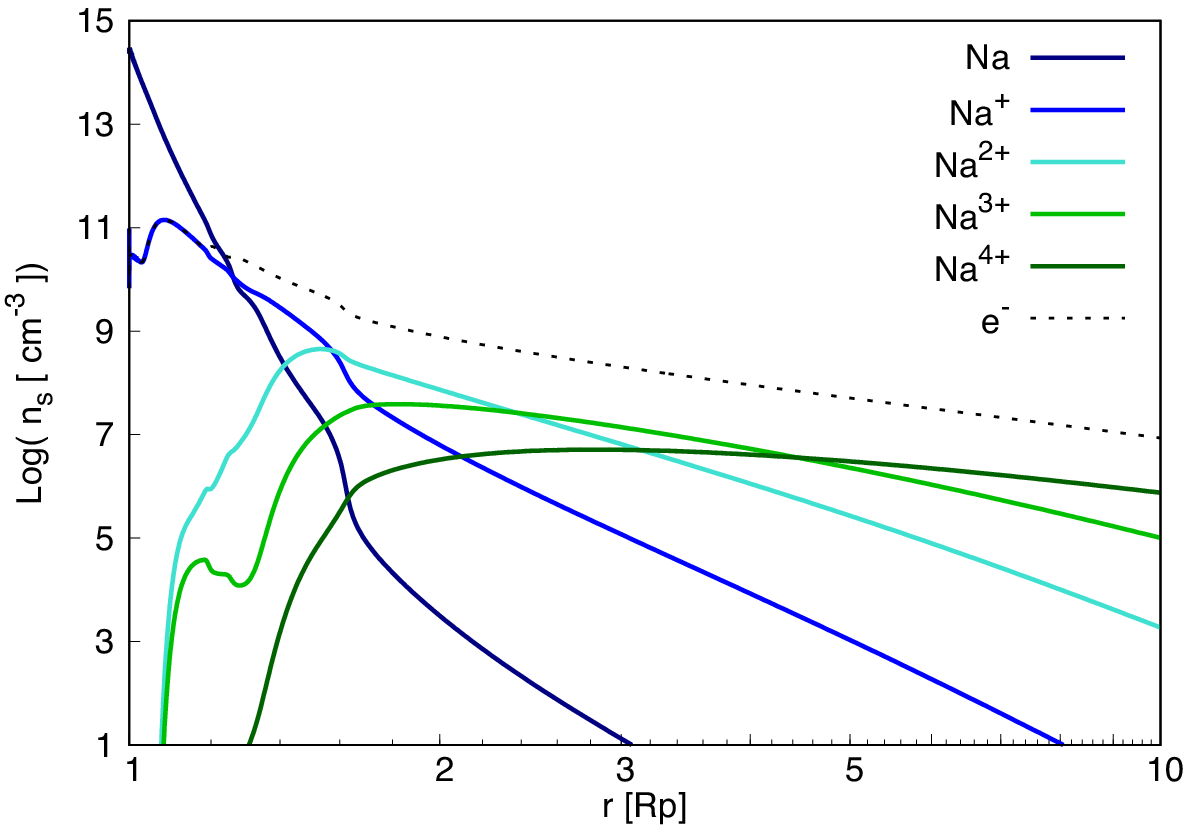}
  \end{center}
 \end{minipage}
 \begin{minipage}{\columnwidth}
  \begin{center}
   ($c$) Oxygen \\
   \includegraphics[width=1.0\columnwidth]{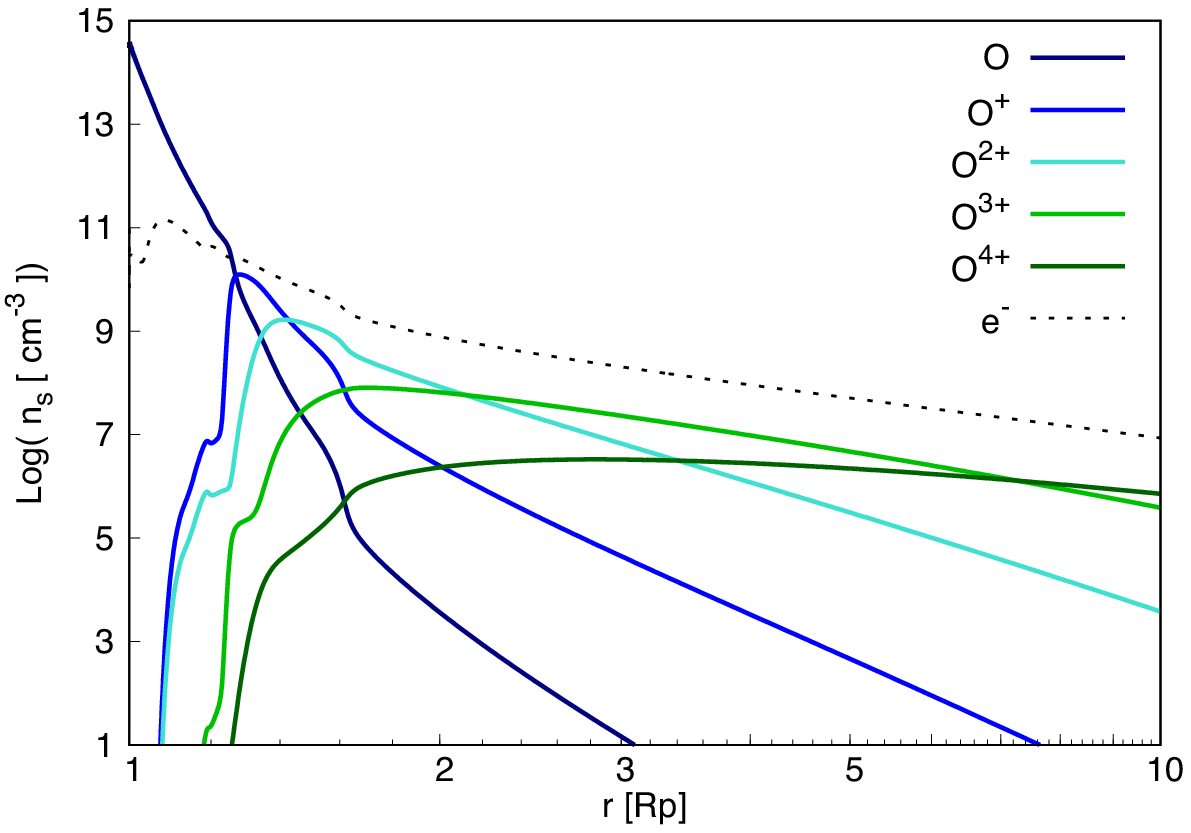}
  \end{center}
 \end{minipage} 
  \begin{minipage}{\columnwidth}
  \begin{center}
    ($d$) Magnesium \\
   \includegraphics[width=1.0\columnwidth]{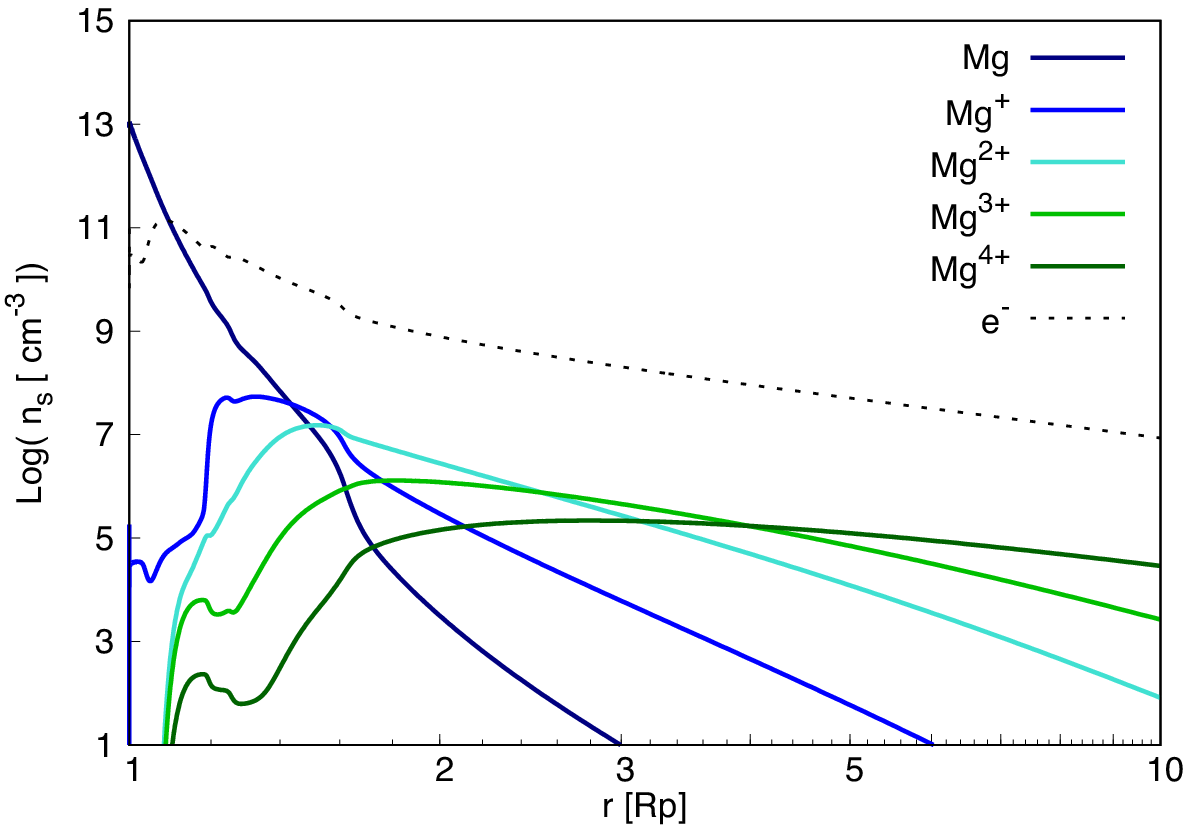}
  \end{center}
 \end{minipage}
 \begin{minipage}{\columnwidth}
  \begin{center}
   ($e$) Silicon \\
   \includegraphics[width=1.0\columnwidth]{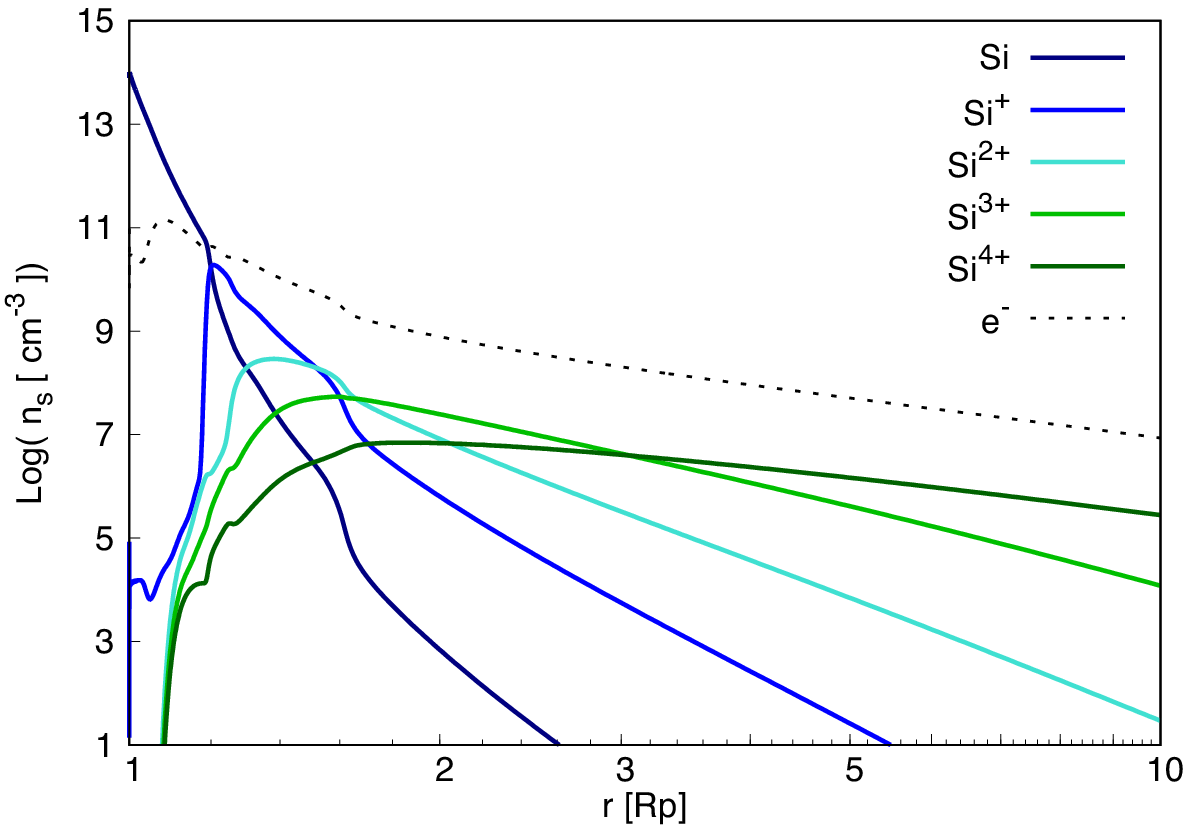}
  \end{center}
 \end{minipage}
  \caption{
Distributions of the gaseous species in the mineral atmosphere on the hot rocky exoplanet (HRE) with mass of 1~$\Mearth$ and radius of 1~$\Rearth$ that is orbiting at 0.02 AU from the solar-type host-star with age of 0.1~Gyr. 
The horizontal axis is the distance from the planetary center in the unit of planetary radius $R_p$.
\textit{Panel (a)}---the mass densities of all the neutrals (navy) and all the singly-charged (blue), doubly-charged (cyan), triply-charged (green), and quadruply-charged (dark green) ions. The total mass density is also shown by the thick black line. 
\textit{Lower panels} --- the number densities of the neutrals and the singly- to quadruply-charged ions of ($b$) Na, ($c$) O, ($d$) Mg, and ($e$) Si. Color coding is the same as in the top panel except the black dotted lines for electrons. 
  }
 \label{fig:COMP1-1}
\end{figure*}

\begin{figure}
 \begin{minipage}{\columnwidth}
    ($a$) 
    \begin{center}
  \includegraphics[width=\columnwidth]{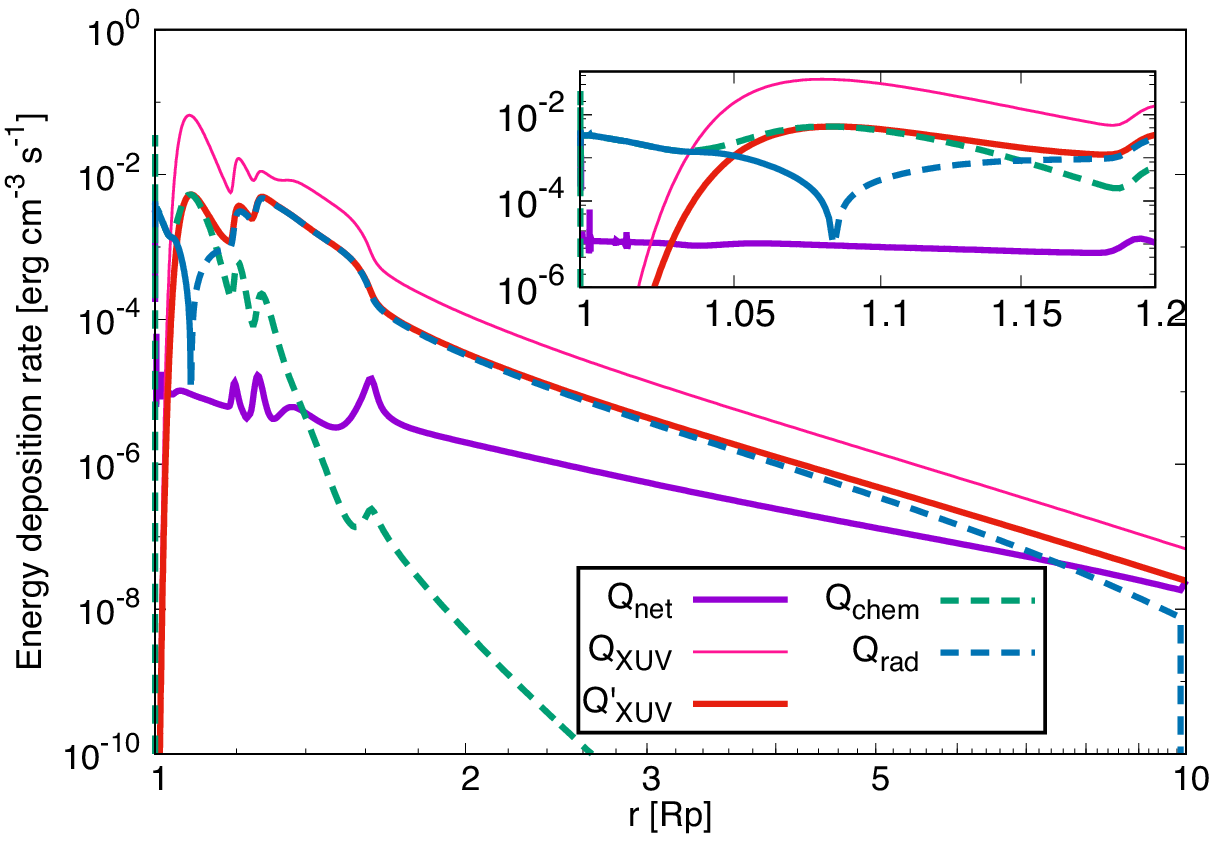}
    \end{center}
 \end{minipage}
     \begin{minipage}{\columnwidth}
    ($b$) 
   \begin{center}
  \includegraphics[width=\columnwidth]{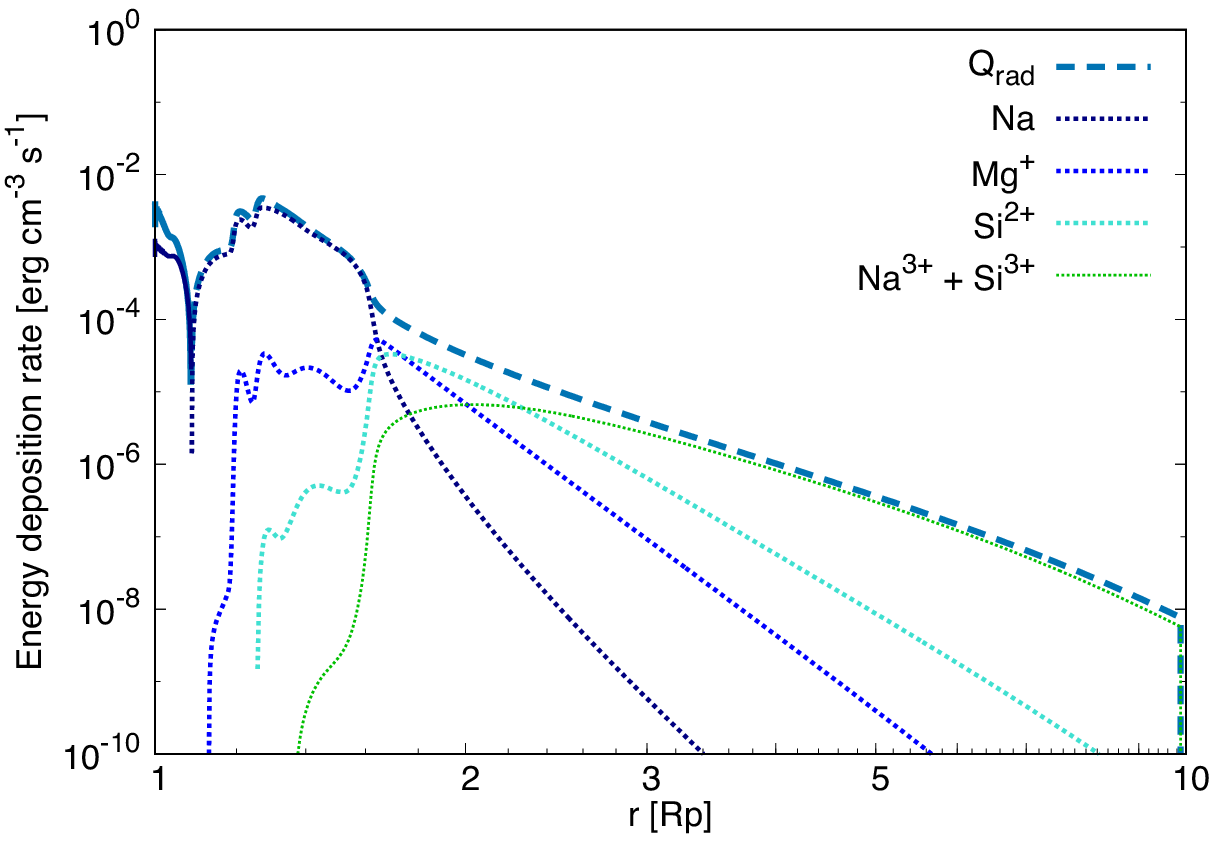}
   \end{center}
 \end{minipage}
 \caption{
Energy budget in the mineral atmosphere on the hot rocky exoplanet (HRE) with mass of 1~$\Mearth$ and radius of 1~$\Rearth$ that is orbiting at 0.02 AU from the solar-type host star with age of 0.1~Gyr. 
The horizontal axis is the radial distance from the planetary center in the unit of planetary radius $R_p$.
\textit{Panel ($a$)}---Heating (solid line) and cooling (dashed line) rates including 
the net energy deposition rate $Q_{\rm{net}}$  (violet),  
the total X+UV energy absorption rate $Q_{\rm{X+UV}}$ (dark pink), 
the primary X+UV heating ($Q_\mathrm{X+UV}$ minus the energy loss rate via photoionization $Q_\mathrm{Ei}$ and via characteristic X-ray emission $Q_\mathrm{X}$; see Eq.~[\ref{eq:qdash}]) $Q'_{\rm{X+UV}}$ (red), 
the heat generation rate due to chemical reactions $Q_{\rm{chem}}$ (green), and the rate of radiative absorption/emission by energy level transitions $Q_{\rm{rad}}$ (royal blue). 
The inset is an enlarged view of the energy budget for 1 $\leq r / R_p \leq$ 1.2.
\textit{Panel ($b$)}---Contributions of the dominant coolant species Na (navy), Mg$^+$ (blue), Si$^{2+}$ (cyan), and  Na$^{3+}$ plus Si$^{3+}$(green) to the radiative heating and cooling by atomic lines, $Q_{\rm{rad}}$ (royal blue).
    }
 \label{fig:Q1-1}
 \end{figure}
 
   \begin{figure}
 \begin{minipage}{\columnwidth}
     ($a$) 
    \begin{center}
  \includegraphics[width=\columnwidth]{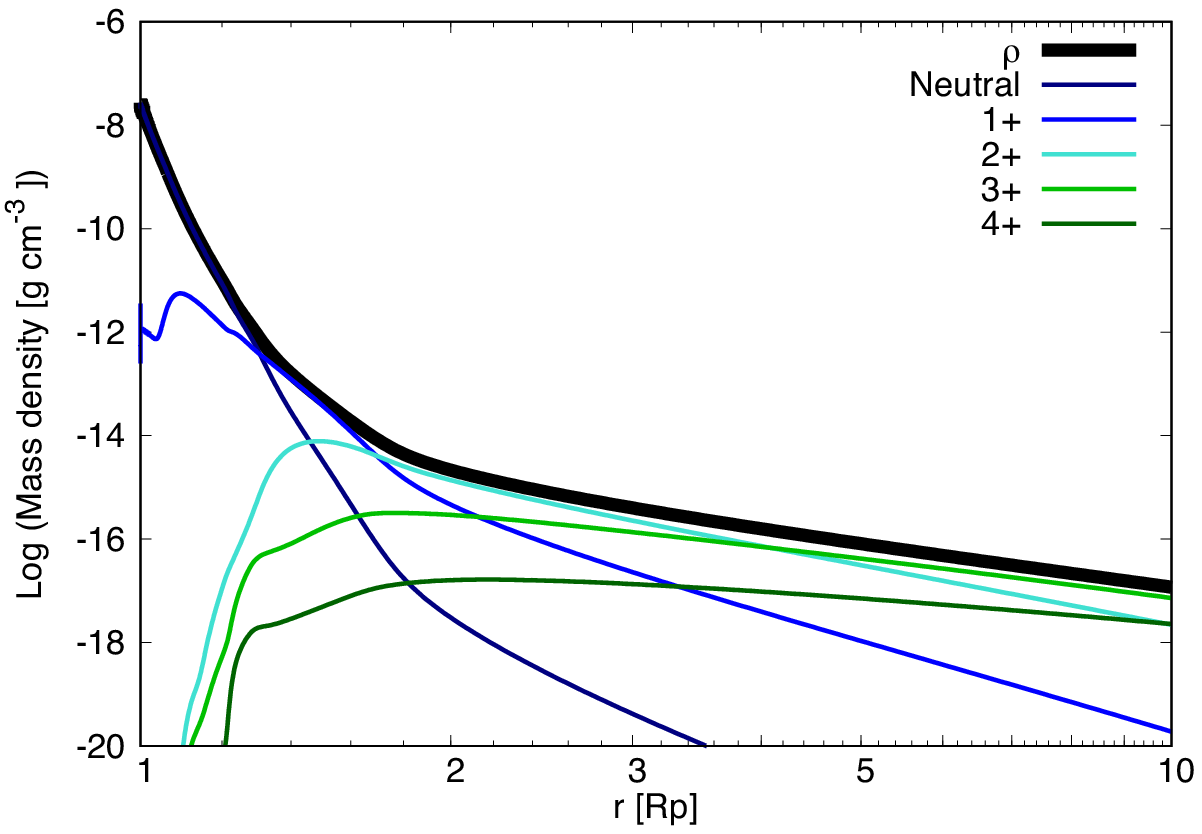}
    \end{center}
 \end{minipage}
 \\
     \begin{minipage}{\columnwidth}
         ($b$) 
   \begin{center}
  \includegraphics[width=\columnwidth]{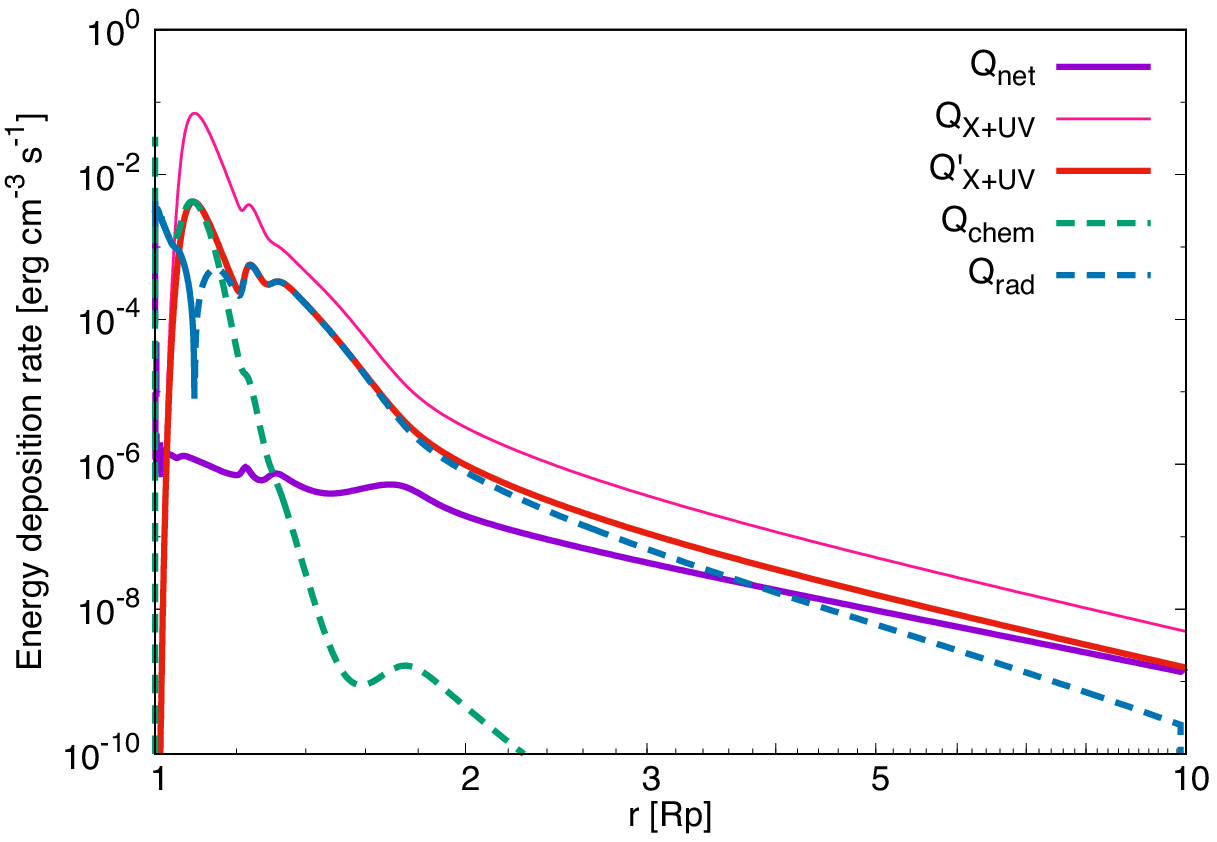}
   \end{center}
 \end{minipage}
 \caption{
Case of a 1-$\Mearth$ hot rocky exoplanet (HRE) orbiting a 1-Gyr-old star. 
 \textit{Panel ($a$)} --- the total mass density (thick black) and the mass densities of all the neutrals (navy) and all the singly-charged (blue), doubly-charged (cyan), triply-charged (green), and quadruply-charged (dark green) ions.
 \textit{Panel ($b$)} --- Heating (solid) and cooling (dotted) rates including 
the net energy deposition rate $Q_{\rm{net}}$  (violet),  
the total X+UV energy absorption rate $Q_{\rm{X+UV}}$ (dark pink), 
the primary X+UV heating (total X+UV deposition rate minus thermal- and photo-ionization rate) $Q'_{\rm{X+UV}}$ (red), 
the heat generation rate due to chemical reactions $Q_{\rm{chem}}$ (green), and the rate of radiative absorption/emission by energy level transitions $Q_{\rm{rad}}$ (royal blue). 
    }
 \label{fig:COMP1-2}
 \end{figure}

Below, we consider an HRE of mass 1 $\Mearth$ with the mineral atmosphere orbiting at 0.02~AU from a solar-type star. 
In Sections~\ref{sssec: as} and \ref{sssec: eb}, we investigate the atmospheric structure and energy budget, respectively, for stellar age of 0.1~Gyr. 
The stellar age dependence is investigated by comparison with results for stellar age of 1~Gyr in Section~\ref{sssec: ad}. 
In both sections we find that the most volatile gas, sodium, can be removed completely from the mineral atmosphere, which motivates us to evaluate the influence of the absence of sodium on the atmospheric escape in Section~\ref{ssec: inl}. 
Table~\ref{tbl:SBC} lists the simulation settings and calculated mass loss rates.

\subsection{Atmospheric structure} \label{sssec: as}
Figure~\ref{fig:rut1-1}$a$ shows the radial profiles of mass density $\rho$ (black), temperature $T$ (orange) and velocity $u$ (cyan) of the atmospheric gas in the case of the 0.1-Gyr-old host star.
Also, Fig.~\ref{fig:COMP1-1} shows the mass/number density profiles of singly- to quadruply-charged ions for ($a$) all the species, ($b$) Na, ($c$) O, ($d$) Mg and ($e$) Si.
First, as found in Fig.~\ref{fig:rut1-1}$a$, the velocity increases with altitude and then exceeds the local speed of sound at $r \simeq$ 3.2~$R_p$, the point of which is termed the sonic point. Thus, a transonic, hydrodynamic escape of the atmosphere turns out to occur.

The mass density decreases with altitude rapidly below $r$ = 1.6~$R_p$, but 
slowly above  
that altitude,
indicating that the pressure scale height increases rapidly from that altitude on. 
This can be readily understood from the fact that temperature increases greatly beyond $r$ = 1.6~$R_p$ to $1.8\times10^4$~K at $r$ = 10~$R_p$. 
Also, as shown in Fig.~\ref{fig:COMP1-1}$a$, 
all the neutrals are photo-ionized almost completely at $r \simeq 1.6 R_p$ by the X+UV irradiation and, then, electrons dominate in number for $r >$ 1.6~$R_p$. 
This reduces the mean mass of a gas particle $\bar{m}$, leading to an increase in the pressure scale height.
For instance, the scale height is larger approximately by a factor of fifteen  at $r = 1.6 R_p$ than that at $r = 1 R_p$, since the mean mass decreases to one-third of the value at $r = 1 R_p$ by ionization,
the temperature doubles and the gravity approximately decreases to $1/1.6^2$ at $r$ $\simeq$ $1.6 R_p$.   

Above $r$ = 1.6~$R_p$, gas density is too low for recombination to occur efficiently.
Thus, regardless of element, all the atoms are completely ionized in the upper atmosphere, as shown in Fig.~\ref{fig:COMP1-1}$b$--$e$.
Also, the similarity in those profiles indicates that gravitational separation is ineffective.
For instance, the fraction of the lightest element O increases by at most 1~\% throughout the atmosphere. 
This is because the bulk velocity of gas is much higher than the diffusion velocity of each species and these ionic gases diffuse along with electrons due to the ambipolar electric force.

Note that a hydrodynamic description is invalid for such a low-density gas that collisions between gaseous particles occur on timescales longer than hydrodynamic one.  
In other words, results shown in this study are invalid above the exobase of the atmosphere, the altitude $r_\mathrm{exo}$ of which is defined as
   \begin{eqnarray}
 \int_{r_{\rm{exo}}}^{\infty} n \sigma dr =1. 
 \label{eq:exos}
\end{eqnarray}
In the present case, $r_{\rm{exo}}$ is estimated to be $\sim$ 9.8 $R_p$ (grey diamond in Fig.~\ref{fig:rut1-1}$a$), using the typical collision cross section $\sigma=3\times10^{-15}$ cm$^2$ and the calculated number densities of ions and neutrals.
Since the sonic point ($r \sim$ 3.2~$R_p$) is located well below the exobase, the hydrodynamic description for the lower atmosphere is valid. 
Although the hydrodynamic description is invalid for $r \gtrsim 9.8 R_p$, the structure of the upper atmosphere has no influence on that of the lower atmosphere, because the fluid motion is already supersonic above the exobase.
We also have confirmed that hydrodynamic description is valid in the other simulations of this study.

\subsection{Energy budget} \label{sssec: eb}
Figure~\ref{fig:Q1-1} shows the energy budget in the mineral atmosphere: Panel~($a$) shows the heating and cooling rates (represented by solid and dashed lines, respectively), which include $Q_{\rm{net}}$, $Q_{\rm{X+UV}}$, $Q'_{\rm{X+UV}}$, $Q_{\rm{chem}}$ and $Q_{\rm{rad}}$; 
panel~($b$) shows the contribution of the effective coolants, Na, Mg$^+$, Si$^{2+}$, Na$^{3+}$ and Si$^{3+}$to $Q_{\rm{rad}}$. 
The radial profiles of those heating/cooling rates are found to be qualitatively different from each other, 
while
all the rates decrease with altitude because of the decrease in gas density (see  Fig.~\ref{fig:rut1-1}$a$). 
The stellar X+UV is completely absorbed in the atmosphere without reaching the lower boundary ($r$ = 1~$R_p$) and energy deposition (i.e., heating) occurs mainly below $r$ = 1.6~$R_p$ (see the pink line, $Q_{\rm{X+UV}}$, and the red line, $Q'_{\rm{X+UV}}$).
The net effect of the thermo-chemical reactions ($Q_{\rm{chem}}$; green dashed line) is
cooling everywhere in the atmosphere and the net cooling rate decreases rapidly with altitude.
Because thermal ionization, which converts thermal energy to ionization energy, dominates over thermal recombination, it acts as a net cooling.
Transition between energy levels occurs by absorption of external radiation (excitation) and radiative emission (de-excitation). Its net effect turns out to be heating in the lower atmosphere below $r \sim 1.08 R_p$ and cooling for $r>1.08 R_p$ (see $Q_{\rm{rad}}$ represented by the cyan solid and dashed lines, respectively, in the inset of Fig.~\ref{fig:Q1-1}$a$).

Atmospheric escape is controlled by energy budget.
As shown in Fig.~\ref{fig:Q1-1}$a$, 
the primary X+UV heating, $Q'_{\rm{X+UV}}$, is almost in balance with the radiative cooling, $Q_{\rm{rad}}$, which is due mainly to line emission, in the region between $r$ $\sim 1.2 R_p$ and $\sim 7.5 R_p$. 
Consequently, the net energy deposition rate, $Q_\mathrm{net}$, is much smaller than $Q_\mathrm{X+UV}^\prime$ and $Q_\mathrm{rad}$.
(Note that the radiative heating due mainly to external-radiation absorption is almost balanced  with the energy consumption via chemical reactions,  $Q_{\rm{chem}}$, below $\sim 1.08 R_p$.)
This indicates that advective energy transfer is ineffective below $7.5 R_p$.
To be more specific, 
the Na-D emission makes a major contribution to the radiative cooling below 1.6~$R_p$ (see the navy line in Fig.~\ref{fig:Q1-1}$b$). 
For $r >$ 1.6~$R_p$, Na is rapidly photo-ionized and, then, the energy budget becomes dominated by emission due to energy-level transition of outermost electrons of multiply-charged ions such as Mg$^+$, Si$^{2+}$,  Na$^{3+}$ and Si$^{3+}$, as shown in Fig.~\ref{fig:Q1-1}$b$. 
Above $7.5 R_p$, instead, $Q_{\rm{net}}$ is larger than $Q_{\rm{rad}}$, indicating that the absorbed X+UV energy is used to drive the advection (i.e., escape) of the atmosphere.

We define the net X+UV heating efficiency, $\epsilon_\mathrm{X+UV}$, as 
\begin{eqnarray}
	\epsilon_\mathrm{X+UV} = \frac{\int_{r \le r_s} r^2 Q_{\rm{net}} dr}{\int_{r \le r_s}  r^2 Q_{\rm{X+UV}} dr},
	\label{eq:epsi}
\end{eqnarray}
where $r_s$ is the radial distance of the sonic point. 
Here we have considered only the region below $r = r_s$ since information cannot propagate from supersonic regions down to subsonic ones.
Integrating Eq.~(\ref{eq:epsi}) over the range where the X+UV heating rate is higher than the line absorption rate (i.e., $Q_{\rm{X+UV}}\geq Q_{\rm{rad}}$), we estimate that $\epsilon_\mathrm{X+UV}$
$\simeq 1.6\times10^{-3}$. 
The mass loss rate comes out to be 0.3~$\Mearth \, \mathrm{Gyr}^{-1}$ or $5.7 \times 10^{10}$~g s$^{-1}$.

\subsection{Stellar age dependence} \label{sssec: ad}

   \begin{figure}
 \begin{minipage}{\columnwidth}
     ($a$) 
    \begin{center}
  \includegraphics[width=\columnwidth]{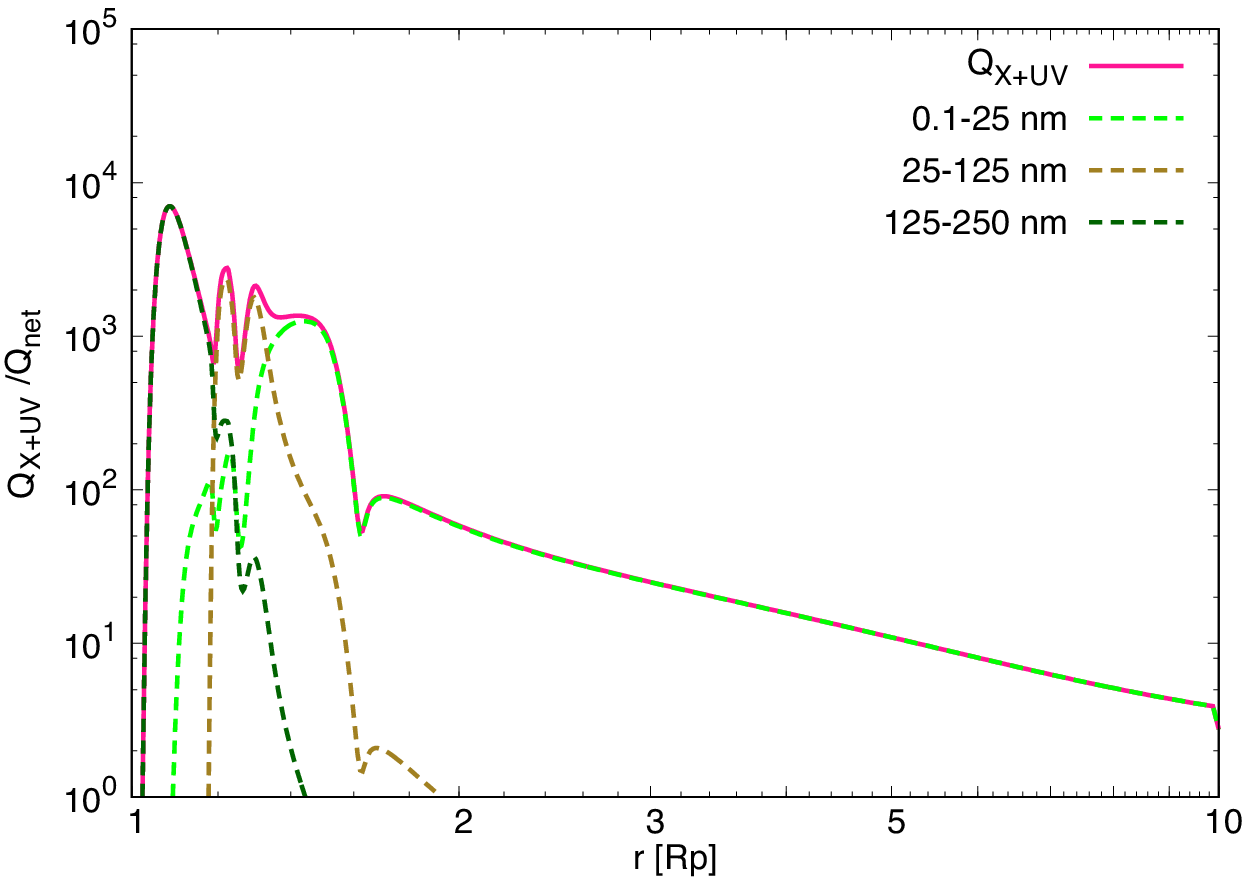}
    \end{center}
 \end{minipage}
 \\
     \begin{minipage}{\columnwidth}
         ($b$) 
   \begin{center}
  \includegraphics[width=\columnwidth]{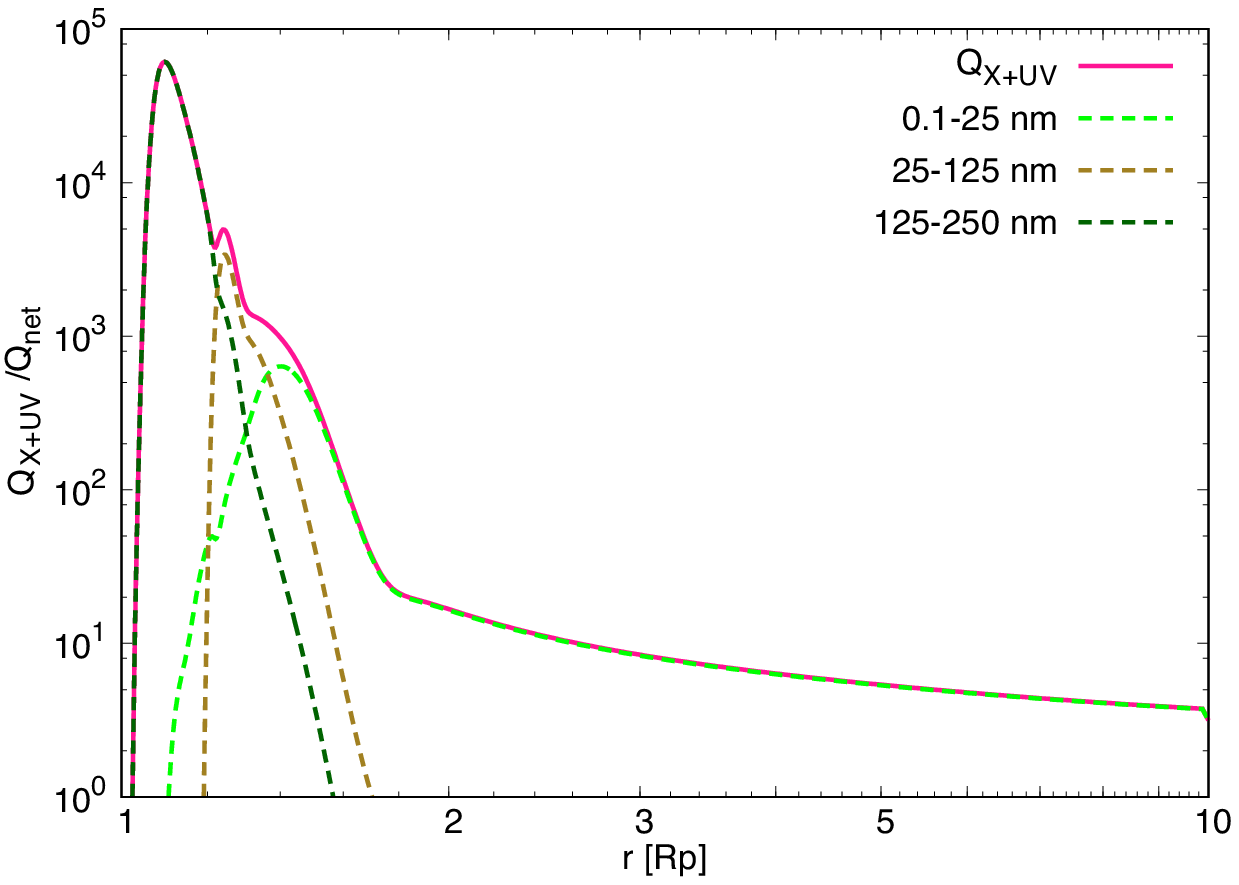}
   \end{center}
 \end{minipage}
 \caption{
Inverse of the local heating efficiency, $(Q_{\rm{net}}/Q_{\rm{X+UV}})^{-1}$ (see Eq.~[\ref{eq:epsi}]), at each altitude in the mineral atmosphere of a 1-$\Mearth$ hot rocky exoplanet (HRE) orbiting a G-type host-star with age of ($a$) 0.1~Gyr and ($b$) 1~Gyr.
The total X+UV energy absorption rate ($Q_{\rm{X+UV}}$; solid dark-pink) and the contributions of absorption rates at $\lambda=$0.1--25~nm (X-ray; dashed light-green); $\lambda=$25--125~nm (EUV; dashed brown); and $\lambda=$125--250~nm (FUV; dashed dark-green) to $Q_{\rm{X+UV}}$ are normalized by the net energy deposition rate, $Q_{\rm{net}}$.
    }
 \label{fig:Qabs_wave}
 \end{figure}

Stellar age affects the structure and escape of the atmosphere, since the stellar emission spectrum including X-ray and UV ($\lesssim$~250~nm) differs with age (see Fig.~\ref{fig:SUV}). 
Figure~\ref{fig:rut1-1}$b$ is the same as Fig.~\ref{fig:rut1-1}$a$, but for the stellar age of 1~Gyr.
Comparing both panels, one finds that each quantity undergoes a qualitatively similar change, although being slightly smaller as a whole in the old-star case than in the young-star case. 
Such similarity and small difference can be understood from the energy budget as follows.

Figure~\ref{fig:COMP1-2} shows ($a$) the mass density profiles of all the neutrals and ions and ($b$) the energy budget in the atmosphere.
Comparing Fig.~\ref{fig:COMP1-2}$a$ with Fig.~\ref{fig:COMP1-1}$a$, one finds that the ionization degree is slightly lower in the old-star case than in the young-star case. 
For example, at $r=3R_p$, 
the major species is
secondary-charged ions
in the former case, while being thirdly-charged ones in the latter case. 
Such a difference in ionization degree is due simply to that in the incident X+UV flux, which can ionize those atoms.
The older star emits $\sim$10 times lower flux of X+EUV than the younger star, whereas
the fluxes of FUV and visible light scarcely differ between the two stars.

The mass loss rate $\dot{m}$ is estimated to be $3.7 \times 10^{-2}$~$\Mearth$~Gyr$^{-1}$ in the old-star case, which is approximately ten times lower than that in the young-star case. 
Also, the net X+UV heating efficiency, $\epsilon_\mathrm{{X+UV}}$, decreases from $1.6 \times 10^{-3}$ in the young-star case to $2.2\times10^{-4}$ in the old-star case.
At first glance, it may sound strange that 
the X+UV heating efficiency is greatly changed despite of similarity in the dominant heating/cooling processes (compare Figs.~\ref{fig:Q1-1}$a$ and \ref{fig:COMP1-2}$b$)  and in the X+UV irradiation energy integrated over $\lambda \lesssim$ 250~nm ($\simeq 4.9 \times 10^6$~erg~cm$^{-2}$~s$^{-1}$ for the younger case and $\simeq 3.4 \times 10^6$~erg~cm$^{-2}$~s$^{-1}$ for the older case, Fig.~\ref{fig:SUV}).
Such an interpretation is, however, incorrect.
Figure~\ref{fig:Qabs_wave} shows the ratio of the X+UV heating rate to the net energy deposition rate ($Q_{\rm{net}}$), equivalent to the inverse of the local heating efficiency, where we include the contributions of respective wavelength regions
 $\lambda$ = 0.1--25~nm (``X-ray''; dashed light-green), 25--125~nm (``EUV''; dashed brown) and 125--250~nm (``FUV''; dashed dark-green) for the young-star ($a$) and old-star ($b$) cases. Note that we show the inverse of the local heating efficiency in order to clarify the contributions of respective wavelength regions.
The inverse of the local heating efficiencies for X-ray and EUV wavelengths change with stellar age only slightly, whereas
the peak value for FUV wavelengths is lower by a factor of 10 in the old-star case than in the young-star case. 
In addition, the X-ray contribution to the inverse of the local heating efficiency is lower than for both EUV and FUV for both stellar ages. Therefore,
the X-ray and EUV make the dominant contribution to the atmospheric escape, 
while FUV accounts for a large proportion of $F_\mathrm{X+UV}$; 
for example, in the young-star case, 
$\int F_\mathrm{X+UV} d\lambda \simeq 3.9 \times 10^5$, $1.4 \times 10^6$, and $3.1 \times 10^6$~erg~cm$^{-2}$~s$^{-1}$ for X-ray, EUV, and FUV, respectively.

To gain a deeper understanding of the sensitivity of mass loss rate to stellar spectrum, we perform the simulations by artificially raising or lowering the young-star emission flux by 10-fold in specific wavelength regions (X-ray, EUV, or FUV).
Table~\ref{tbl:SML} shows the calculated mass loss rates for the different cases. 
The mass loss rate is found to be relatively insensitive to difference in FUV.
This means that the X-ray and EUV effectively drive the atmospheric motion rather than FUV. 
Thus, the mass loss rate in the old-star case is approximately ten times lower than that in the young-star case because the X-ray and EUV fluxes are lower. 
Note that although FUV is less important for the mass loss rates,  the difference in the net X+UV heating efficiencies, $\epsilon_\mathrm{{X+UV}}$, between the young and old cases is due to that in FUV heating efficiency
because $F_\mathrm{X+UV}$ accounts for a large proportion of FUV.
In Section~\ref{sssec: em}, we discuss why such a short-wavelength radiation dominates the escape of the mineral atmosphere.

      \begin{table}
          \caption{Calculated mass loss rates for different X+UV irradiation fluxes}
    \label{tbl:SML} 
      \begin{center}
  \begin{tabular}{ l c c c} 
    \hline \hline
    Case  & Wavelength of changed photon flux & Mass loss rate  \\
     &  [nm]  & [$\Mearth/$Gyr] \\
    \hline \hline
    0.1 $\times$ X-ray & 0.1--25 & 1.3$\times10^{-1}$  \\ 
 10 $\times$ X-ray & 0.1--25 & 1.0  \\ 
  0.1 $\times$ EUV  & 25--125 & 1.9$\times10^{-1}$  \\ 
   10 $\times$ EUV  & 25--125 & 1.1  \\ 
    0.1 $\times$ FUV & 125--250 & 2.9$\times10^{-1}$  \\ 
     10 $\times$ FUV & 125--250 & 3.8$\times10^{-1}$  \\ 
    \hline
  \end{tabular}  
  \begin{flushleft}
Note: In Section~\ref{sssec: ad}, we have done a sensitivity test to stellar spectra by artificially raising/lowering the emission flux by 10-fold in specific wavelength regions for the 0.1-Gyr-old star (Case~A). Here we call the radiation with wavelength of 0.1--25~nm, 25--125~nm, and 125--250~nm X-ray, EUV, and FUX, respectively.
\end{flushleft}
    \end{center}
    \end{table}

\subsection{Impact of sodium on atmospheric escape}\label{ssec: inl}

   \begin{figure}
   \begin{center}
  \includegraphics[width=\columnwidth]{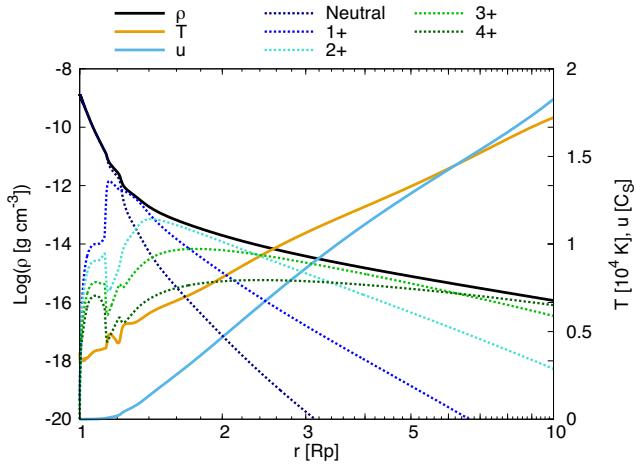}
   \end{center}
   \caption{
   Profiles of mass density $\rho$ (black), temperature $T$ (orange), and velocity $u$ 
(cyan)  in the Na-free mineral atmosphere on the  hot rocky exoplanet (HRE) with mass of 1~$\Mearth$ and inner radius of 1~$\Rearth$ that  is orbiting at 0.02 AU around the solar-type host-star with age of 0.1 Gyr. 
The velocity is given in the unit of the local sound velocity, namely, the Mach number.
Dotted lines show the mass densities of all the neutrals (navy) and all the singly-charged (blue), doubly-charged (cyan), triply-charged (green), and quadruply-charged (dark green) ions.
}
 \label{fig:rut1-1wona}
 \end{figure}
   
   \begin{figure}
 \begin{minipage}{\columnwidth}
 ({\it a})
    \begin{center}
  \includegraphics[width=\columnwidth]{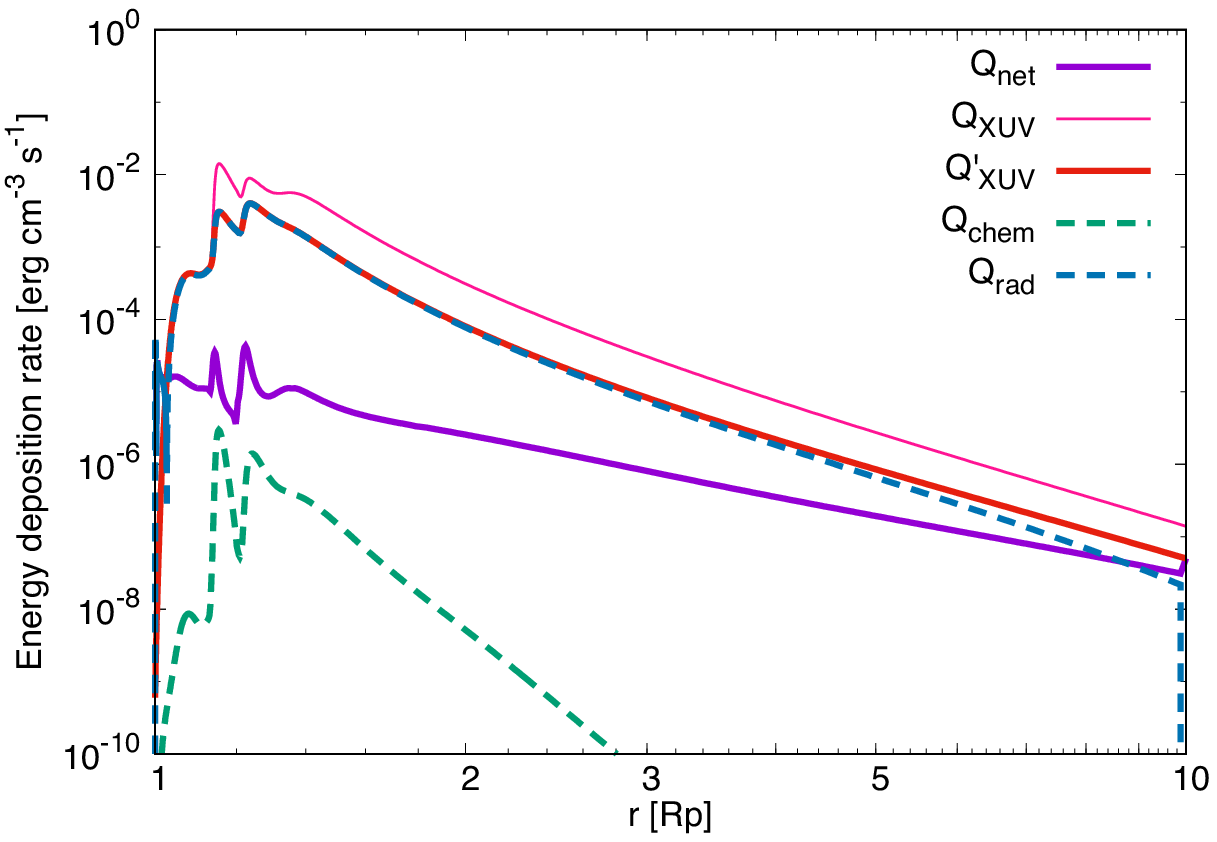}
    \end{center}
 \end{minipage}
     \begin{minipage}{\columnwidth}
 ({\it b})
   \begin{center}
  \includegraphics[width=\columnwidth]{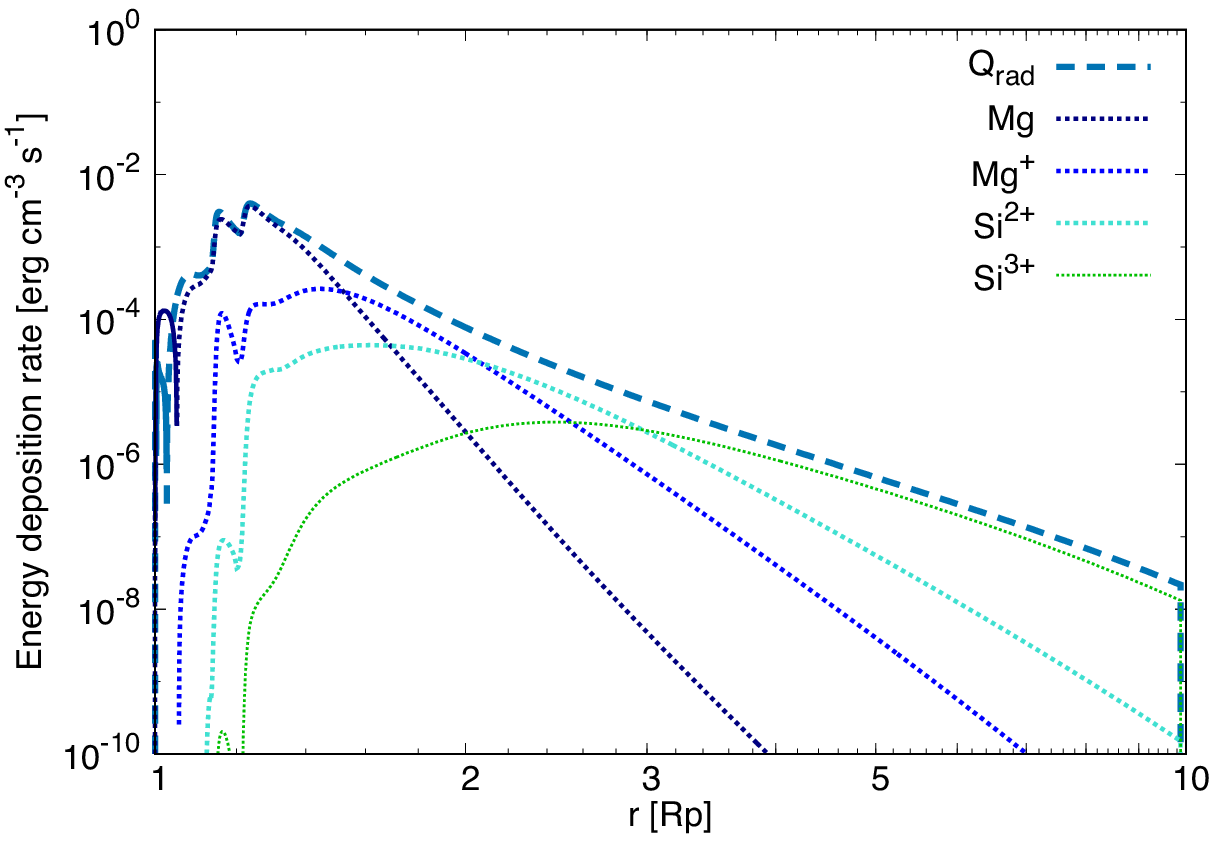}
   \end{center}
 \end{minipage}
 \caption{
 Energy budget in the Na-free mineral atmosphere on the hot rocky exoplanet (HRE) with mass of 1~$\Mearth$ and inner radius of 1~$\Rearth$ that is orbiting at 0.02 AU around the solar-type host-star with age of 0.1 Gyr.
\textit{Panel (a)} --- Heating (solid) and cooling (dotted) rates including 
the net energy deposition rate $Q_{\rm{net}}$  (violet),  
the total X+UV energy absorption rate $Q_{\rm{X+UV}}$ (dark pink), 
the primary X+UV heating ($Q_\mathrm{X+UV}$ minus the energy loss rate via photo-ionization $Q_\mathrm{Ei}$ and via characteristic X-ray emission $Q_\mathrm{X}$; see Eq.~[\ref{eq:qdash}]) $Q'_{\rm{X+UV}}$ (red), 
the heat generation rate due to chemical reactions $Q_{\rm{chem}}$ (green), and the rate of radiative absorption/emission by energy level transitions $Q_{\rm{rad}}$ (royal blue). 
\textit{Panel (b)} --- Contributions of the dominant coolant species Mg (navy), Mg$^+$ (blue), Si$^{2+}$ (cyan), and Si$^{3+}$(green) to the radiative heating and cooling by atomic lines, $Q_{\rm{rad}}$ (royal blue).
    }
 \label{fig:Q1-1wona}
 \end{figure}

Alkali metals such as Na and K are minor elements in normal rocky planets (e.g., $\sim$ 0.1~wt\% in the bulk silicate Earth composition), although they are major components in mineral atmospheres \citep{Schaefer+09,Miguel+11,Ito+2015}. 
The mass loss rate derived above (see Table~\ref{tbl:SBC}) is massive enough to remove completely Na from the atmosphere and interior.
Indeed, one can estimate that the Na of 0.1~wt\% of 1~$\Mearth$ is removed in $\sim10$ Myr, 
provided convection in the magma ocean is vigorous enough to supply Na quickly to the surface. 
After such a selective loss of Na, the major atmospheric components become SiO, Si, Mg, O and O$_2$ \citep{Schaefer+09}.

Here, we investigate the impact of absence of Na on the escape of the mineral atmosphere.
Figure~\ref{fig:rut1-1wona} shows the radial profiles of mass density, temperature, and velocity (solid lines) and those of mass densities of all the neutral atoms and ions (dotted lines) in the Na-free mineral atmosphere on an Earth-mass HRE.
Also, Fig.~\ref{fig:Q1-1wona} shows ($a$) the energy budget in the atmosphere and ($b$) the contribution of the effective coolants such as Mg, Mg$^+$, Si$^{2+}$ and Si$^{3+}$ to $Q_{\rm{rad}}$. 
In those calculations, we have
artificially set the molar fraction of Na to be zero, and have re-normalized the mole fractions of the other species from those set above so that the sum of the mole fractions becomes unity.
Note that we have set the lower boundary at a lower pressure (= 20 dyne cm$^{-2}$) in this case than in the cases above (= 300 dyne cm$^{-2}$). 
This is because the first ionization energy of Na and Mg corresponds to ~250~nm and 160~nm in terms of wavelength, respectively, and, thus, the Na-free atmosphere absorbs the incident X+UV photons only of wavelength $\lesssim$ 160~nm, which are completely absorbed above the pressure.

Comparing Figs.~\ref{fig:COMP1-1}$a$ and \ref{fig:rut1-1wona}, 
one finds that the number of singly-charged ions is much smaller near the lower boundary in the Na-free atmosphere than that in the Na-containing atmosphere. 
This is because the other atoms have higher ionization energies than Na. 
Because of such a low number density of electrons deep in the atmosphere, the cooling via thermo-chemical reactions ($Q_{\rm{chem}}$) makes a tiny contribution to the energy budget throughout the atmosphere, as shown in Fig.~\ref{fig:Q1-1wona}$a$.
Also, the Na-free atmosphere is warmer below 1.6~$R_p$ than the Na-containing atmosphere in which Na is the dominant coolant  (see Fig.~\ref{fig:Q1-1}$b$). 
Instead, in this region of the Na-free atmosphere, line emission by Mg occurs effectively because of high temperature; Namely, Mg is the dominant coolant instead of Na, as shown in Fig.~\ref{fig:Q1-1wona}$b$. 
The mass loss rate is estimated to be $\dot{m}=4.4 \times 10^{-1}$ $\Mearth$~Gyr$^{-1}$, which is $\sim$ 50~\% higher than that for the Na-containing atmosphere. 
In summary, even in the atmosphere without the strongest coolant Na, Mg acts as a substitute to cool the atmosphere efficiently and, consequently, brings about a similar mass loss rate with that in the Na-containing atmosphere.

\section{Discussion}\label{sec: dis}
\subsection{XUV heating efficiency} \label{ssec: ee}
In Section~\ref{sec: r}, our hydrodynamic simulations have shown that the net X+UV heating efficiency for the mineral atmosphere, $\epsilon_\mathrm{X+UV}$, is on the order of $10~^{-4}$--$10^{-3}$. 
Such values are much larger than the simple estimates ($\sim 10^{-13}$--$10^{-6}$) that we made in Section~\ref{sec: int}, taking Na-D line cooling into account, under the assumption that the mineral atmosphere is optically thin in LTE for $T$ = 3000--5000~K. On the other hand, the typical value of heating efficinty is as large as $\sim$ 0.1
for hydrogen-dominated atmospheres and oxidized
terrestrial atmospheres \citep[e.g.,][and references therein]{Tian2015}.
Here, we provide an interpretation of the cooling process and escape mechanism of the X+UV-irradiated mineral atmosphere that control the X+UV heating efficiency.

\subsubsection{Dominant cooling processes}\label{sssec: dp}
The dominant cooling process in the mineral atmosphere is the line emission during electronic transitions in the coolant species such as Na, Mg, Mg$^+$, Si$^{2+}$, Na$^{3+}$ and Si$^{3+}$, as shown in Figs.~\ref{fig:Q1-1} and \ref{fig:Q1-1wona}. 
Among them, Na is the most effective coolant, because of the permitted transition of the peripheral electron from the 3p level to the 3s level (D line) with large Einstein coefficient (equivalent to $6.23\times10^7$~s$^{-1}$) and low excitation energy (equivalent to $2.4\times10^4$~K). 
To be more specific, the peripheral electron of Na (3s$\rightarrow$3p) is readily excited through collision with electrons produced through photo-ionization of metals, 
but is de-excited (3p$\rightarrow$3s) immediately through the D-line emission.
In general, alkali metals such as Na and alkaline-earth metals such as Mg, which have one and two peripheral electrons, respectively, can generate line emission in similar ways. 
Among the species considered in this study, 
Na, Mg, Mg$^+$, Si$^{2+}$, and Si$^{3+}$ 
have such electron configurations, namely, [Ne] 3s$^1$, [Ne] 3s$^2$, [Ne] 3s$^1$, [Ne] 3s$^2$, and [Ne] 3s$^1$, respectively.
Thus, almost all of the absorbed X+UV energy ends up being lost by such efficient emission. 
This is why the X+UV heating efficiency is sufficiently low in the mineral atmosphere of the 1-$\Mearth$ HRE.

Indeed, as shown with the royal-blue solid lines ($Q_\mathrm{rad}$) in Figs.~\ref{fig:Q1-1} and \ref{fig:Q1-1wona},
the net effect of stellar irradiation is heating
in the deep atmosphere where temperature is below the radiative equilibrium temperature determined mainly by atomic-line cooling.
Provided radiative absorption and emission are in equilibrium at a wavelength $\nu_0$,
the equilibrium temperature $T'$ is given by 
\citep[e.g.,][]{Chatzikos+2013} 
\begin{eqnarray}
\frac{c^2J_\mathrm{ext, \nu_0}}{2h\nu_0^3}=\frac{1}{\exp(h \nu_0 /k_bT')-1}.
\label{eq:eqT}
\end{eqnarray}
For the Na-D, $T^\prime$ is estimated to be $\sim$~3300~K under the conditions considered in this study 
($J_{\mathrm{ext}, \nu_0}\sim1.6\times10^{-6}$~erg cm$^{-2}$ Hz$^{-1}$ and $\nu_0\sim 5.1 \times 10^{14}$~Hz).
Since the temperature near $r = 1 R_p$ is about 3000~K (see Fig.~\ref{fig:rut1-1}) and relatively lower than the above value of $T^\prime$,
the deep atmosphere is heated via the absorption by Na and other neutrals. 
 Whereas the thermal energy equivalent to such an equilibrium temperature ($\sim 2 \times 10^{10}$~erg \, g$^{-1}$) is sufficiently smaller than the planetary gravitational potential of HREs with mass of 1~$M_\oplus$ ($\sim 6 \times 10^{11}$~erg \, g$^{-1}$), 
the former is comparable to the latter for sub-Earth-mass HREs.
If such heating processes occur in their atmosphere, sub-Earth-mass HREs are predicted to lose a substantial fraction of planetary mass via the escape of the mineral atmosphere. 
Detailed investigation will be done in our forthcoming paper.

\begin{figure}
   \begin{center}
  \includegraphics[width=\columnwidth]{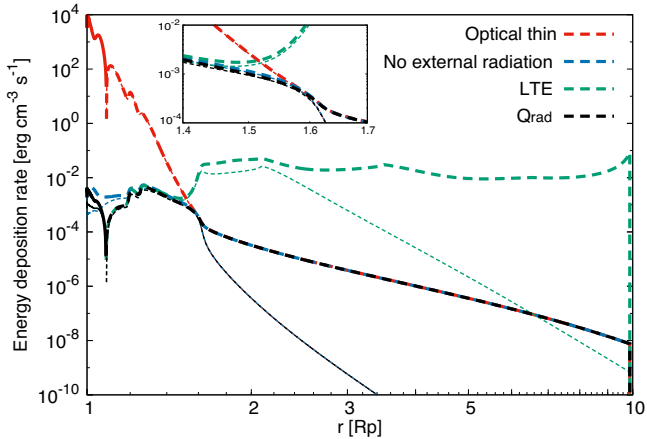}
   \end{center}
   \caption{Radiative line heating (solid) and cooling (dashed) rates, $Q_{\rm{rad}}$ (black), in the mineral atmosphere on the hot rocky exoplanet (HRE) with mass of 1~$\Mearth$ and radius of 1~$\Rearth$ that is orbiting at 0.02 AU from the solar-type host star with age of 0.1~Gyr. 
   For each contribution to be clarified, $Q_{\rm rad}$ calculated under the assumption of zero optical thickness ($P_e=1$), no external radiation ($J_\mathrm{ext}=0$) or LTE are also shown by red, royal blue and green lines, respectively. 
   Note that in the LTE case, we have calculated $Q_{\rm{rad}}$ by imposing a sufficiently high number density of electrons (10$^{15}$cm$^{-3}$) everywhere.
   Also, the contribution of Na to the energy deposition rate in each case is shown by a thin line. The inset is an enlarged view of the energy budget for $1.4\leq r/R_p\leq 1.7$.
}
 \label{fig:radcomp}
 \end{figure}

The effects of optical thickness and non-LTE have great influence on
the radiative heating and cooling that occur through energy transitions. 
Figure~\ref{fig:radcomp} shows the radiative line heating/cooling rate,  $Q_{\rm{rad}}$ (black), 
for the Na-bearing atmosphere illuminated by the young host star (Case~A in Table~\ref{tbl:SBC}).
To quantify the significance of those effects, we also show $Q_{\rm rad}$ calculated under the assumption of
zero optical thickness (red; $P_e=1$), no external radiation (royal blue; $J_\mathrm{ext}=0$) or LTE (green).
It turns out that the optical thickness reduces $Q_{\rm{rad}}$ by up to nine orders of magnitude at $r$~$\lesssim1.6R_p$, while 
the non-LTE effect does so by up to seven orders of magnitude at $r$~$\gtrsim1.4R_p$. 
At $r = 1.6 R_p$, the non-LTE effect is found to be already important, while
the optical depths for some of the radiative lines are still larger than unity and also the LTE condition starts to be broken below the altitude (see the inset). 
For example, although the optical depth for the Na-D is as large as $\sim300$ at $r=1.6R_p$, the Na cooling rate (black thin line) is as almost the same as that for zero optical thickness (red thick line). This is due to the non-LTE effect limited by the collisional transition (green thin dashed line).
Thus, the non-LTE effect rather than the effect of optical thickness suppresses the cooling at $r \geq 1.6 R_p$.
This is why the dominance of these two effects on $Q_{\rm{rad}}$ is switched at $r=1.4$--1.6~R$_p$. 
Finally, the impact of external radiation is smaller than the other two effects throughout the atmosphere, while it changes the net radiative effect from cooling to heating in the deep atmosphere, as we explained in the previous paragraph.

\subsubsection{Escape mechanism}\label{sssec: em}
\begin{figure}
   \begin{center}
  \includegraphics[width=\columnwidth]{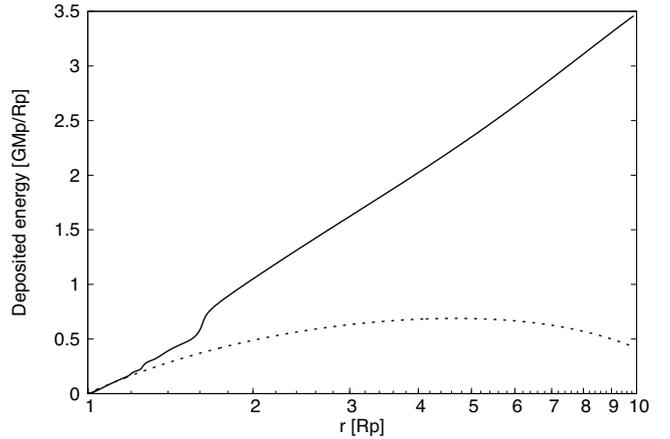}
   \end{center}
   \caption{The sum of enthalpy, kinetic energy, and gravitational potential per gas particle (in the unit of $GM_p/R_p$) relative to that at $r$ = 1~R$_p$ in the mineral atmosphere on the hot rocky exoplanet (HRE) with mass of 1~$\Mearth$ and radius of 1~$\Rearth$ that is orbiting at 0.02 AU from the solar-type host star with age of 0.1~Gyr. The contribution of the gravitational potential is plotted by the dotted line separately. 
}
 \label{fig:HV}
 \end{figure}

The escape of the mineral atmosphere is driven by the following mechanism.
As demonstrated in Section~\ref{sec: r}, almost all of the primary X+UV heating energy is lost by the radiative emission over a wide region below several planetary radii. 
In the sense that advection accounts for only a small fraction of the energy budget, 
the atmospheric thermal structure 
is almost in the radiative equilibrium where the heating by X+UV and the radiative absorption/emission by energy level transitions balance with each other.
The small amount of residual energy is partitioned into the enthalpy, $\gamma P / [(\gamma-1)\rho]$, kinetic energy, $u^2/2$, and potential (gravity plus tide) of the atmospheric gas, $V(r)$, which is given by
\begin{equation}
V(r)= - \frac{GM_p}{{r}}-\frac{GM_*}{a-r}
-\left(\frac{M_*}{M_*+M_p} a-r  \right)^2 \frac{G(M_*+M_p)}{2a^3}
\label{eq:Vext}
\end{equation}
(see Eqs.~(\ref{eq: total energy}) and (\ref{eq: fext})). 
Figure~\ref{fig:HV} shows the sum of the kinetic energy, enthalpy 
and potential per gas particle measured from that at $r$ = 1~$R_p$ in Case A. 
To clarify each contribution, we also plot only the potential by the dotted line. 
The potential $V (r)$ peaks at $r \simeq$ 4.7~$R_p$ (or the L1 point) and its peak value, $V_{\rm max}$, is $\sim$ 0.7  $GM_p/R_p$.

For the gas at an altitude to end up escaping from the planet, its kinetic energy plus enthalpy (or $E + P/\rho$ hereafter denoted by $W$) must be larger than $V_{\rm max} - V$. 
At $r \lesssim$ 1.2~$R_p$, $W \ll (V_{\rm max} - V)$.
At $r=1.2$--$1.6R_p$, 
$W$ increases up to $0.1 (V_{\rm max}-V)$ mainly because of an increase in number density of electrons by
the ionization of firstly-charged ions and a slight rise in the temperature (see Fig.~\ref{fig:rut1-1}a).
Beyond 1.6~R$_p$, the 
temperature increases greatly with altitude because of inefficient non-LTE cooling, which is limited by collisional transitions. 
The temperature increase and photo-ionization lead to reducing the mean mass of a gas particle, resulting in an increase in $W$. 
Consequently $W$ exceeds $V_{\rm max}-V$, so that the atmosphere transitions from a hydrostatic state to a hydrodynamic one.

Outward motion occurs below $r=1.2R_p$, since otherwise the steady state is never maintained. 
This means that the outward motion is driven not by the X+UV heating, but by
the pressure gradient caused by the advection of the hydrodynamic region. 
The energy that the atmosphere receives via X+UV irradiation is almost entirely lost by spontaneous emission (or radiative cooling) caused by the alkali metals and alkaline-earth-metal-like ions,
as described in Section~\ref{sssec: dp}.
Thus, such radiative cooling, which occurs over a wide region, controls the escape of the mineral atmosphere.
This is obviously different from the energy-limited escape of hydrogen-dominated atmospheres which is controlled by the radiative cooling in a narrow region of the deep atmosphere (for example, as for hot Jupiters' hydrogen-dominated atmospheres, the cooling of H$^{3+}$ is dominant in the narrow region between $1.0$ and $1.1 R_p$; \citet{GM2007b}). 
In addition,
at the sonic point 
($r$~$\sim3.1R_p$; see Fig.~\ref{fig:rut1-1}), more of the deposited energy is partitioned into the enthalpy and kinetic energy than into the potential energy.
In such a case, the escape flux is not necessarily proportional to the gravitational potential \citep[see also Sec.~3.5 of][]{GM2007b}). 
This is also different from the energy-limited escape.
Note that the gravity dependence of the mass loss rate of the mineral atmosphere  will be investigated in our forthcoming paper.

Also, 
such a mechanism for the atmospheric motion naturally explains the dependence of the  escape rate on stellar X+UV spectra shown in Section~\ref{sssec: ad}. 
Since neutrals are completely ionized deep in the atmosphere, 
ions are the dominant absorbers in the region above 
$r=1.2R_p$ where the absorbed X+UV energy
drives the atmospheric motion effectively, as shown in Fig.\ref{fig:HV}.
Those ions absorb the incident stellar radiation of $\lambda\leq$ 90~nm, 
while the atoms except O absorb longer waves of $\lambda > 125$~nm. 
This is why 
the short-wavelength radiation of $\lambda\leq$ 125~nm makes the dominant contribution to the atmospheric escape, as shown in Section.~\ref{sssec: ad}.
Therefore, the mass loss rate of the mineral atmosphere strongly depends on 
the flux of X+UV with energy high enough to ionize the 
ions, which are the major gas species at high altitudes.

\subsection{Caveats} \label{ssec: c}

\subsubsection{Eddy diffusion} \label{sssec: ioe}
We have ignored eddy diffusion in this study, for simplicity.
Eddy diffusion carries neutral species upwards,
modifying the profiles of mass density, temperature and velocity.
Three-dimensional simulations of atmospheric circulation are necessary to know the efficiency of eddy diffusion.
The effect of eddy diffusion itself is, however, negligible at altitudes higher than $r\sim1.1R_p$, where the optical thickness of the Na gas is unity
because the photo-ionization occur more rapidly than the eddy diffusion.
For instance, at such high altitudes,  the eddy diffusion timescale is estimated to be of the order of $10^3$~s 
with use of the typical diffusion coefficient, 
10$^{11}$ cm$^2$ s$^{-1}$, from \citet{Parmentier+2013}, while the photo-ionization timescale of Na and also advection timescale are of the order of $10^2$~s. 
 
 \subsubsection{Inhibition of ion escape: magnetic field and stellar wind} \label{sssec: iie}
As demonstrated in Section~\ref{sec: r}, the escaping gas is completely ionized and charged. 
Thus, its motion is subject to magnetic fields.
In particular, escape of ions is inhibited by a planetary magnetic field, 
provided the magnetic pressure dominates over the ions' thermal pressure.
While magnetic fields of HREs have not been detected, 
if HREs have large closed magnetospheres, 
then the escaping ions in mineral atmospheres would be bound by the magnetic fields and return to the planets.
On the other hand, if the magnetic fields of HREs are open to the stellar wind at sufficiently high altitudes, the escaping ions go out along the magnetic field lines \citep[see][for the cases of the Earth and Mars]{Terada+2009}. 
Detailed investigation of interaction between the stellar wind and planetary magnetic field is beyond the scope of this study and will be done in our future study.

 \subsubsection{Horizontal variation
 }
 \label{sssec: as}
In this study, we have considered a radially one-dimensional flow of the atmosphere above the sub-stellar point and assumed spherical symmetry when estimating the mass loss rate. 
However, HREs tidally locked to their host star obviously have axially-asymmetric surface environments.
The mineral atmosphere 
of $\lesssim 0.1$~bar
is so tenuous that 
horizontal heat transport via atmospheric winds is inefficient.  Consequently, at each location, the
atmospheric pressure is almost determined by the vapour pressure of magma for the local equilibrium temperature \citep{Casten+2011,Ding+2018}. 
This means that the magma ocean is localized to some areas of the dayside; 
for the equilibrium temperature of 3000~K, the dayside is covered entirely with magma, while the magma ocean is localized to limited areas for lower equilibrium temperatures.
In addition,
since the atmosphere with vapor pressure near the edge of the 
magma ocean ($\lesssim$ 1500~K) is optically thin against X-ray and UV 
radiation \citep[see Fig.~2 of][]{Kite+2016}, 
photo-evaporation would never take place near the terminator.
To quantify
the effects of asymmetry in detail, three-dimensional atmospheric models are needed. In any case, such effects do not affect our conclusion that the escape of the mineral atmospheres of Earth-mass HREs is inefficient.

Also, \citet{Kite+2016} propose that 
gases vaporizing at
the sub-stellar point are transported by atmospheric winds horizontally 
and then  
will condense mostly before reaching the edge of the 
magma ocean on the dayside.
This process may cause compositionally variegated atmosphere and magma ocean, depending on the efficiency of mixing in the magma ocean and mantle. 
This may also change the main gas components to CaO and AlO, which are the most refractory gases. 
Then, the atmospheric pressure is much lower than the Na-containing atmosphere and may be too small to absorb X-ray and UV photons, depending on surface temperature (see their Fig.~2).
If thick enough, 
the atmosphere may photo-evaporate with
a similar mass loss rate 
to the Na-containing atmosphere. 
This is because CaO and AlO, after dissociation and ionization,
possibly become strong coolants as their ions such as Ca$^+$ and Al$^{2+}$ behave like alkali metal atoms in the sense that a single electron exists in the outermost shell.

 \subsubsection{Cooling rates of other main gas components} \label{sssec: kfe}
 Although gas-melt equilibrium calculations \citep{Schaefer+09, Miguel+11, Ito+2015} show that the mineral atmosphere contains various elements,
 this study has considered only Na, O, Si and Mg and ignored the relatively abundant elements K and Fe. 
 For example, according to \cite{Ito+2015}, 
 the abundances of K and Fe 
 are similar to each other and are about twice as 
large as that
 of Mg which is about one thirtieth
 of the Na abundance for the substellar-point equilibrium temperate of 3000~K.
 
\begin{figure}
   \begin{center}
  \includegraphics[width=\columnwidth]{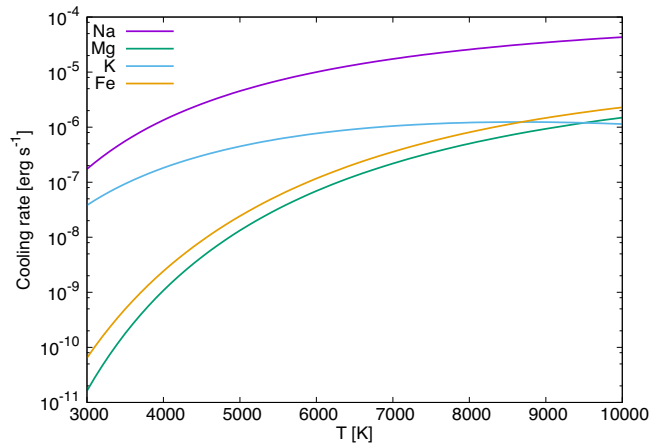}
   \end{center}
   \caption{Cooling rates of Na (violet), Mg (green), K (light blue) and Fe (yellow) in a optically thin and local thermal equilibrium (LTE)  condition with different temperatures. Note that, based on the molar fraction difference between Na and the others  in the mineral atmosphere shown in Fig,~\ref{fig:HSE_sub}, 
   the cooling rate of Na is represented as it of one molecule but the others are
   divided by thirty for Mg and divided by fifteen for K and Fe.
}
 \label{fig:kfe}
 \end{figure}

In Fig.~\ref{fig:kfe}, we compare
 the cooling rates of Na (violet), Mg (green), K (light blue) and Fe (yellow), assuming the collisional equilibrium 
 (i.e., optically thin and LTE conditions). We have calculated
 those of K and Fe 
 using their line properties shown in NIST Database\footnote{\url{https://www.nist.gov/pml/atomic-spectra-database}} and their partition functions \citep{Irwin81}. 
 The cooling rates of K and Fe are found to be lower
 than that of Na but higher
 than that of Mg. This indicates that addition of K and Fe has little effect on
 the mass loss rate of the Na-containing mineral atmosphere.  
 On the other hand, Fe may contribute to cooling in the atmosphere to a certain extent,
 if Na and K are removed from the atmosphere via photo-evaporation (see Sec.~\ref{ssec: inl}). 
 However, given that the mass loss rate of the Na-free atmosphere  
 is only slightly different from that of
 the Na-containing atmosphere (see Sec.~\ref{ssec: inl}), Fe would have no large effect on the mass loss rate, since the difference in collisional-equilibrium cooling rate between Fe and Mg is much smaller than that between Na and Mg  (see Fig.~\ref{fig:kfe}). 
 
 \subsubsection{Sensitivity 
 to rate and diffusion coefficients}
 \label{sssec: st}
 This model includes
 various processes such as photo-ionization heating, non-LTE atomic line cooling and diffusion. 
 Although  
 the rate coefficients 
 of some gas species in the mineral atmosphere have not been experimentally measured, we have used 
 the values from open databases and published literature if available, but from classic theories if not. 
 Below, we discuss the sensitivity of the results to the rate and diffusion coefficients.

For the binary diffusion coefficients, 
the values derived based on the classic theory \citep{Burgers1969,Schunk+2000} may differ from their actual values. However, as a sensitivity test, we have performed the Case~A simulations with
the diffusion coefficients multiplied by 10 or 100 and then found that
the fraction of the lightest element O increases by at most 2~\% or 15~\%, respectively, throughout the atmosphere. 
The reason for such small differences is that
the bulk velocity of gas is much higher than the diffusion velocity of each species, as mentioned in Sec.~\ref{sssec: as}. 
Thus, these coefficients would not be important for the gravitational separation, as long as the actual values differ 
by a factor of up to $\sim$ 100 from those considered in this model.

Also, we have performed the Case~A simulations by artificially lowering thermal reaction rates by a factor of 10, which has resulted in only a small difference in the mass loss rate (at most 1~\%).
In the atmosphere, the lowered thermal reaction rates reduce the heat generation rate, $Q_{\rm{chem}}$, 
leading to an increase in temperature at $r$ = 1.0--1.2~$R_p$ only by about 200~K.  
Then, the radiative cooling rates, $Q_{\rm{rad}}$, increases due to the higher temperature.
As a consequence, the net X+UV energy deposition rate and mass loss rate are hardly changed because the increase in $Q_{\rm{rad}}$ compensates for the decrease in the heat generation rate, $Q_{\rm{chem}}$.
On the other hand, the values of rate coefficients related to X+UV heating and non-LTE cooling are relatively important because no other processes can compensate for any change in those rates.
When the collisional transition rates are reduced artificially by ten or photo-ionization cross sections are reduced by two in Case~A, the mass loss rates respectively increase or decrease by about a factor of two. 
Although the cross sections of some neutral atoms considered in this model agree very well with experimental data \citep[see][]{Verner+1995}, many of the cross sections and collisional transition rates have not been well studied experimentally. 
Future studies of such key processes will be quite helpful.

 \subsection{Implication for observations}\label{ssec: imp}
This study has revealed that the hydrodynamic escape of the mineral atmosphere on a 1$M_\oplus$ HRE is massive enough to remove  Na and K from the atmosphere and interior (see also Section~\ref{ssec: inl}), although its flux is too weak to change the bulk planet mass and composition.
According to \citet{Ito+2015}, emission spectra of the mineral atmosphere contain prominent features due to Na and K at $\sim$0.6 and 0.8~$\mu$m, respectively, and due to SiO at 4 and 10~$\mu$m.
Thus, detection of SiO and non-detection of Na and K would yield a piece of the evidence that the HRE has experienced the escape of the mineral atmosphere.

Transit observation in the UV is useful for detection and characterization of escaping planetary atmospheres \citep[e.g.,][]{Vidal-Madjar+2003,Ehrenreich+2015}.
Our models have shown that the major components in the upper part of the flow of the mineral atmosphere are multiply-charged ions produced via X+UV photo-ionization. 
Since those ions have strong absorption power in the UV (e.g., the Einstein coefficient for the transition between the 3s and the 3p level of Si$^{3+}$ at $\sim$ 139nm $\sim$ 8 $\times10^8$ s$^{-1}$), such expanding mineral atmospheres possibly bring about large transit depths in the UV than in the optical.
Indeed,
\citet{Bourrier+2018} reported on the detection of variations in line emission intensity in the FUV from the G-type star 55 Cnc including O, C$^{+}$, C$^{2+}$, N$^{4+}$, Si$^{+}$, Si$^{2+}$ and Si$^{3+}$.
Such an observation might have detected the escape of the mineral atmosphere because a relatively high-density super-Earth, 55 Cnc e, which has a mass of 8.09~$\Mearth$ and a radius of 1.99~$\Rearth$ \citep{Dragomir+2014,Nelson+2014}, orbits very close to ($0.015$~AU) the host star and would have a molten rocky surface \citep{Demory+2016}.
Although the observed variations of 55 Cnc are too complex to conclude that they originate purely from 55 Cnc e or from the host-star, our study suggests that the variations of the Si ions' lines could be induced by the escaping Si ions from the planet.

\subsection{Mass loss of hot rocky planets}
We have found that the mass loss rate of the mineral atmosphere $\dot{m}$ is as low as 3.7$\times10^{-2}$--3.0$\times10^{-1}$ $\Mearth$/Gyr in Section~\ref{sec: r}.
This indicates that $1 M_\oplus$ HREs  hardly lose their mass via the atmospheric escape.
Below is a rough estimation of the lost mass.

Given the observational fact that the flux of X-ray and UV from solar-type stars is constant until the stellar age is $\sim$ 0.1~Gyr ($\equiv t_0$) and, then, decreases with stellar age \citep{Ribas+2005},
we assume that $\dot{m}$ (t) = $\dot{m}_0$ for $t \leq t_0$ and $\dot{m}_0 \, (t/t_0)^{-\alpha}$ for $t > t_0$ with a constant power index $\alpha$. 
Integrating this equation, one obtains 
the total mass lost in $t$ as 
\begin{eqnarray}
M_{\rm{esc}} (t) =  \frac{1}{4} \dot{m_0} t_0 \left[ 1+ \frac{(t/t_0)^{1-\alpha}-1} {1-\alpha}  \right],
\label{eq: tml}
\end{eqnarray}
where 
the $1/4$ is the geometric reduction factor of $\dot{m}$ to account for a difference in the received X+UV cross-sectional area over the planet's surface.
From Cases A and B, $\alpha \sim 0.9$.
By use of $\alpha=0.9$, $t_0=0.1$ Gyr and $\dot{m_0}=3.0\times10^{-1}$ $\Mearth$ Gyr$^{-1}$ in Eq. ~(\ref{eq: tml}), 
it is estimated that $M_{\rm esc}$~(10~Gyr) $\sim 5\times10^{-2} \Mearth$ for HREs, 
which is much smaller than $1 M_\oplus$.
Note that Na depletion would yield a slight increase in $M_{\rm esc}$ (see Sec.~\ref{ssec: inl}).

As mentioned in Introduction, \citet{Valencia+10} estimated that 
CoRoT-7~b might have lost
a significant fraction of its initial mass through a massive escape of the mineral atmosphere, base on an energy-limited escape approximation with the X+UV heating efficiency of 0.4 (see their Fig.~6). Our finding from the detailed heating/cooling calculations is, however,
that the X+UV heating efficiency, $\epsilon_{\rm X+UV}$, is as low as $10~^{-4}$--$10^{-3}$ for the mineral atmosphere considered in this study,
because of the efficient radiative cooling of the gas species such as Na and Mg. 
Although only a 1~$M_\oplus$ HRE is considered in this study, the values of $\epsilon_{\rm X+UV}$ are similar or less for super-Earths.
Such a low heating efficiency leads to reducing the lost mass by a few order of magnitude relative to the previous estimate. 
Thus, our results suggest that HREs survive in such strong stellar X+UV environments. 
This is consistent with the detection of several hundred close-in exoplanets whose radii are less than 2 Earth radii \citep[e.g.,][]{Fluton+2017}.

\section{Conclusions} \label{sec: c}
In this study we have developed a 1D hydrodynamic model of the highly UV-irradiated, mineral atmosphere on a 1$M_\oplus$ rocky planet covered with a magma ocean, including the effects of molecular diffusion, thermal conduction, photo-/thermo-chemistry, X-ray and UV heating, and radiative line cooling 
To detail the radiative cooling, which is key to understanding of the energy balance that controls photo-evaporation, we have adopted the formulae of the cooling rates based on the radiative/collisional transitions of gas species and the escape probably method for taking the effect of radiative transfer into account approximately.
Our results have demonstrated that alkali metals such as Na, alkaline-earth metals such as Mg, and ions with the same electron configurations as them act as strong coolants in the mineral atmosphere.
Thereby, almost all of the incident X-ray and UV energy from the host-star is converted into and lost by the radiative emission of the coolant gas species. 
We have estimated the net X+UV heating efficiency $\epsilon_\mathrm{X+UV}$ to be on the order of $10~^{-4}$--$10^{-3}$.
Thus, we conclude that the photo-evaporation of the mineral atmospheres on 1$M_\oplus$ HREs is not intense enough to exert a great influence on the mass and bulk composition but efficient enough to remove sodium completely. 
In this paper we have focused on describing our methodology. Future experimental works for determining rocky vapor properties such as collisional transition rates and photo-ionization cross sections will be quite helpful for improving the hydrodynamic escape models of mineral atmospheres.
 We will explore the dependence on gravity and other parameters in our forthcoming paper.

\section*{Acknowledgements}
We appreciate Antonio Garc{\'{\i}}a-Mu{\~n}oz for giving us fruitful comments about numerical schemes for fluid dynamics and also for providing data from his simulations. 
We also appreciate the referee, Laura Schaefer, for her careful reading and valuable comments which helped to improve this paper greatly.
We also thank Naoki Terada and Kanako Seki for helpful discussions.
This work was supported by JSPS KAKENHI JP18H05439 and JP18H01265
and JSPS Core-to-core Program “International Network of Planetary Sciences (Planet$^2$).”

     \begin{table*}
  \caption{Atomic chemistry in mineral atmosphere}
    \label{tbl:CHEMT} 
    \begin{tabular}{ l l r l r r } 
    \hline \hline
    Reaction &\ &\ &\ & Rate coefficient & Ref.  \\ 
    \hline \hline
        PI1&Na + $h\nu$ &$\to$ & Na$^+$ \ +\ e & -  & A, B \\ 
   PI2& & $\to$ & Na$^{2+}$ +\ 2e & - & A, B \\
    PI3&  & $\to$ & Na$^{3+}$ +\ 3e   & - & A, B \\ 
    PI4&Na$^+$ + $h\nu$ &$\to$ & Na$^{2+}$ +\ e & -  & A, B \\ 
   PI5& & $\to$ & Na$^{3+}$ +\ 2e & - & A, B \\
    PI6&Na$^{2+}$ + $h\nu$ &$\to$ & Na$^{3+}$ +\ e & -  & A, B \\ 
   PI7& & $\to$ & Na$^{4+}$ +\ 2e & - & A, B \\
     PI8& Na$^{3+}$ + $h\nu$ & $\to$ & Na$^{4+}$ +\ e & - & A, B \\
    PI9&O + $h\nu$ &$\to$ & O$^+$ \ +\ e & -  & A, B \\ 
   PI10& & $\to$ & O$^{2+}$ +\ 2e & - & A, B \\
    PI11&  & $\to$ & O$^{3+}$ +\ 3e   & - & A, B \\ 
    PI12&O$^+$ + $h\nu$ &$\to$ & O$^{2+}$ +\ e & -  & A, B \\ 
   PI13& & $\to$ & O$^{3+}$ +\ 2e & - & A, B \\
    PI14&O$^{2+}$ + $h\nu$ &$\to$ & O$^{3+}$ +\ e & -  & A, B \\ 
   PI15& & $\to$ & O$^{4+}$ +\ 2e & - & A, B \\
      PI16& O$^{3+}$ + $h\nu$ & $\to$ & O$^{4+}$ +\ e & - & A, B \\
          PI17&Mg + $h\nu$ &$\to$ & Mg$^+$ \ +\ e & -  & A, B \\ 
   PI18& & $\to$ & Mg$^{2+}$ +\ 2e & - & A, B \\
    PI19&  & $\to$ & Mg$^{3+}$ +\ 3e   & - & A, B \\ 
        PI20&  & $\to$ & Mg$^{4+}$ +\ 4e   & - & A, B \\ 
    PI21&Mg$^+$ + $h\nu$ &$\to$ & Mg$^{2+}$ +\ e & -  & A, B \\ 
   PI22& & $\to$ & Mg$^{3+}$ +\ 2e & - & A, B \\
      PI23& & $\to$ & Mg$^{4+}$ +\ 3e & - & A, B \\
    PI24&Mg$^{2+}$ + $h\nu$ &$\to$ & Mg$^{3+}$ +\ e & -  & A, B \\ 
   PI25& & $\to$ & Mg$^{4+}$ +\ 2e & - & A, B \\
      PI26& Mg$^{3+}$ + $h\nu$ & $\to$ & Mg$^{4+}$ +\ e & - & A, B \\
                PI27&Si + $h\nu$ &$\to$ & Si$^+$ \ +\ e & -  & A, B \\ 
   PI28& & $\to$ & Si$^{2+}$ +\ 2e & - & A, B \\
    PI29&  & $\to$ & Si$^{3+}$ +\ 3e   & - & A, B \\ 
        PI30&  & $\to$ & Si$^{4+}$ +\ 4e   & - & A, B \\ 
    PI31&Si$^+$ + $h\nu$ &$\to$ & Si$^{2+}$ +\ e & -  & A, B \\ 
   PI32& & $\to$ & Si$^{3+}$ +\ 2e & - & A, B \\
      PI33& & $\to$ & Si$^{4+}$ +\ 3e & - & A, B \\
    PI34&Si$^{2+}$ + $h\nu$ &$\to$ & Si$^{3+}$ +\ e & -  & A, B \\ 
   PI35& & $\to$ & Si$^{4+}$ +\ 2e & - & A, B \\
      PI36& Si$^{3+}$ + $h\nu$ & $\to$ & Si$^{4+}$ +\ e & - & A, B \\
      TI1 &Na +\ e &$\to$ & Na$^+$ \ +\ 2e & 1.01$\times10^{-7}\left(\frac{1+U^{1/2}}{0.275+U}\right)U^{0.23}\exp(-U),\ 
       U=5.1/T\rm{[eV]}$ & C\\
      TI2&Na$^+$ +\  e &$\to$ & Na$^{2+}$ +\ 2e & 7.35$\times10^{-9}\left(\frac{1+U^{1/2}}{0.056+U}\right)U^{0.35}\exp(-U),\ 
       U=47.3/T\rm{[eV]}$ & C \\ 
      TI3&Na$^{2+}$ +\  e &$\to$ & Na$^{3+}$ +\ 2e & 8.1$\times10^{-9}\left(\frac{1+U^{1/2}}{0.148+U}\right)U^{0.32}\exp(-U),\ 
       U=71.6/T\rm{[eV]}$ & C \\ 
      TI4&Na$^{3+}$ +\  e &$\to$ & Na$^{4+}$ +\ 2e & 1.14$\times10^{-8}\left(\frac{1}{0.553+U}\right)U^{0.28}\exp(-U),\ 
       U=98.9/T\rm{[eV]}$ & C \\ 
       
        TI5 &O +\ e &$\to$ & O$^+$ \ +\ 2e & 3.59$\times10^{-8}\left(\frac{1}{0.073+U}\right)U^{0.34}\exp(-U),\ 
       U=13.6/T\rm{[eV]}$ & C\\
      TI6&O$^+$ +\  e &$\to$ & O$^{2+}$ +\ 2e & 1.39$\times10^{-8}\left(\frac{1+U^{1/2}}{0.212+U}\right)U^{0.22}\exp(-U),\ 
       U=35.1/T\rm{[eV]}$   & C \\ 
      TI7&O$^{2+}$ +\  e &$\to$ & O$^{3+}$ +\ 2e & 9.31$\times10^{-9}\left(\frac{1+U^{1/2}}{0.27+U}\right)U^{0.27}\exp(-U),\ 
       U=54.9/T\rm{[eV]}$   & C \\ 
      TI8&O$^{3+}$ +\  e &$\to$ & O$^{4+}$ +\ 2e & 1.02$\times10^{-8}\left(\frac{1}{0.614+U}\right)U^{0.27}\exp(-U),\ 
       U=77.4/T\rm{[eV]}$   & C \\ 
       
       TI9 &Mg +\ e &$\to$ & Mg$^+$ \ +\ 2e & 6.21$\times10^{-7}\left(\frac{1}{0.592+U}\right)U^{0.39}\exp(-U),\ 
       U=7.6/T\rm{[eV]}$ & C\\
      TI10&Mg$^+$ +\  e &$\to$ & Mg$^{2+}$ +\ 2e & 1.92$\times10^{-8}\left(\frac{1}{0.0027+U}\right)U^{0.85}\exp(-U),\ 
       U=15.2/T\rm{[eV]}$   & C \\ 
      TI11&Mg$^{2+}$ +\  e &$\to$ & Mg$^{3+}$ +\ 2e & 5.56$\times10^{-9}\left(\frac{1+U^{1/2}}{0.107+U}\right)U^{0.30}\exp(-U),\ 
       U=80.1/T\rm{[eV]}$   & C \\ 
      TI12&Mg$^{3+}$ +\  e &$\to$ & Mg$^{4+}$ +\ 2e & 4.35$\times10^{-9}\left(\frac{1+U^{1/2}}{0.159+U}\right)U^{0.31}\exp(-U),\ 
       U=109.3/T\rm{[eV]}$   & C \\ 
       
        TI13 &Si +\ e &$\to$ & Si$^+$ \ +\ 2e & 1.88$\times10^{-7}\left(\frac{1+U^{1/2}}{0.376+U}\right)U^{0.25}\exp(-U),\ 
       U=8.2/T\rm{[eV]}$ & C\\
      TI14&Si$^+$ +\  e &$\to$ & Si$^{2+}$ +\ 2e & 6.43$\times10^{-8}\left(\frac{1+U^{1/2}}{0.632+U}\right)U^{0.20}\exp(-U),\ 
       U=16.4/T\rm{[eV]}$   & C \\ 
      TI15&Si$^{2+}$ +\  e &$\to$ & Si$^{3+}$ +\ 2e & 2.01$\times10^{-8}\left(\frac{1+U^{1/2}}{0.473+U}\right)U^{0.22}\exp(-U),\ 
       U=33.5/T\rm{[eV]}$   & C \\ 
      TI16&Si$^{3+}$ +\  e &$\to$ & Si$^{4+}$ +\ 2e & 4.94$\times10^{-9}\left(\frac{1+U^{1/2}}{0.172+U}\right)U^{0.23}\exp(-U),\ 
       U=54.0/T\rm{[eV]}$   & C \\ 
        RR1 &Na$^+$ +\ e &$\to$ & Na\  + $h\nu$ & \small{$f(5.095\times10^{-12},  0,  3.546\times10^{2},  2.310\times10^{6},  0.9395,  4.297\times10^{5}) $}& D\\
      RR2&Na$^{2+}$ +\  e &$\to$ & Na$^{+}$ + $h\nu$ &  \small{$f(5.176\times10^{-11},  0.4811,  7.751\times10^{1},  1.351\times10^{7},  0.3467,  3.140\times10^{5})$ }& D \\ 
          \end{tabular}
  \end{table*}  
    \begin{table*}
  \contcaption{Atomic chemistry in mineral atmosphere}
    \begin{tabular}{ l l r l r r } 
    \hline \hline
    Reaction &\ &\ &\ & Rate coefficient & Ref.  \\ 
    \hline \hline
      RR3&Na$^{3+}$ +\  e &$\to$ & Na$^{2+}$   + $h\nu$ & \small{$f(3.192\times10^{-10},  0.6726,  1.640\times10^{1},  1.263\times10^{7},  0.1232,  3.725\times10^{5})$  }& D \\ 
      RR4&Na$^{4+}$ +\  e &$\to$ & Na$^{3+}$  + $h\nu$& \small{$f(1.087\times10^{-9},  0.7284,  6.132,  1.088\times10^{7},  0.0629,  4.559\times10^{5})$  }& D \\ 
      RR5 &O$^+$ +\ e &$\to$ & O\  + $h\nu$ & \small{$f(6.622\times 10^{-11}, 0.6109, 4.136, 4.214\times 10^{6}, 0.4093, 8.770\times 10^{4}) $}& D\\
      RR6&O$^{2+}$ +\  e &$\to$ & O$^{+}$ + $h\nu$ & \small{$f(2.096\times 10^{-9}, 0.7668, 1.602\times 10^{-1}, 4.377\times 10^{6}, 0.1070, 1.392\times 10^{5})$ }& D \\ 
      RR7&O$^{3+}$ +\  e &$\to$ & O$^{2+}$   + $h\nu$ & \small{$f(2.501\times 10^{-9}, 0.7844, 5.235\times 10^{-1}, 4.470\times 10^{6}, 0.0447, 1.642\times 10^{5})$  }& D \\ 
      RR8&O$^{4+}$ +\  e &$\to$ & O$^{3+}$  + $h\nu$& \small{$f(3.955\times 10^{-9}, 0.7813, 6.821\times 10^{-1}, 5.076\times 10^{6}, 0,  0)$  }& D \\ 
      RR9 &Mg$^+$ +\ e &$\to$ & Mg\  + $h\nu$ & \small{$f(5.452\times 10^{-11}, 0.6845, 5.637, 1.551\times 10^{6}, 0.3945, 8.360\times 10^{5}) $}& D\\
      RR10&Mg$^{2+}$ +\  e &$\to$ & Mg$^{+}$ + $h\nu$ & \small{$f(1.345\times 10^{-11}, 0.1074, 7.877\times 10^{2}, 7.925\times 10^{7}, 0.4631, 5.027\times 10^{5})$ }& D \\ 
      RR11&Mg$^{3+}$ +\  e &$\to$ & Mg$^{2+}$   + $h\nu$ & \small{$f(1.249\times 10^{-10}, 0.5600, 7.748\times 10^{1}, 2.015\times 10^{7}, 0.1917, 5.139\times 10^{5})$  }& D \\ 
      RR12&Mg$^{4+}$ +\  e &$\to$ & Mg$^{3+}$  + $h\nu$& \small{$f(4.031\times 10^{-10}, 0.6803, 3.205\times 10^{1}, 1.626\times 10^{7}, 0.0764, 5.399\times 10^{5})$  }& D \\ 
      RR13 &Si$^+$ +\ e &$\to$ & Si\  + $h\nu$ & $5.9\times10^{-13}\left( {T{\rm{[K]}}}/{10^4}\right)^{-0.601} $& E\\
      RR14&Si$^{2+}$ +\  e &$\to$ & Si$^{+}$ + $h\nu$ &  $1\times10^{-12}\left( {T{\rm{[K]}}}/{10^4}\right)^{-0.786} $ & E \\ 
      RR15&Si$^{3+}$ +\  e &$\to$ & Si$^{2+}$   + $h\nu$ & \small{$f(6.739\times 10^{-11}, 0.4931, 2.166\times 10^{2}, 4.491\times 10^{7}, 0.1667, 9.046\times 10^{5})$  }& D \\ 
      RR16&Si$^{4+}$ +\  e &$\to$ & Si$^{3+}$  + $h\nu$& \small{$f(5.134\times 10^{-11}, 0.3678, 1.009\times 10^{3}, 8.514\times 10^{7}, 0.1646, 1.084\times 10^{6})$  }& D \\ 
    \hline
    \end{tabular}
    \\
  \begin{flushleft}  
  Notes.  Rate coefficients in cgs units.  References are  A, \citet{Kaastra+1993}; B, \citet{Verner+1995}; C, \citet{Voronov1997};D, \citet{Badnell2006}; E, \citet{Aldrovandi+1973}. The function of the rate coefficient from \citet{Badnell2006} is given by $f(x_1, x_2, x_3, x_4, x_5, x_6)=x_1[\sqrt{T/x_3} (1+\sqrt{T/x_3} )^{1-Y} (1+\sqrt{T/x_4} )^{1+Y}]^{-1}, Y=x_2+x_5\exp(-x_6/T)$.  And, based on the Saha ionization equation, the rate of thermal recombination is calculated.
  \end{flushleft}
  \end{table*}

      \begin{table}
          \caption{Energy levels of atoms}
    \label{tbl:ELG} 
  \begin{tabular}{ l r c c r } 
    \hline \hline
    Spices & Configuration & Statistical weight  & Excitation energy [cm$^{-1}$]  \\ 
    \hline \hline
    Na & 3s & 2 & 0  \\ 
     & 3p & 6 & 17106.03  \\ 
      Na$^{3+}$    & $2p^4(3P)$ & 9 & 540.8122  \\ 
      & $2p^4(1D)$  & 5 & 31105.89  \\ 
      & $2p^4(1S)$  & 1 & 65514.89  \\ 
     O &   $2p^4(3P)$  &9  & 76.83111 \\ 
      &  $2p^4(1D)$  & 5 & 15868.34 \\ 
      &  $2p^4(1S)$  & 1 & 33792.22  \\ 
      & $2p^33s(5S^\circ)$ & 5 & 73767.79 \\ 
      & $2p^33s(3S^\circ)$ &3  & 76795.82 \\ 
       O$^{+}$& $2p^3(4S^\circ)$ & 4 & 0 \\ 
      & $2p^3(2D^\circ)$ & 10 & 26818.61 \\ 
      & $2P^3(2P^\circ)$ & 6 & 40469.22 \\ 
      O$^{2+}$& $2s^22p^2(3P)$ & 9 & 207.46 \\ 
      & $2s^22p^2(1D)$ & 5 & 20273.11 \\ 
      & $2s^22p^2(1S)$ & 1 & 43186.75  \\ 
      & $2s^12p^3(5S^\circ)$ & 5 & 60324.71 \\ 
      O$^{3+}$&  $2p^2(2P^\circ)$ & 6 & 258.14 \\ 
      & $2p^2(4P)$ & 12 & 71724.05 \\ 
      Mg& $3s^2$ & 1 & 0 \\ 
      & $3s3p(3P^\circ)$ & 9 & 21890.85 \\ 
      & $3s3p(1P^\circ)$ & 3 & 35052.25 \\ 
      Mg$^+$& 3s & 2 & 0 \\ 
      & 3p & 6 & 35863.03 \\
       
      Si& $3s^23p^2(3P)$ & 9 & 125.54 \\ 
      & $3s^23p^2(1D)$ & 5 & 6386.84 \\ 
      & $3s^23p(1S)$ & 1 & 15535.64 \\ 
      & $3s3p^3$ & 5 & 32236.95 \\ 
      & $3s^23p4s(3P^\circ)$ & 9 & 39890.38 \\ 
      
      Si$^+$& $3s^23p$ &  6& 186.65 \\ 
      & $3s3p^2(4P)$ & 12 & 41817.71 \\ 
      & $3s3p^2(2D)$ & 10 & 54540.73 \\ 
      
      Si$^{2+}$& $3s^2$ & 1 &  0 \\ 
      & $3s3p(3P^\circ)$ & 9 & 52985.88  \\ 
      & $3s3p(1P^\circ)$ & 3 &  82885.54 \\ 
      
      Si$^{3+}$& $3s$ &2  & 0 \\ 
      & $3p$ & 6 & 71693.39 \\ 

    \hline
  \end{tabular}  
  \begin{flushleft}
Those data are calculated from the dataset in MCHF/MCDHF database (\url{http://nlte.nist.gov/MCHF/}).
For grouping the fine-structure levels shown in MCHF/MCDHF database, we sum the statistical weight, and average the excitation energy by weighting the statistical weight of each level.
\end{flushleft}
    \end{table}

      \begin{table*}
          \caption{Radiative and collisional transition between levels}
    \label{tbl:RCT} 
  \begin{tabular}{ l r c c r } 
    \hline \hline
    Spices & Transition & Einstein coefficient [1/s] & Effective collision strength & Ref.  \\ 
    \hline \hline
    Na & $3s$ - $3p$ & 6.23$\times10^7$ & 12.1 & A*  \\ 
    
     Na$^{3+}$ & $2p^4(3P)$ - $2p^4(1D)$ & 0.814 & 1.09 &B  \\ 
       & $2p^4(3P)$ - $2p^4(1S)$  &  7.30 & 0.178 & B  \\ 
        & $2p^4(1D)$ - $2p^4(1S)$  &  3.25 & 0.211 & B  \\ 
        
      O  & $2p^4(3P)$ - $2p^4(1D)$ & $8.57\times10^{-3}$ & 0.293 & C \\ 
        & $2p^4(3P)$ - $2p^4(1S)$  & $7.87\times10^{-2}$ & 3.23$\times10^{-2}$ & C  \\ 
        & $2p^4(3P)$ - $2p^33s(5S^\circ)$ & $1.84\times10^{3}$ & 0.232 & C  \\ 
        & $2p^4(3P)$ - $2p^33s(3S^\circ)$ & $5.64\times10^{8}$ & 0.353 & C  \\ 
        & $2p^4(1D)$ - $2p^4(1S)$  & $1.26$ & 8.83$\times10^{-2}$  &  C \\ 
        &  $2p^4(1D)$ - $2p^33s(5S^\circ)$ & $1.36$ & 0.05 & **  \\ 
        & $2p^4(1D)$ - $2p^33s(3S^\circ)$ & $1.75\times10^{3}$ & 8.23$\times10^{-4}$  & C  \\ 
        &  $2p^4(1S)$ - $2p^33s(5S^\circ)$& $ 0 $ & 0.05 & **  \\ 
        & $2p^4(1S)$ - $2p^33s(3S^\circ)$ & $6.2\times10^{-2}$ & 0.05 &**\\ 
        & $2p^33s(5S^\circ)$ - $2p^33s(3S^\circ)$ & $ 0 $ & 0.05 & **  \\ 
        
        O$^{+}$& $2p^3(4S^\circ)$ - $2p^3(2D^\circ)$ & 7.68$\times10^{-5}$ & 1.33  & D  \\ 
        & $2p^3(4S^\circ)$ - $2p^3(2P^\circ)$ & 4.51$\times10^{-2}$ & 0.406 & D  \\ 
        & $2p^3(2D^\circ)$ - $2p^3(2P^\circ)$ & 9.68$\times10^{-2}$ & 1.70 & D \\ 
        
        O$^{2+}$ & $2s^22p^2(3P)$ - $2s^22p^2(1D)$ & 2.71$\times10^{-2}$ & 2.28 & E  \\ 
        & $2s^22p^2(3P)$ - $2s^22p^2(1S)$  & 2.25$\times10^{-1}$ & 0.292 & E  \\ 
        & $2s^22p^2(3P)$ - $2s^12p^3(5S^\circ)$ & 8.07$\times10^{2}$ & 1.20 & E  \\ 
        
        & $2s^22p^2(1D)$ - $2s^22p^2(1S)$ & $1.68$ & 0.581 & E \\ 
        & $2s^22p^2(1D)$ - $2s^12p^3(5S^\circ)$ & 5.77$\times10^{-3}$ & 0.05 & **  \\ 
        
        & $2s^22p^2(1S)$ - $2s^12p^3(5S^\circ)$ & 3.76$\times10^{-11}$ & 0.05 & **  \\ 
        
          O$^{3+}$      & $2p^2(2P^\circ)$ - $2p^2(4P)$ & 1.20$\times10^{2}$ & 1.12 & F  \\ 
          
        Mg& $3s^2$ - $3s3p(3P^\circ)$  & 8.45$\times10^{1}$ & 2.97 & G\\ 
        & $3s^2$ - $3s3p(3P^\circ)$  & 4.66$\times10^{8}$ & 1.47 &   G \\ 
                & $3s3p(3P^\circ)$ - $3s3p(1P^\circ)$  & 0 & 1.98 &  G  \\ 
        
        Mg$^+$& 3s-3p & 2.59$\times10^{8}$ & 17.5 & H  \\ 
        
        Si& $3s^23p^2(3P)$ - $3s^23p^2(1D)$ & 2.53$\times10^{-3}$ & 0.478 & I  \\         
        & $3s^23p^2(3P)$ - $3s^23p(1S)$ & 2.49$\times10^{-2}$ & 3.79$\times10^{-3}$ & I  \\ 
       & $3s^23p^2(3P)$ - $3s3p^3$  & 7.46$\times10^{2}$ & 1.76$\times10^{-6}$ & I  \\         
        & $3s^23p^2(3P)$ - $3s^23p4s(3P^\circ)$ & 2.30$\times10^{8}$ & 1.99 & I  \\ 
        & $3s^23p^2(1D)$ - $3s^23p(1S)$ & 8.88$\times10^{-1}$ & 2.65$\times10^{-2}$ & I  \\         
        & $3s^23p^2(1D)$ - $3s3p^3$  & 3.42$\times10^{-2}$ & 2.37$\times10^{-3}$ & I  \\ 
        & $3s^23p^2(1D)$ - $3s^23p4s(3P^\circ)$ & 4.85$\times10^{5}$ & 1.53$\times10^{-2}$ & I  \\         
        & $3s^23p(1S)$ - $3s3p^3$  & 2.16$\times10^{-11}$ & 1.1$\times10^{-2}$ & I  \\ 
        & $3s^23p(1S)$ - $3s^23p4s(3P^\circ)$ & 2.00$\times10^{4}$ & 3.27$\times10^{-3}$ & I  \\         
        &  $3s3p^3$ - $3s^23p4s(3P^\circ)$ & 0 & 0.768 & I \\ 
        
        Si$^+$&  $3s^23p$ - $3s3p^2(4P)$& 2.81$\times10^{3}$ & 5.63 & J  \\         
        & $3s^23p$ - $3s3p^2(2D)$& 3.04$\times10^{6}$ & 17.4 & J  \\ 
        & $3s3p^2(4P)$ - $3s3p^2(2D)$& 1.50$\times10^{-2}$ & 10.7 & J  \\ 
        
        Si$^2+$&$3s^2$ - $3s3p(3P^\circ)$& 5.82$\times10^{3}$ & 5.45 & K  \\         
        &$3s^2$ - $3s3p(1P^\circ)$ & 2.45$\times10^{9}$ & 5.59 & K  \\ 
        &$3s3p(3P^\circ)$ - $3s3p(1P^\circ)$ & 0 & 8.29 & K  \\         
        
        Si$^3+$&3s - 3p& 8.66$\times10^{8}$ & 15.6 & H  \\ 
    \hline
  \end{tabular}  
  \begin{flushleft}
  The values of Einstein coefficient are calculated from MCHF/MCDHF database (http://nlte.nist.gov/MCHF/).
References of effective collision strength are  A, \citet{Igenbergs+2008}; B, \citet{Bulter+1994}, C, \citet{Zatsarinny+2003}($10\times10^3$); D, \citet{Pradhan+1976}; E, \citet{Lennon+1994}; F, \citet{Liang+2012}; G, \citet{Merie+2015}, H, \citet{Liang+2009}; I, \citet{Cincunegui+2001}; J, \citet{Tayal+2008}; K, \citet{Dufton+1989}. 
The values of effective collision strength for O$^3+$ and Mg$^+$ are assumed as those at T[K]  $=8\times10^3$, and the others are assumed as those at T[K] $=10^4$.
\\
* Calcurated value of effective collision strength from the cross section shown in the reference.\\
**  Approximated values of effective collision strength as forbidden transition, following \citet{Allende+2003}.
\end{flushleft}
    \end{table*}

\section*{Data Availability}

The data underlying this article will be shared on reasonable request to the corresponding author.


\bibliographystyle{mnras}
\bibliography{ref} 

\bsp	
\label{lastpage}
\end{document}